\documentclass[%
 aip,
 amsmath,amssymb,
 reprint,%
]{revtex4-1}

\usepackage{graphicx}
\usepackage{dcolumn}
\usepackage{bm}

\usepackage[utf8]{inputenc}
\usepackage[T1]{fontenc}
\usepackage{mathptmx}
\usepackage{etoolbox}
\usepackage{xcolor}

\usepackage[table]{xcolor}
\newcommand{\yellowrow}{\rowcolor{purple!10}}
\newcommand{\bluerow}{\rowcolor{blue!10}}
\newcommand{\grayrow}{\rowcolor{gray!15}}
\newcommand{\redrow}{\rowcolor{orange!15}}

\makeatletter
\def\@email#1#2{%
 \endgroup
 \patchcmd{\titleblock@produce}
  {\frontmatter@RRAPformat}
  {\frontmatter@RRAPformat{\produce@RRAP{*#1\href{mailto:#2}{#2}}}\frontmatter@RRAPformat}
  {}{}
}%
\makeatother
\begin{document}

\preprint{AIP/123-QED}

\title[]{Longitudinal and Transverse Thermoelectric Properties in Topological Quantum
Materials}

\author{Jyoti Duhan}
\affiliation{Department of Physics, Indian Institute of Technology Kanpur, UP 208016, India}

\author{Anil Kumar B.M.}
\affiliation{Department of Chemistry, Birla Institute of Technology and Science, Pilani – Hyderabad Campus, Jawahar Nagar, Kapra Mandal, Hyderabad 500078, India}

\author{Satya N. Guin} 
\email{satyanarayan.g@hyderabad.bits-pilani.ac.in}
\affiliation{Department of Chemistry, Birla Institute of Technology and Science, Pilani – Hyderabad Campus, Jawahar Nagar, Kapra Mandal, Hyderabad 500078, India}

\author{Koushik Pal}
\email{koushik@iitk.ac.in}
\affiliation{Department of Physics, Indian Institute of Technology Kanpur, UP 208016, India}

\date{\today}

\begin{abstract}
Thermoelectric materials, which can be either topologically trivial or non-trivial, enable reversible inter-conversion of heat and electricity using a solid-state technology. Incidentally, few of the first experimentally discovered three-dimensional topological materials, namely, bismuth, antimony, tin and lead based chalcogenides (Bi$_2$Te$_3$, Bi$_2$Se$_3$, Sb$_2$Te$_3$, SnTe, Pb$_{x}$Sn$_{1-x}$Te), were long known as good thermoelectric materials. In these compounds, the heavy atoms which give rise to strong spin-orbit coupling often become essential for band inversion and non-trivial band topology. At the same time, they lead to low lattice thermal conductivity,  linearly dispersing Dirac-like bands induce high electrical mobility, and band-inverted electronic structure with multiple band extrema gives rise to high Seebeck coefficients, which are prerequisites for high-efficiency thermoelectrics. In materials, where the applied temperature gradient induces electrical voltage (i.e., Seebeck effect) in the same direction, they are termed as longitudinal thermoelectrics. On the other hand, when the generated electrical voltage and applied thermal current remain orthogonal to each other, the materials are termed as transverse thermoelectrics. The transverse response can be driven by an additional external magnetic field (ordinary Nernst effect), intrinsic remanent magnetization or non-zero Berry curvature of a material in the absence of an external field (anomalous Nernst effect). In the last few years, several topological quantum materials have been shown to exhibit giant ordinary  or anomalous  Nernst effect which is induced by the non-zero Berry curvature that arises from their non-trivial topology. Few examples in this category include NbP, Cd$_3$As$_2$,  Co$_3$Sn$_2$S$_2$, YMn$_6$Sn$_6$, Mn$_3$Sn,  which are topological semimetals. We discuss another class of quantum materials, known as goniopolar materials, that exhibits axis-dependent conduction polarity and hence transverse thermoelectric effects which is based on mechanisms of longitudinal thermoelectric properties. We termed such materials as goniopolar thermoelectrics and examples include topologically non-trivial materials such as WSi$_2$ and NaSn$_2$As$_2$. In this review, we cover both types of thermoelectricity – longitudinal and transverse – focusing on materials which possess non-trivial topological order. We make connections between their structures, topology and transport properties and suggest tunability of thermoelectric properties going beyond conventional mechanisms such as through electron-phonon interactions. Finally, we suggest possible applications of topological thermoelectrics in energy management devices, quantum science and technology.
\end{abstract}

\maketitle


\section{Introduction}
The thermoelectric (TE) effect was first discovered in 1821, which describes the direct conversion of heat to electric voltage, and vice versa. TE materials came into practical focus during the 1950s, however, significant advancements have been made over the past two decades, largely driven by the growing focus on sustainable energy resources \cite{bell2008cooling, snyder2008complex}. The efficiency of a TE material is characterized by the figure of merit, defined as $zT = S^2 \sigma T/ (\kappa_e + \kappa_l)$, where $S$ is the Seebeck coefficient, $\sigma$ is the electrical conductivity, $T$ is the temperature, $\kappa_e$ is the electrical thermal conductivity, and $\kappa_l$ is the lattice thermal conductivity. The overall  efficiency of a thermoelectric device  depends on the temperature difference of the TE materials, and is given by
\[\eta = \frac{T_h - T_c}{T_h} \cdot \frac{\sqrt{1 + zT} - 1}{\sqrt{1 + zT} + \frac{T_c}{T_h}}\]

Where, $T_h$  and $T_c$  are the temperatures of the hot and cold sides, respectively. The performance of TE materials has been improved using various strategies, such as band convergence \cite{haldar2025conduction}, nanostructuring \cite{biswas2011strained}, alloying \cite{slade2020contrasting}, chemical bond engineering \cite{pal2018bonding, pei2012band}, and various intrinsic approaches \cite{zhao2014ultralow, pal2021microscopic, jana2017intrinsic}. We note that all these approaches have led to considerable progress in longitudinal TEs. For many years, application-based research primarily focused on longitudinal TEs \cite{xu2017topological, perumal2016high, yan2022high, kim2017bridge, toriyama2024topological}. It is only in more recent years that transverse thermoelectrics have begun to attract considerable research interest \cite{feng2024transverse, pan2024topological, adachi2025fundamentals, kumar2023topological}.

In Section 2 of this review article, we briefly discuss the effects on which longitudinal and transverse TEs are based on, namely the Seebeck and Nernst effects. The Nernst effect is further subdivided into two categories: the ordinary Nernst effect and the anomalous Nernst effect. We present the complete theoretical framework underlying these effects. In section 3, we explain how and why the topological materials are good candidates for thermoelectric applications. This section covers topological insulators, Weyl semimetals, Dirac semimetals, and Nodal line semimetals. In addition, we discuss another emerging class of materials known as goniopolar materials, which exhibit axis-dependent conduction polarity. In Section 4, we discuss prominent examples of the longitudinal and transverse TEs. Although we have made every effort to include key studies, this review is not exhaustive and some important contributions may not have been covered. We also examine the influence of electron–phonon interactions (EPI) on thermoelectric performance, supported by selected examples in Section 5. The connection between topology and thermoelectric transport properties is discussed in Section 6. The TE performance of topologically non-trivial materials can be attributed to the topology-independent (i.e., conventional), topology-dependent factors, or a combination of both. These aspects are summarized in Section 6 with several material examples. Finally, we conclude the article by discussing future applications in quantum energy and information technologies. The overall aim of this review is to highlight TE research in topological materials and to contrast their transverse thermoelectric properties from the conventional longitudinal thermoelectric properties.


In addition to the Seebeck and Nernst effects, a variety of longitudinal and transverse phenomena can be defined. By establishing various relations between electric field, temperature gradient, heat and electrical current, and the magnetic field, R. Fieschi  et al. proposed that a total of 560 local thermomagnetic and galvanomagnetic effects are theoretically possible \cite{FIESCHI1954259, goupil2016thermodynamics}. All these effects are not experimentally measurable, and we will discuss six most physically significant phenomena,  as shown in Figs. 1(a) and 1(b).

When a magnetic field is applied in the direction perpendicular to the direction of electrical current, a transverse voltage, $V_H$ is observed. This effect is known as Hall effect \cite{hall1879new}. There is no temperature gradient in any direction. This phenomenon occurs due to the deflection of charge carriers by the the Lorentz force, causing them to accumulate in transverse direction, shown in Fig. 1(b). The sign of the Hall coefficient indicates the dominant charge carriers in a semiconductors, while the magnitude is related to the carrier concentrations. When two-dimensional systems subject to strong magnetic fields and low temperatures, the Hall conductivity of a metal can be quantized, leading to quantum Hall effect \cite{laughlin1981quantized}. On the other hand, the accumulation of spins can occur on the lateral surfaces of an electric current carrying sample in the absence of any magnetic field. This is known as the spin Hall effect \cite{dyakonov1971current}, which can be observed in non-magnetic materials having strong spin-orbit coupling (SOC). In magnetic materials, spontaneous magnetization and SOC, generate a Berry curvature in the momentum space, which acts as an effective magnetic field and gives rise to the anomalous Hall effect. \cite{karplus1954hall}. 


When an electric current is flowing through the junction connecting two different materials, heat will be absorbed at one junction (cooling) and released (heating) at the other. This creates a temperature gradient as shown in Fig. 1(b), known as the Peltier effect, discovered in 1834 \cite{peltier1834nouvelles}.
This effect is the basis for many modern-day thermoelectric refrigeration devices \cite{shilpa2023systematic}.


In the Ettingshausen effect, a transverse temperature gradient $(\nabla_yT)$ develops in a material when an electric current flows through it, perpendicular to a magnetic field \cite{paranjape1960theory}. The number of collision between the electrons increase due to accumulation along one side of the sample. This lead to heating of the material on one end. This is the transverse analogue of the longitudinal Peltier effect. Due to the flow of current through a resistive medium, irreversible Joule heating coexists with the Ettingshausen effect, and limits the efficiency of Ettingshausen coolers. Therefore, devices are designed to reduce Joule heating via geometrical optimization and cascading.


The Righi-Leduc effect is observed when a temperature gradient and a magnetic field are applied perpendicular to each other, producing a transverse temperature gradient, $\nabla_yT$ \cite{leduc1888modifications}. A transverse electric field is also observed along with the transverse temperature gradient. This is the thermal analog of the electric Hall effect. The Righi-Leduc effect was initially observed in materials like HgSe \cite{whitsett1961righi}, In \cite{stephan1972righi}, In-Pb alloys and Pb-In alloys \cite{sichel1971righi}. More recently, the anomalous and planar Righi-Leduc effects are observed in ferromagnets \cite{madon2016anomalous} and topological materials \cite{zhou2022anomalous}. The spin effects also significantly influence heat transport in ferromagnetic metals through the anomalous Righi-Leduc effect \cite{zhou2020identification}.

The Seebeck effect is reverse phenomenon of the Peltier effect and the Nernst effect can be seen as the transverse equivalent of the Seebeck effect in the presence of a magnetic field. Discussing every effect in detail is beyond the scope of this review article, and we will mainly focus on the Seebeck, Nernst, and anomalous Nernst effects.

\section{Thermoelectric Effects}

\begin{figure*}[t]
\centering
\includegraphics[width=1.0\textwidth]{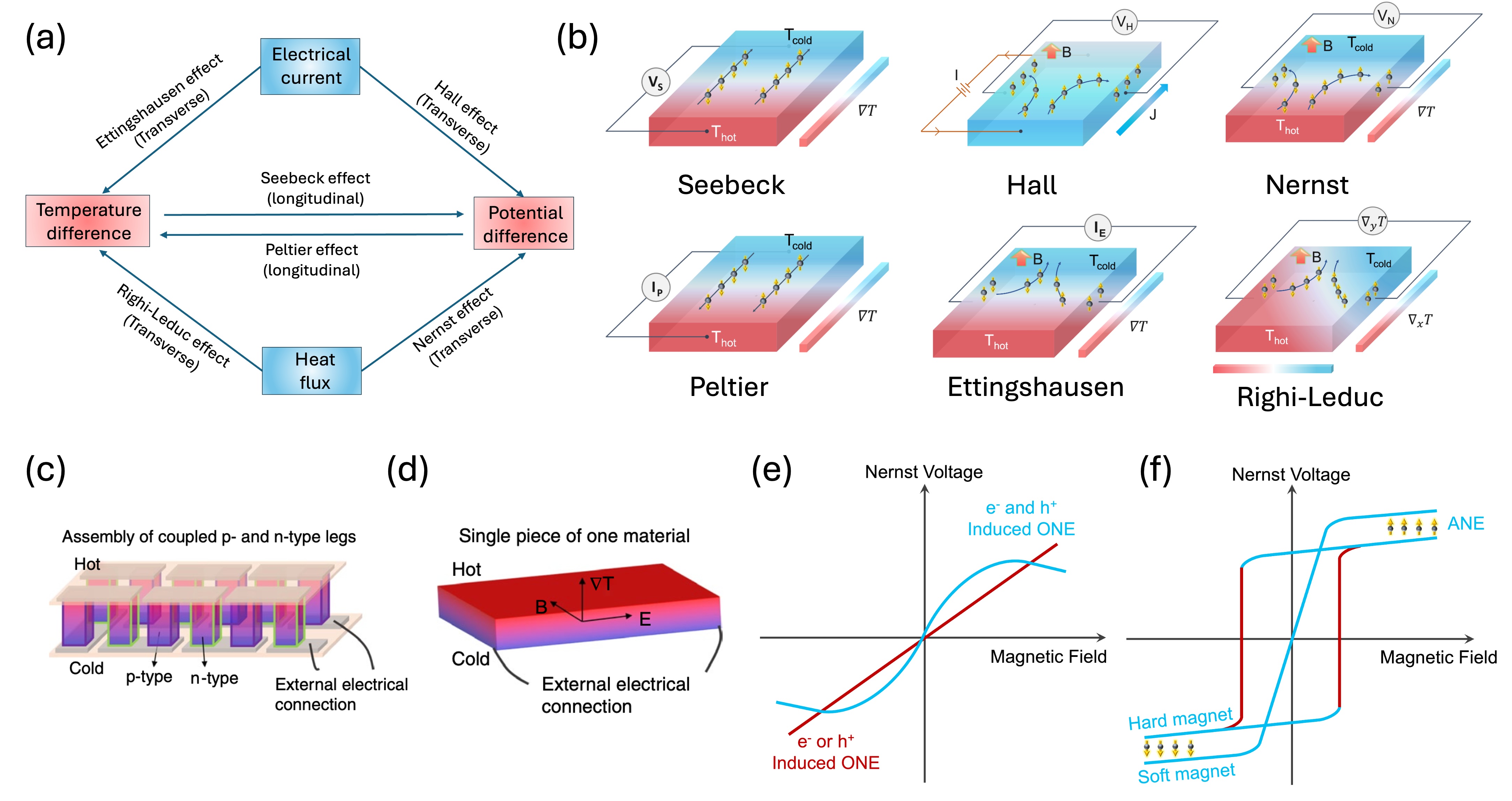}
\caption{(a) Schematic representation of the coupling between six major thermomagnetic and galvanomagnetic effects. (b) Schematic showing the flow of charges according to the Seebeck effect, Hall effect, Nernst effect, Peltier effect, Ettingshausen, and Righi-Leduc effect \cite{yadav2023topological}.  Reproduced from ref. 36 under the terms of the Creative Commons Attribution 4.0 International License (CC BY), 2023. (c and d) Schematic of thermoelectric devices based on the Seebeck effect and Nernst effect \cite{pan2022giant}. Devices based on Seebeck effect, have $n-$ and $p-$type semiconductors alternately arranged and connected in series. Nernst effect based devices have single material, eliminates the complex electrical connections required for Seebeck devices. Here, B is the magnetic field, $\nabla T$ is the temperature gradient, and E is the electrical field. Reproduced from ref. 37 under the terms of the Creative Commons Attribution 4.0 International License (CC BY), 2021. (e and f) Magnetic-field-dependent ordinary and anomalous Nernst signal response curves for non-magnetic and magnetic materials, respectively \cite{yadav2023topological}. The red curve shows the Nernst signal when a single type of charge carrier dominates. In some materials, multiple valence and/or conduction bands can cross the Fermi energy, resulting in comparable concentrations of both electron ($e^-$) and hole ($h^+$) charge carriers coexisting simultaneously. In such cases, the Nernst signal response to magnetic field is shown in cyan. Reproduced from ref. 36 under the terms of the Creative Commons Attribution 4.0 International License (CC BY), 2023.}
\end{figure*}

\subsection{Seebeck Effect}
The longitudinal TEs are based on the Seebeck effect, wherein an electrical voltage is generated along the direction of the applied temperature gradient [as shown in Fig. 1(b)]. The electric current density $J$ in the presence of an electric field $E$ and a temperature gradient ($\nabla T$) is given by  

\[\vec{J} = \hat{\sigma} \cdot \vec{E} - \hat{\alpha} \cdot \vec{\nabla} T\]

Where, $\hat{\sigma}$ is the electrical conductivity tensor, and $\hat{\alpha}$ is the thermoelectric tensor. Under open-circuit conditions, $\vec{J} = 0$, and hence 

\[ \vec{E} = \hat{\sigma}^{-1} \cdot \hat{\alpha} \cdot \vec{\nabla} T \]

For the Seebeck effect, 
\[E_x = \alpha_{xx}/\sigma_{xx} \cdot \nabla_x T \]

The proportionality factor $\alpha_{xx}/\sigma_{xx}$ is referred to as the Seebeck coefficient ($S_{xx}$), where the applied temperature gradient and the induced electrical voltage appear in the same direction.  Most of the known high-$zT$ TE semiconductors exhibit longitudinal thermoelectricity, where the absolute $S_{xx}$ is generally in the range of 100–500 $\mu\mathrm{V\,K^{-1}}$ \cite{chang20183d, kim2012effect, roychowdhury2021enhanced, duhan2025strong}.  The expression for the Seebeck coefficient in terms of the electrical conductivity can be derived by solving the Boltzmann transport equation \cite{heremans2008enhancement} and is given by:

\[ S_{xx} = \frac{\pi^2 k_B^2 T}{3e} \left. \frac{\partial \ln \sigma_{xx}(\epsilon)}{\partial \epsilon} \right|_{\epsilon = \epsilon_F} \]

The Seebeck power factor is given by the product of the square of the Seebeck coefficient and the longitudinal electrical conductivity, $PF_{xx} = S_{xx}^2 \sigma_{xx}$, and the corresponding thermoelectric figure of merit for the Seebeck effect is expressed as follows:

\[
zT = \frac{S_{xx}^{2}\sigma_{xx}\ T}{\kappa_{xx}}
\]


A TE module based on the Seebeck effect [as illustrated in Fig. 1(c)] consists of multiple $p$-type and $n$-type high-performance longitudinal TE semiconductors alternately arranged and connected in series, requiring the contact between the metal-electrode and TE material of low electrical resistivity, strong mechanical bonding, and high thermal stability. These contacts add electrical and thermal contact resistances, which can significantly reduce the overall device efficiency relative to the intrinsic material performance. As the hot side of the TE module is part of the electrical circuitry, the contacts are subject to thermal degradation. Therefore, simultaneous optimization of both the intrinsic material properties and the reliability of the electrical contacts is needed to achieve superior TE device performance.

The above drawbacks of the longitudinal TEs are overcome by using transverse thermoelectric materials, in which the electric voltage is generated (typically in the presence of a perpendicular external magnetic field) in the direction perpendicular to the temperature gradient. This is known as the Nernst effect [as shown in the right-most panel in Fig. 1(b)]. Unlike Seebeck devices, which require both $p$- and $n$-type materials and often face issues such as thermal expansion mismatch and material compatibility, transverse thermoelectric devices operate using a single material. The number of electrical contacts is reduced to two, both positioned on the colder side of the device, which not only simplifies the module design but also prevents interfacial degradation associated with hot-side contacts [as illustrated in Fig. 1(d)]. The voltage and power generated by transverse TE effects can be enhanced by increasing the length and area perpendicular to the temperature gradient, in a single material without the need for a multi-element structure.

\subsection{Ordinary Nernst Effect}
The transverse TE materials work based on the Nernst effect. In the ordinary Nernst effect, a transverse voltage is generated in the direction perpendicular to both the applied magnetic field and the temperature gradient. In this setup [illustrated in Fig. 1(b)], the charge carriers accumulate in the transverse direction, while phonons move in the longitudinal direction, decoupling the electrical and thermal transport properties. In ordinary Nernst effect (ONE), the induced voltage due to the Lorentz force is linearly proportional to the applied magnetic field [as shown in Fig. 1(e)]. Consider a temperature gradient \( \nabla T \) along the \( x \)-direction and the magnetic field is applied along the \( z \)-direction, an electric field develops along the \( y \)-direction. The Nernst thermopower is defined as:
\[S_{xy} = -\frac{E_y}{\nabla_x T}\]

Under open-circuit conditions, where the electric current densities \( J_{x} = J_{y} = 0 \), the resulting transverse electric field \( E_y \) is given by:

\[E_y = \frac{\sigma_{xx} \alpha_{xy} - \sigma_{xy} \alpha_{xx}}{\sigma_{xx}^2 + \sigma_{xy}^2} \cdot \nabla_x T\]

In the presence of a weak magnetic field, the Seebeck and Nernst signals are coupled through the following relation:

\[S_{xy} \approx \frac{\alpha_{xy}}{\sigma_{xx}} - \frac{\sigma_{xy}}{\sigma_{xx}} \cdot S_{xx}\]

The ordinary Nernst thermopower, $S_{xy}$ is larger than Seebeck thermopower in presence of strong magnetic field, goes upto 800–1000 $\mu\mathrm{V\,K^{-1}}$ \cite{watzman2018dirac, wu2025dirac, liang2018anomalous}. The Nernst power factor is defined as, $PF_{xy} = S_{xy}^2 \sigma_{yy} $ and the figure of merit corresponding to Nernst effect is given as 
\[
z_{N}T = \frac{S_{xy}^{2} \sigma_{yy}\ T}{\kappa_{xx}}
\]

where, electrical conductivity, $\sigma_{yy}$ is measured in the direction of induced electric field, and the thermal conductivity, $\kappa_{xx}$ is measured in the direction of applied temperature gradient.

\subsection{Anomalous Nernst Effect}

\begin{figure*}[t]
\centering
\includegraphics[width=0.98\textwidth]{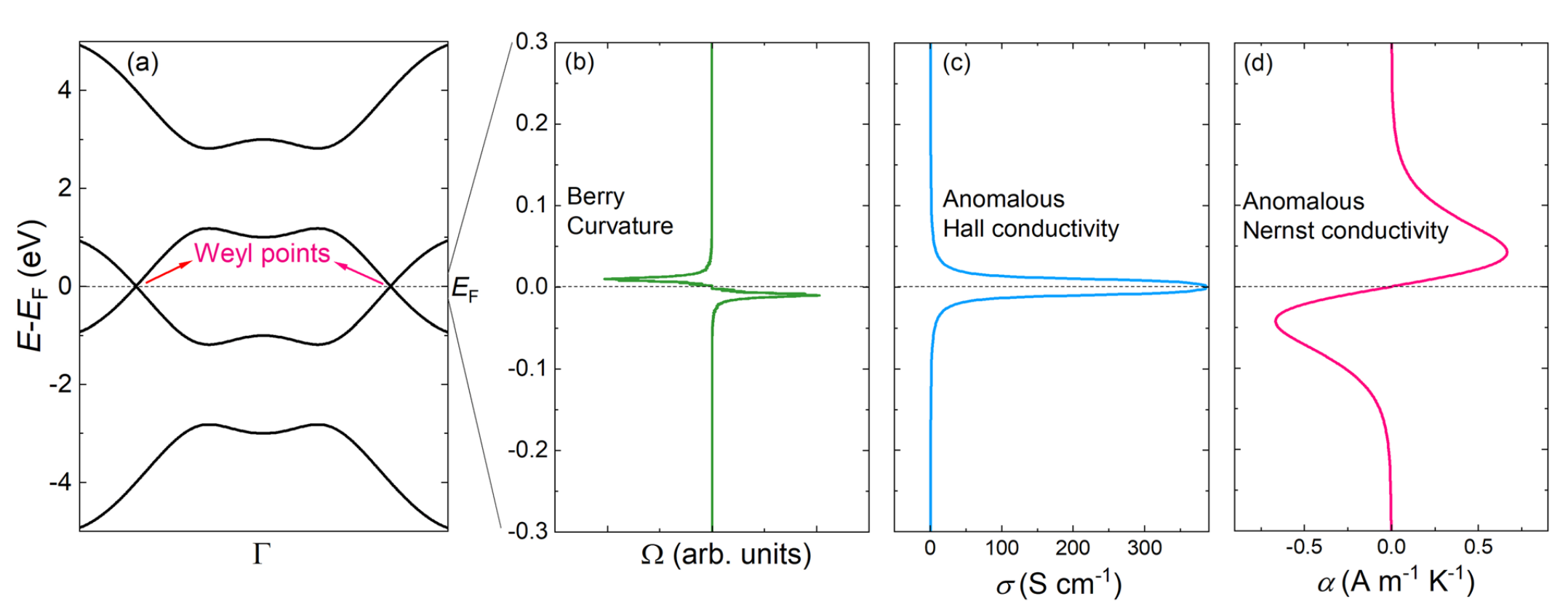}
\caption{(a) Band structure with Weyl points on the Fermi level. (b) Berry curvature at fixed energy integrated over the whole Brillouin zone. (c) Anomalous Hall conductivity reaches the peak value when $E_F$ locates at the Weyl points. (d) Anomalous Nernst conductivity reaches the peak value when $E_F$ shifts away from Weyl points \cite{fu2020topological}. Reproduced from ref. 46 under the terms of the Creative Commons Attribution 4.0 International License (CC BY), 2020.}
\end{figure*}

We need an external magnetic field to observe ONE, however, there are materials which have intrinsic remanent magnetization and show deviation from the linear Nernst response, known as the anomalous Nernst effect (ANE). In such cases, the field dependence closely resembles the hysteresis-like behavior observed in the anomalous Hall effect. The Nernst signal as a function of the magnetic field shows a non-linear behavior, increasing at low fields and saturating at higher fields [see Fig. 1(f)]. The anomalous Nernst conductivity ($\alpha_{xy}^A$) is related to the energy derivative of the anomalous Hall conductivity ($\sigma_{xy}^A$) by the Mott relation \cite{xiao2006berry} and can be expressed as

 \[ \alpha_{xy}^A = \frac{\pi^2 k_B^2 T}{3e} \left. \frac{\partial \sigma_{xy}^A(\epsilon)}{\partial \epsilon} \right|_{\epsilon = \epsilon_F} \]

ANE arises in materials with strong spin-orbit coupling and spontaneous magnetization, often found in ferromagnets, ferrimagnets, and topological quantum materials. In some topological materials, the non-zero Berry curvature acts like a magnetic field in momentum space and gives rise to anomalous transport phenomena. When a temperature gradient is applied, the carriers are deflected in transverse direction by the Berry curvature. The $z$-component of the Berry curvature, $\Omega^z_n(k)$, i.e., the Berry phase per unit area in $\textbf{k}$ space, is related to the anomalous Hall conductivity \cite{weng2015quantum} as 

\[\sigma_{xy}^A = -\frac{2\pi e^2}{h} \int_{\text{BZ}} \frac{d^3k}{(2\pi)^3} \sum_n f_n(k) \, \Omega^z_n(k)\]

in which $f_n(k)$ is the Fermi-Dirac distribution function with band index $n$ and wave vector $k$.  The Berry curvature can be derived from the Kubo-formula \cite{weng2015quantum},

\[\Omega_{n}^{z}(\mathbf{k}) = -2 \, \text{Im} 
\sum_{m \ne n} 
\frac{\langle \Psi_{n\mathbf{k}} | v_{x} | \Psi_{m\mathbf{k}} \rangle 
      \langle \Psi_{m\mathbf{k}} | v_{y} | \Psi_{n\mathbf{k}} \rangle}
     {(E_{m}(\mathbf{k}) - E_{n}(\mathbf{k}))^{2}}\]

where $v_{x(y)}$ is the velocity operator. From the above equations, we see that a large $\Omega_n^z(k)$ leads to a large $\sigma_{xy}^A$, while a strong energy dependence of $\sigma_{xy}^A$ near the Fermi level is essential to obtain a large anomalous Nernst conductivity, $\alpha_{xy}^A$. The AHE is obtained by integration of the Berry curvature of occupied electronic bands over the entire BZ, while the ANE provide a measure of the Berry curvature only at the Fermi energy. A large Berry curvature is expected from the non-trivial band structure involving Weyl or Dirac nodes near the Fermi level. The relationship between the Weyl points, $\Omega_n^z(k)$, $\sigma_{xy}^A$, and $\alpha_{xy}^A$ is demonstrated in Fig. 2 \cite{noky2018characterization}. The presence of two Weyl crossings, as shown in Fig. 2(a), act as monopoles (sources or sinks) of Berry curvature in momentum space. As a result, the Berry curvature is strongly enhanced in the vicinity of the Weyl points [Fig. 2(b)]. This leads to a peak value of anomalous Hall conductivity, $\sigma_{xy}^A$ at the Fermi energy $E_F$ [Fig. 2(c)]. Since, the anomalous Nernst conductivity, $\alpha_{xy}^A$ depends on energy derivative of $\sigma_{xy}^A$, the maximum $\alpha_{xy}^A$ occurs when $E_F$ shifts away from the Weyl points [Fig. 2(d)]. The anomalous Nernst thermopower, $S_{xy}^A$ is observed to be very less than Seebeck thermopower, ranging between 3–8 $\mu\mathrm{V\,K^{-1}}$ \cite{sakai2018giant, feng2022giant}. To the best of our knowledge, the highest reported value is 23 $\mu\mathrm{V\,K^{-1}}$ for UCo$_{0.8}$Ru$_{0.2}$Al \cite{asaba2021colossal}.



\section{Materials Classification}
\subsection{Topological Insulators}
Topological materials share several key features with thermoelectric materials, such as heavy elements and narrow band gaps \cite{pal2015thermoelectric}. Narrow band gap materials with high carrier mobility are good for increasing the electrical conductivity, while the presence of heavy atoms suppresses the phonon frequencies, contributing to low lattice thermal conductivity. Heavy elements are also desired to facilitate strong spin orbit-coupling, which is necessary for certain nontrivial topological states. Several materials with good thermoelectric properties such as Sb$_2$Te$_3$, Bi$_2$Te$_3$, Bi$_2$Se$_3$, BiSe, BiTe, SnTe, and their alloys were found to exhibit non-trivial electronic topology \cite{osterhage2014thermoelectric, tan2014high, slade2020contrasting, zhang2009topological, majhi2017emergence, samanta2018localized, samanta2020intrinsically, tanaka2012experimental}.

Topological insulators (TIs) are a class of materials that exhibit insulating behavior in the bulk while the surface states are conducting. Typically, the surface states cross linearly within the bulk band gap, forming Dirac cones  \cite{fu2007topological, kane2005quantum, chen2009experimental, kumar2020topological} within the bulk band gap (Fig. 3).  These topological surface states arise from the nontrivial topology of the bulk and are linked to intrinsic bulk properties such as the spin–orbit coupling, band inversion, and band extrema of opposite parities. Owing to the topological protection of the conducting surface states, the spins of Dirac fermions are tightly locked to their momentum, preventing back-scattering of charge carriers from impurities or defects. As a result, the boundary states of TIs provide low-resistance conduction channels. TIs are considered strong if they have an odd number of Dirac cones on their surfaces, leading to gapless, topologically protected states on all crystal surfaces. Sb$_2$Te$_3$, Bi$_2$Te$_3$, Bi$_2$Se$_3$ are examples of strong TIs.  Weak TIs, on the other hand, exhibit conducting states only on certain surfaces and are characterized by an even number of band crossings in their bulk electronic structures \cite{fu2007topological}. While BiSe and BiTe have been shown to exhibit weak topological insulating phase, SnTe is established as a topological crystalline insulator.

Both weak and strong TIs are highly robust against crystal disorder, enabling high electrical conductivity via protected surface states while reducing phonon thermal conductivity through suppressed phonon transport induced by intentional impurity doping or crystal structure disorder \cite{poudel2008high, szczech2011enhancement}. Due to the linear crossing of electronic bands, the surface carriers in the bulk band gap, have a longer mean free path than surface carriers in the bulk bands, leads to high mobility at the surfaces. Thus, the thermoelectric performance in a TIs can be effectively optimized and greatly improved. The geometry of the structure and size of the material determine the corresponding contributions from the boundary and the bulk states. Typically, the effects of TI surface states are pronounced when the sample dimensions are on the order of a few nanometers, having higher surface-to-volume ratio. In this low-dimensional limit, the bulk contribution to conductivity diminishes significantly, allowing the surface states to become the dominant conduction channel \cite{shin2016surface}.

\begin{figure*}[t]
\centering
\includegraphics[width=0.90\textwidth]{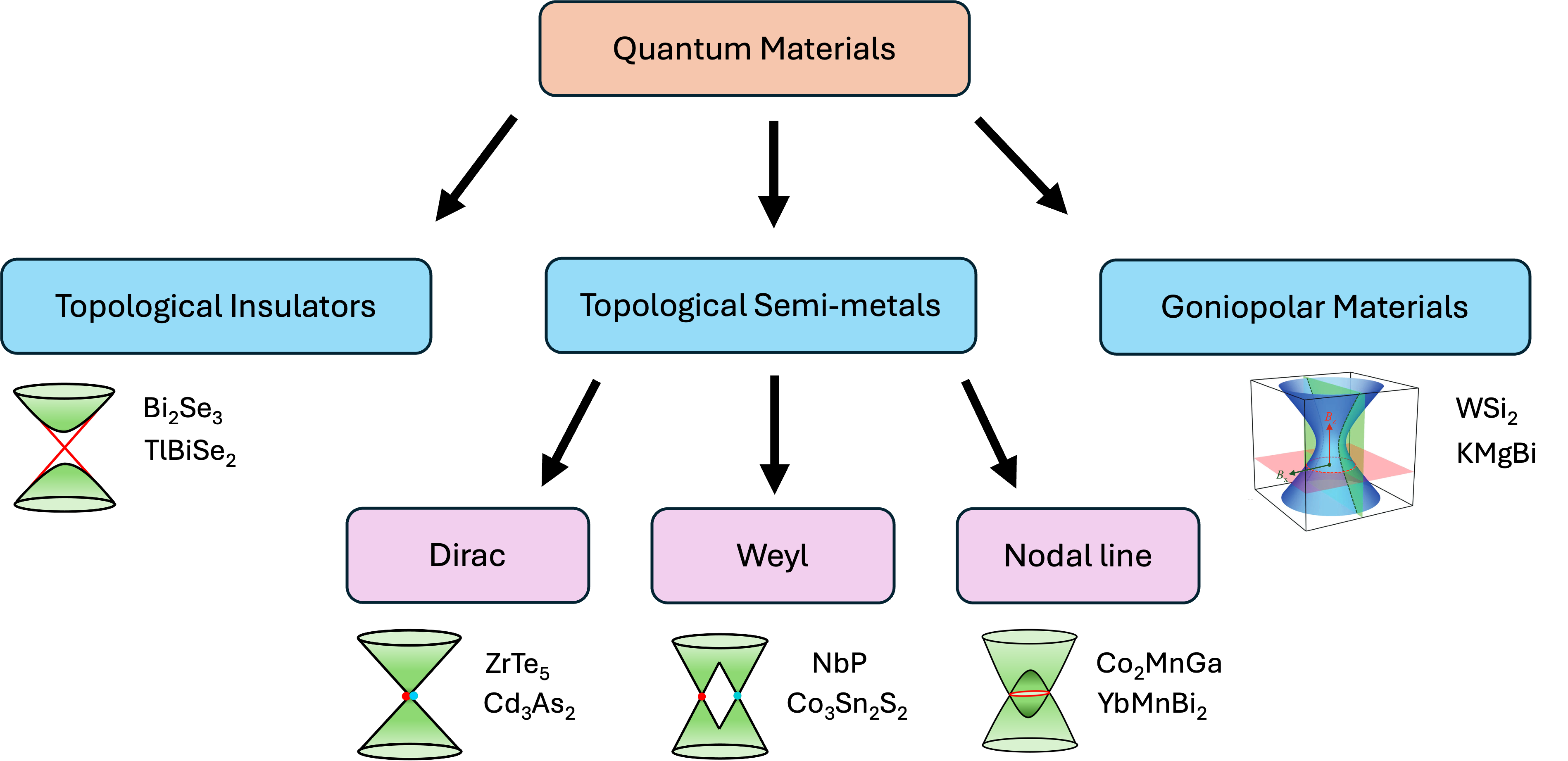}
\caption{Different types of topological quantum materials relevant to thermoelectric applications, showing schematic band crossings and respective examples.}
\end{figure*}

\subsection{Topological Semimetals}

The nontrivial topological phases in which symmetry-protected, linearly dispersing valence and conduction bands appear at the surface states of TIs have also been realized in some semimetals and metals, where linear band crossings occur in the bulk \cite{burkov2016topological, kumar2020topological}. Depending on the nature of band crossings, topological semimetals are classified as Dirac, Weyl, or nodal-line semimetals. Dirac semimetals are materials in which two-fold degenerate bands intersect to form four-fold degenerate Dirac points (Fig. 3). Such states arise in centrosymmetric (inversion-symmetric) and nonmagnetic (time-reversal-symmetric) systems. Time-reversal symmetry guarantees Kramers degeneracy ensuring that all bands are doubly degenerate with spin-up and spin-down states. When these doubly degenerate bands cross linearly, a robust Dirac cone appears in the bulk electronic structure. Along with these symmetries, further details of the crystal structures, such as the presence of $C_3$, $C_4$, or $C_6$ uniaxial rotational symmetries, are also vital for stabilizing Dirac points and protecting the linear band crossings against small perturbations.

In Weyl semimetals, the spin-up and spin-down bands split due to symmetry breaking, forming non-degenerate bands. These non-degenerate bands cross linearly in the bulk electronic structure to form two-fold degenerate Weyl points. Weyl points always appear in pairs having opposite chirality (Fig. 3). 
The Dirac point can be thought of as a combination of two Weyl points of opposite chirality; results in zero net chirality. The Dirac point can be split into Weyl points by introducing a magnetic element into the crystal, by applying an external magnetic field, or removing the center of inversion in the crystal. In addition to Dirac and Weyl semimetals, there is another class of semimetals in which the valence and conduction bands cross along a line or a ring in momentum space, known as nodal-line semimetals (Fig. 3). The nodal lines generally occur in crystals with mirror reflection symmetry, such as Co$_2$MnGa \cite{guin2019anomalous} , Fe$_3$X (X = Ga, Al) \cite{sakai2020iron} and Co$_3$Sn$_2$S$_2$ \cite{guin2019zero}. The Weyl or  Dirac points sometimes originate from these nodal lines owing to spin–orbit coupling (SOC). In the absence of SOC, the two crossing bands have opposite and real mirror eigenvalues $(\pm 1)$, allowing them to cross freely within the mirror-invariant plane. Once SOC is included, the mirror eigenvalues become imaginary $(\pm i)$, allowing the states to hybridize and opening a gap everywhere, except at a few points preserved by certain symmetries. Unlike TIs, the transport properties in topological semimetals are dominated by bulk states. The unique feature of topological semimetals having two types of charge carriers with comparable concentration and high mobility from linear band dispersions, significantly increase the ordinary Nernst effect.

These materials can exhibit longitudinal TE effect due to the band convergence, where multiple band degeneracy can effectively increase the density of states effective mass favors large thermopower, along with high carrier mobility \cite{fu2020topological, pal2015thermoelectric, samanta2020intrinsically}. In addition, they can also show a large anomalous Nernst effect owing to the topology of their electronic wavefunctions in the presence of large Berry curvature near the Fermi level.

\begin{figure*}[t]
\centering
\includegraphics[width=0.60\textwidth]{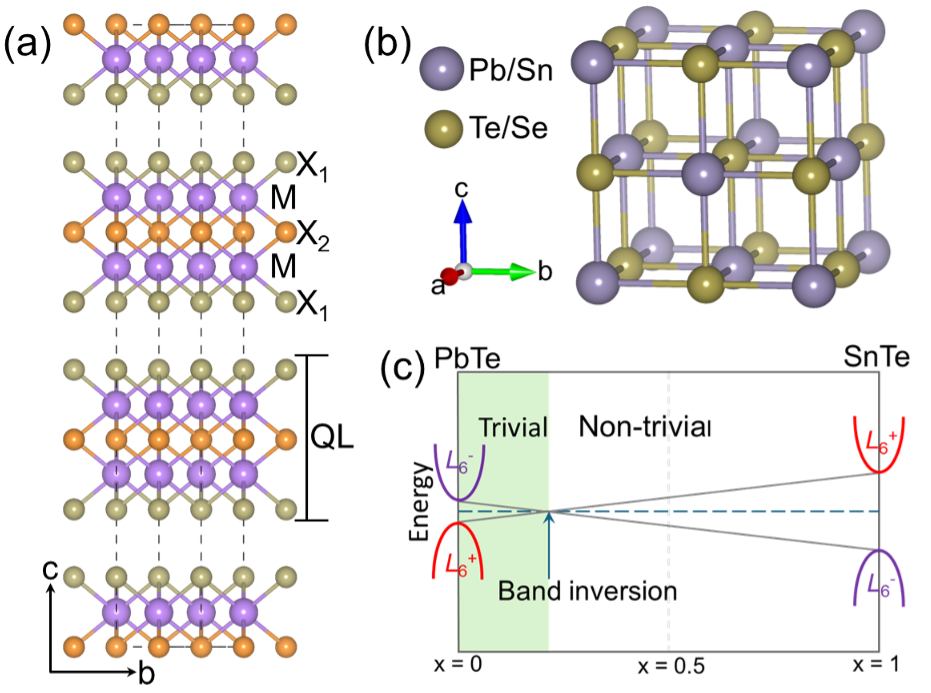}
\caption{(a) Crystal structure of tetradymites ($R\overline{3}m$), (X= Te or Se) anions reside on X$_1$ and X$_2$ sites, while (M = Bi, Sb) cations occupy the metal sites, (b) the crystal structure of Pb$_{1-x}$Sn$_x$Te/Se, and (c) a schematic illustration of transition of PbTe from topologically trivial to non-trivial upon substitution of Pb sites with Sn.}
\end{figure*}

\subsection{Goniopolar Materials}
There is a class of materials, known as goniopolar materials or $p \times n-$type conductors, which exhibit axis-dependent conduction polarity (ADCP) and can be used in transverse thermoelectric devices that operates at zero magnetic field \cite{goto2024band, yang2026axis}. This phenomenon appears in anisotropic materials that exhibit opposite signs of the Seebeck coefficient along different crystallographic directions. When the $x$-axis is rotated by an angle $\theta$ relative to the crystallographic $i$- and $j$-axes, and a temperature gradient $\nabla_xT$ is applied along $x$, the transverse electric field is given by

\[E_y = S'_{xy}\,\nabla_x T 
= \left[(S_{ii} - S_{jj}) \cos\theta \sin\theta \right] \nabla_x T \]

Where $S_{ii}$ and $S_{jj}$ are the thermopower along the crystallographic $i$ and $j$ axes, respectively. The transport of charge carriers in gonioploar materials is shown in Fig. 9(a). The origin of gonipolarity can be a single band with negative Gaussian curvature along one direction and a positive Gaussian curvature along another, or a multi band effect, in which conduction electron have a higher mobility along one direction and holes have higher mobility along another \cite{he2019fermi}. Here, we will discuss the materials having an ADCP based on mixed-dimensionality of the Fermi surface \cite{manako2024large}. The rotation angle $\theta$ of the heat current from the $i$-axis is optimized to maximize the transverse thermopower, $S'_{xy}$, using the relation

\[\cos^2 \theta_{opt} = \left(1 + \sqrt{\frac{\kappa_{jj}/\kappa_{ii}}{\rho_{jj}/\rho_{ii}}}\right)^{-1} \]

where $\kappa_{ii}$ and $\kappa_{jj}$ are the lattice thermal conductivities and $\rho_{ii}$ and $\rho_{jj}$ are the electrical resistivities along the $ii$ and $jj$ directions, respectively. In an ideal gonipolar materials with isotropic electrical and thermal conductivity, this angle is 45$^o$. The transverse dimensionless figure of merit of goniopolar materials is defined as 

\[
z_{G}T
= \frac{4 S_{xy}'^{\,2} T}
{\left(\sqrt{\rho_{jj}\kappa_{jj}} + \sqrt{\rho_{ii}\kappa_{ii}}\right)^2}
\]

Since goniopolar materials exhibit simultaneous $p$- and $n$-type conductivities, these materials can be used to create transverse thermoelectric modules \cite{uchida2022thermoelectrics}, eliminating the need for maintaining multiple metal-semicondutor contacts at the hot side that are essential while making modules using longitudinal thermoelectrics. This can improve the long-term thermal stability of the devices \cite{goto2025goniopolar}.

\begin{figure*}[t]
\centering
\includegraphics[width=0.70\textwidth]{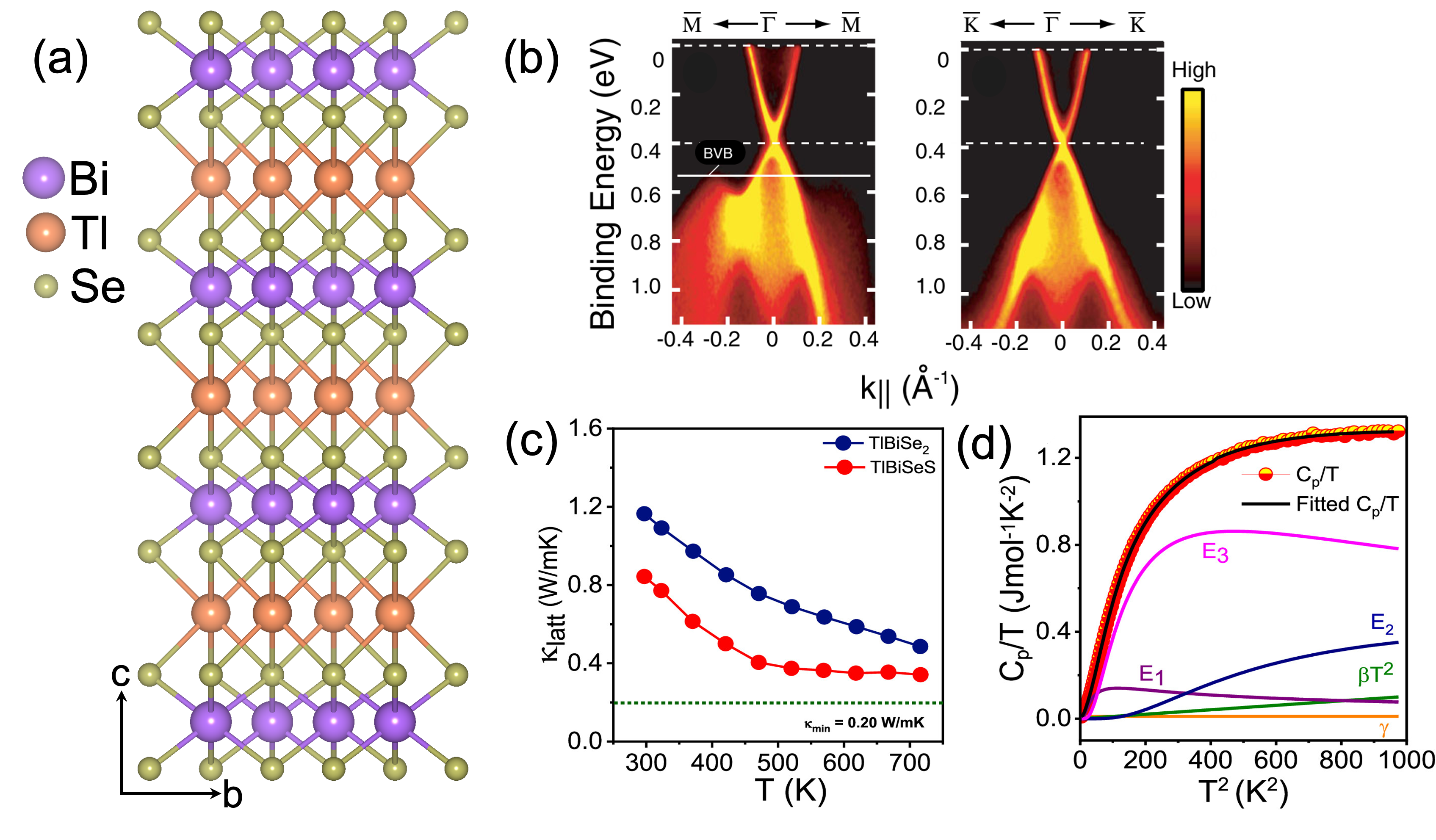}
\caption{(a) Crystal structure of TlBiSe$_2$, (b) ARPES intensities of TlBiSe$_2$ along the $\overline{\Gamma}\,\overline{M}$ (left) and $\overline{\Gamma}\,\overline{K}$ (right) \cite{kuroda2010experimental}. Reproduced from ref. 78 with permission from the American Physical Society, copyright 2010. (c) Temperature-dependent lattice thermal conductivity ($\kappa_l$) of TlBiSe$_2$ and TlBiSeS, and (d) experimental low-temperature heat capacity (C$_p$/T vs. T$^2$) of TlBiSe$_2$, fitted using a Debye-Einstein model \cite{maria2023metavalent}. Reproduced from ref. 79 with permission from the American Chemical Society copyright 2023.}
\end{figure*}

\section{\textbf{Thermoelectric properties}}
\subsection{Longitudinal Thermoelectrics}

For decades, binary and ternary chalcogenides based on heavy $p$-block elements have been extensively researched to optimise their longitudinal thermoelectric properties. The most well-established high-performance thermoelectric materials within this group are topological insulators (TIs), which can be broadly classified structurally into tetradymites and non-tetradymites.

\subsubsection{Tetradymite family of compounds}

The tetradymite family, named after the mineral Bi$_2$Te$_2$S \cite{heremans2017tetradymites}, comprises anisotropic narrow bandgap layered materials crystallising in the $R\bar{3}m$ space group with a general formula of M$_2$X$_3$ (M = Bi, Sb; X = Se, Te). Their structure is characterised by stacking of quintuple-layer (X$_1$-M-X$_2$-M-X$_1$) of approximately 1 nm thick, along the (001) direction as shown in Fig. 4(a) \cite{heremans2017tetradymites, samanta2020layered}. In materials containing two different chalcogens, the more electronegative elements are thermodynamically favoured to occupy the more ionic X$_2$ sites \cite{heremans2017tetradymites, cook2007minerals}. Strong covalent bonding within the layers and weak van der Waals-like interaction between the layers result in highly anisotropic properties. The presence of strong spin-orbit coupling (SOC) leads to band inversion and the generation of topological states \cite{samanta2020layered, bradlyn2017topological, zhang2009topological}. For instance, in Bi$_2$Te$_3$ and Sb$_2$Te$_3$, the leading tetradymites for near-room-temperature power generation, greatly benefit from the strong SOC and band inversion. Band inversion modifies the band curvature, which in many thermoelectric materials result in a lower effective mass and higher carrier mobility, which can enhance the power factor \cite{wang2018nontrivial, xiao2018approaching, xie2024topological, xie2020band}. Higher power factors and low thermal conductivity due to structural anisotropy and heavy elements lead to high thermoelectric performance in these materials. A few of the other important TIs of this family include Bi$_2$Se$_3$, BiSbTe$_3$ and Bi$_2$Te$_2$Se \cite{heremans2017tetradymites, samanta2020layered, pan2019synergistic}. It’s worth mentioning that SnSb$_2$Te$_4$, a member of the Sn$_m$Sb$_{2n}$Te$_{3n+m}$ series, is a strong topological insulator that crystallises in a tetradymite-type structure and has exhibited a peak $zT$ of $\sim$ 0.55 at 673K \cite{das2024enhanced}.

To achieve a promising $zT$ value, the thermoelectric parameters have been optimised through careful modulation of both electronic and thermal transport. For example, substitution of Se into Te sites leads to an increase in band gap, which in turn suppresses bipolar conduction, leading to a peak $zT$ of $\sim$ 1.1 at 340 K in Bi$_2$Te$_{2.85}$Se$_{0.15}$ and a peak $zT$ of $\sim$ 0.86 at 600 K in Bi$_2$Te$_{1.5}$Se$_{1.5}$, respectively \cite{wang2012enhanced}. Further, doping with Cu helps in modulating the Te vacancies \cite{han2011formation, liu2011thermoelectric}. Similarly, a peak $zT$ of $\sim$ 1.86 at 320 K has been achieved in Bi$_{0.5}$Sb$_{1.5}$Te$_3$ through dislocation engineering \cite{ii2015kim}. In Bi$_2$Te$_3$ and Sb$_2$Te$_3$, anti-site defects, where Bi or Sb atoms substitute Te sites and vice versa, naturally occur due to the comparable electronegativity of the constituent elements. Although traditionally viewed as detrimental to achieving high $zT$ values, these defects have recently gained attention for imparting plasticity while preserving excellent thermoelectric performance. This combination is particularly critical for flexible thermoelectrics, another area where TIs show exceptional performance \cite{li2025plasticity}. Anti-site defects are less pronounced in Bi$_2$Se$_3$ due to a larger electronegativity difference \cite{heremans2017tetradymites}. 

A remarkable characteristic of the TIs is the presence of gapless metallic surface states. However, the effect of topological surface states is minimal in the bulk samples and on the associated thermoelectric performance, attributed to the low surface-to-volume ratio. Conversely, in low-dimensional TI materials with large surface-to-volume ratio, the effect of surface states is expected to be significant.  For instance, in Bi$_{1.5}$Sb$_{0.5}$Te$_{1.7}$Se$_{1.3}$ nanowires with a diameter of 180 $nm$, a significant enhancement of $zT$ $\sim$ 0.36, an order of magnitude improvement has been observed at 300 K compared to its bulk counterpart ($zT$ $\sim$ 0.028). The enhancement was primarily attributed to an order of magnitude increase in the electrical conductivity ($\sim$ 238 S cm$^{-1}$) arising from the surface-dominated transport and thermally activated charge carriers in the NWs, compared with their counterpart bulk materials ($\sim$ 25 S cm$^{-1}$) \cite{hsiung2014surface}. Hence, facile robust synthesis methods yielding materials with high surface-to-volume ratios have become essential for harnessing the effects of topological surface states \cite{kumar2025facile, hong2015enhanced}.

\begin{figure*}
\centering
\includegraphics[width=0.80\textwidth]{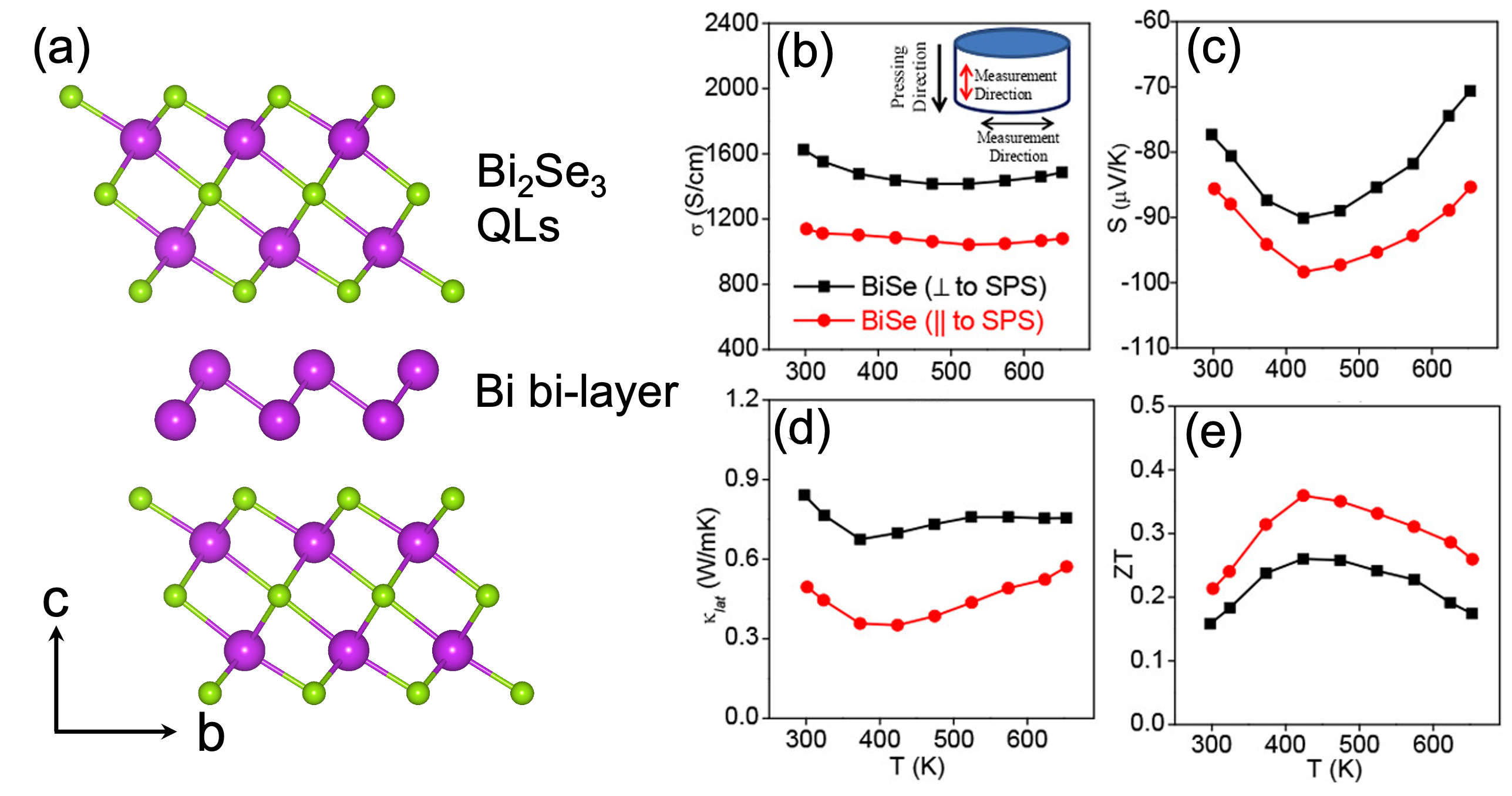}
\caption{(a) Crystal structure of BiSe ($P\overline{3}m1$), (b), (c), (d) and (e) temperature dependence of electrical conductivity ($\sigma$), Seebeck coefficient ($S$), lattice thermal conductivity ($\kappa_l$) and figure of merit ($ZT$) of BiSe, respectively \cite{samanta2018localized}. Reproduced from ref. 58 with permission from the American Chemical Society, copyright 2018.}
\end{figure*}

\begin{figure*}[t]
\centering
\includegraphics[width=0.75\textwidth]{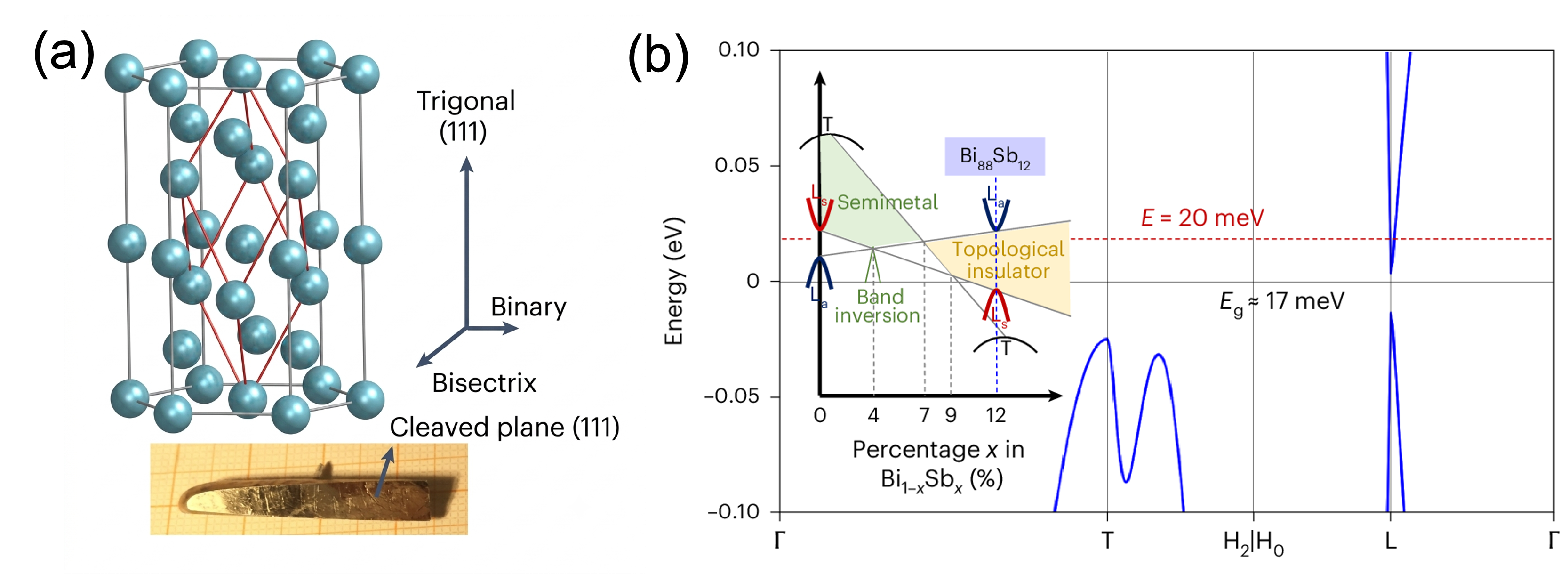}
\caption{(a) Crystal structure of Bi-rich Bi$_{1-x}$Sb$_x$. The bottom panel shows a cleaved Bi$_{88}$Sb$_{12}$ surface, exposing the (111) plane. (b) Band structure near the Fermi level at the $\Gamma$ and L points, highlighting linear dispersion. The inset provides a schematic of the band evolution across different Sb concentrations in Bi$_{1-x}$Sb$_x$ \cite{pan2025magneto}. Reproduced from ref. 98 under the terms of the Creative Commons Attribution 4.0 International License (CC BY), 2025.}
\end{figure*}

\subsubsection{SnTe-based topological crystalline insulators}
Among the non-tetradymites family, the important materials are SnTe \cite{tanaka2012experimental}, Pb$_{1-x}$Sn$_x$Te \cite{xu2012observation}, Pb$_{1-x}$Sn$_x$Se \cite{dziawa2012topological}, TlBiSe$_2$ \cite{kuroda2010experimental}, BiSe \cite{majhi2017emergence} and Bi$_{88}$Sb$_{12}$ \cite{pan2025magneto}.
Unlike tetradymites, where topological states are protected by time reversal symmetry, in SnTe-based materials, topological states are primarily protected by crystal symmetries, hence, they are termed as topological crystalline insulators (TCIs). SnTe, Pb$_{1-x}$Sn$_x$Te, and Pb$_{1-x}$Sn$_x$Se are the widely investigated TCIs crystallising in a simple rock salt structure [Fig. 4(b)], where Sn and Pb occupy the cationic sites and Te and Se occupy the anionic sites. $\mathrm{Sn} \leq 0.4$ is crucial for retaining rock-salt crystal structure in Pb$_{1-x}$Sn$_x$Te and Pb$_{1-x}$Sn$_x$Se systems \cite{tanaka2012experimental, dziawa2012topological, hsieh2012topological, xu2012observation}. More than one type of surface states has been observed in TCIs, unlike one in the tetradymites \cite{liu2013two}. 
PbTe is topologically trivial; however, substituting Sn into Pb sites leads to a transition from trivial to non-topologically trivial [Fig. 4(c)] phase. The key evidence was the SnTe-like surface states on the (001) plane \cite{xu2012observation}. A similar observation of surface states was reported in the related system Pb$_{1-x}$Sn$_x$Se (Sn = 0.23) \cite{dziawa2012topological}. High $zT$ in these materials results from complex band structures where the TCI state contributes to high electronic performance, which is further enhanced by strategies to reduce lattice thermal conductivity \cite{tan2014high, banik2014lead, wang2025advances}.

\begin{figure*}[t]
\centering\includegraphics[width=0.85\textwidth]{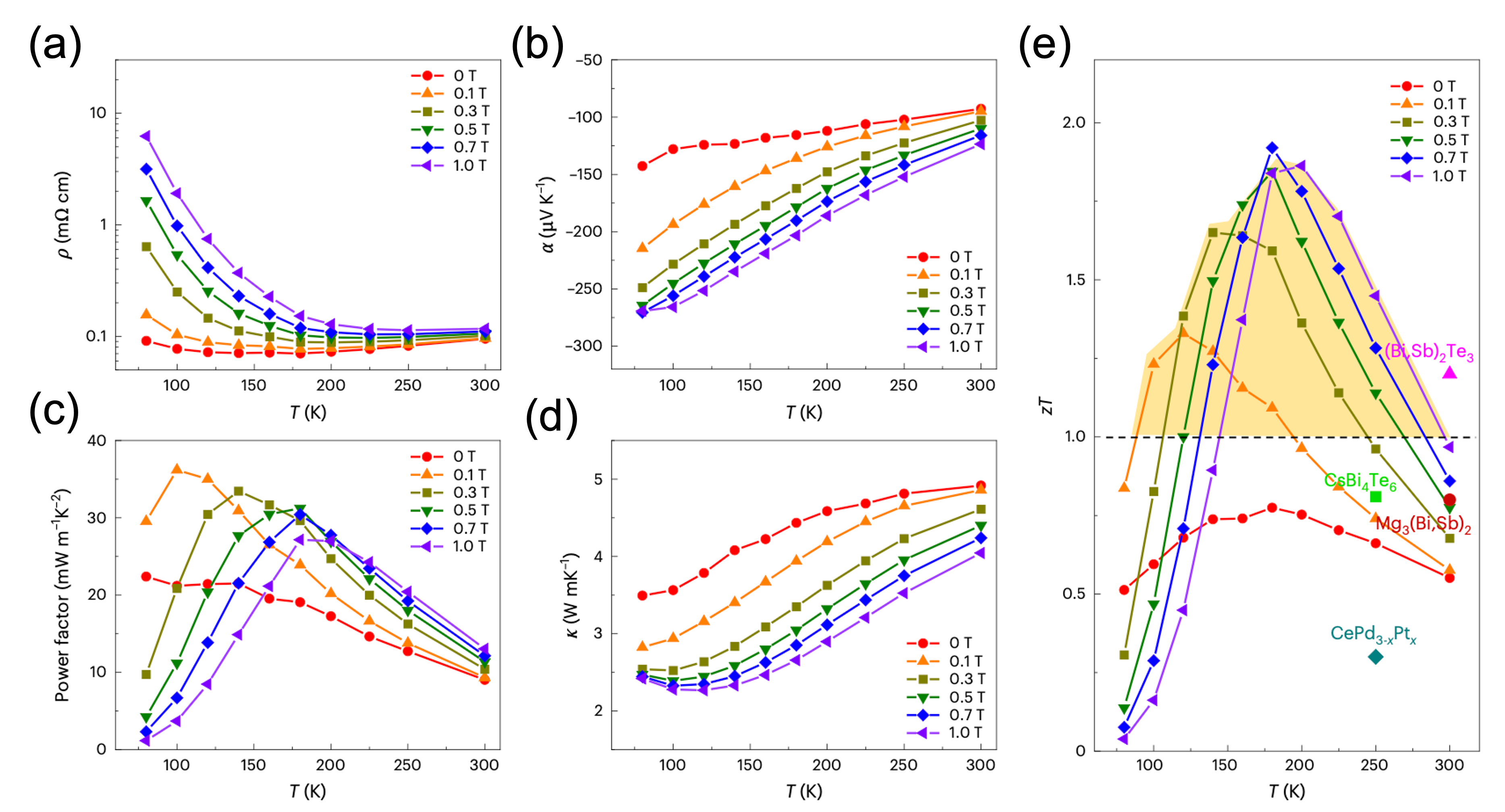}
\caption{Temperature-dependent thermoelectric properties of a Bi$_{88}$Sb$_{12}$ single crystal under magnetic fields from 0 to 1 T. (a) Resistivity ($\rho$), (b) Seebeck coefficient ($S_{xx}$ represented here as $\alpha$), (c) Power factor, (d) Thermal conductivity ($\kappa$), and (e) figure of merit ($zT$). For comparison, (e) also includes the high thermoelectric performance of previously reported materials near and below 300 K \cite{pan2025magneto}.  Reproduced from ref. 98 under the terms of the Creative Commons Attribution 4.0 International License (CC BY), 2025.}
\end{figure*}

\subsubsection{Tl-based topological insulators}
3D topological insulator states have been observed in many thallium-based ternary compounds. For example,  TlBiSe$_2$ crystallises in a three-dimensional (3D) crystal structure derived from the tetradymite family. Tetradymites comprise quintuple layers (X$_1$-M-X$_2$-M-X$_1$) stacked by weak van der Waals-like interaction along the $c$-axis. Integration of thallium (Tl) atoms directly into these layers create a continuous, covalently bonded structure with a [-Se-Tl-Se-Bi-Se-]$_n$ stacking sequence along the $c$-axis with strong interactions rather than van der Waals-like in tetradymites [Fig. 5(a)] \cite{maria2023metavalent, rajaji2023pressure, kuroda2010experimental, sato2010direct}. The presence of robust topological surface states in TlBiSe$_2$ directly manifests the SOC-induced band inversion, which is experimentally realised by ARPES [Fig. 5(b)] \cite{kuroda2010experimental, sato2010direct}. Furthermore, the material can be transitioned into a TCI phase under high pressure \cite{rajaji2023pressure}. Alongside the topological surface states TlBiSe$_2$ hosts metavalent bonding which is characterised by the Bi-Se bond length $\sim$ 2.8 \AA, which is close to the sum of their covalent radii (\(\sim 2.7 \, \text{\AA}\)), and a Tl-Se bond length of $\sim$ 3.2 \AA, near to the sum of their ionic radii ($\sim$ 3.4 \AA). This bonding anharmonicity, driven by the stereochemically active $6s^2$ lone pairs on Bi and Tl, is a promising mechanism for achieving intrinsically low lattice thermal conductivity. TlBiSe$_2$ exhibits an ultralow lattice thermal conductivity ($\kappa_l$), dropping from $\sim$ 1.1 Wm$^{-1}$K$^{-1}$ at 300 K to $\sim$ 0.4 Wm$^{-1}$K$^{-1}$ at 725 K [Fig. 5(c)]. This suppression originates from the material's intrinsic anharmonicity \cite{xia2020particlelike, pathak2024deciphering, semwal2025concerted}. Analysis of low-temperature heat capacity reveals three low-frequency Einstein modes [Fig. 5(d)], corresponding to localised vibrations of Tl and Bi atoms. These optical phonons couple with acoustic phonons, scattering heat-carrying vibrations. Origin of this behaviour is a unique lattice shearing along the $\langle 100 \rangle$ direction, driven by the stereochemical activity of the 6s$^2$ lone pairs on Tl and Bi atoms. Surface states induced enhanced electronic transport  and intrinsic low lattice thermal conductivity lead to a peak $zT$ of $\sim$ 0.6 at 715 K. The thermoelectric performance was further improved by substituting sulphur for selenium to form TlBiSeS, achieving a $zT$ of $\sim$ 0.8. This enhancement results from two key effects: a notable increase in $\sigma$ due to a higher carrier concentration ($2.4 \times 10^{19} \,\text{cm}^{-3} \; \text{vs.} \; 1.7 \times 10^{19} \,\text{cm}^{-3}$ in the pristine material), and a significant reduction in $\kappa_l$ through solid-solution point defect scattering \cite{maria2023metavalent}.

\subsubsection{Bi-based weak topological insulators}
BiSe, a member of the homologous series (Bi$_2$)$_m$(Bi$_2$Se$_3$)$_n$ with a stoichiometry of $m:n = 1:2$, is classified as a weak topological insulator (WTI) \cite{majhi2017emergence}. Its crystal structure consists of a bismuth (Bi) bilayer sandwiched between Bi$_2$Se$_3$ quintuple layers, stacked along the crystallographic $c$-axis [Fig. 6(a)]. By alternately stacking a 2D quantum spin Hall insulator (bismuth bilayer) with a strong 3D topological insulator (Bi$_2$Se$_3$), this architecture creates a natural heterostructure that is a weak topological insulator. This combination is highly attractive for thermoelectric applications, as the heterogeneous interface is predicted to suppress lattice thermal conductivity, while the topological layers could enhance the electronic power factor. Experimental results demonstrate a figure of merit ($zT$) of $\sim$ 0.36 at 425 K [Fig. 6(e)] due to superior charge transport and low thermal conductivity \cite{samanta2018localized}. The ultra-low lattice thermal [Fig. 6(d)] conductivity is attributed to an anisotropic layered structure and optical-acoustic phonon coupling, which is facilitated by localised vibrations within the bismuth bilayer, which give rise to low-energy optical phonon modes \cite{samanta2018localized}. The Te analogue compound BiTe is a natural van der Waals heterostructure compound, crystallised with space group $P\overline{3}m1$ \cite{pathak2025impact}. The compound is a dual topological insulator \cite{eschbach2017bi1te1}, i.e., a combination of weak topological insulators and topological crystalline insulators. Thermoelectric investigations \cite{samanta2020intrinsically} show that the compound is an intrinsically low thermal conducting material with $\kappa_l$ $\sim$ 0.47-0.8 Wm$^{-1}$K$^{-1}$ in the temperature range 300-650 K. The low thermal conductivity originates from the effective coupling of low-energy optical modes, with heat carrying acoustic phonon modes. Additionally, BiTe exhibits high charge carrier mobility due to the presence of metallic surface states \cite{samanta2020intrinsically}.

\begin{figure*}[t]
\centering
\includegraphics[width=0.95\textwidth]{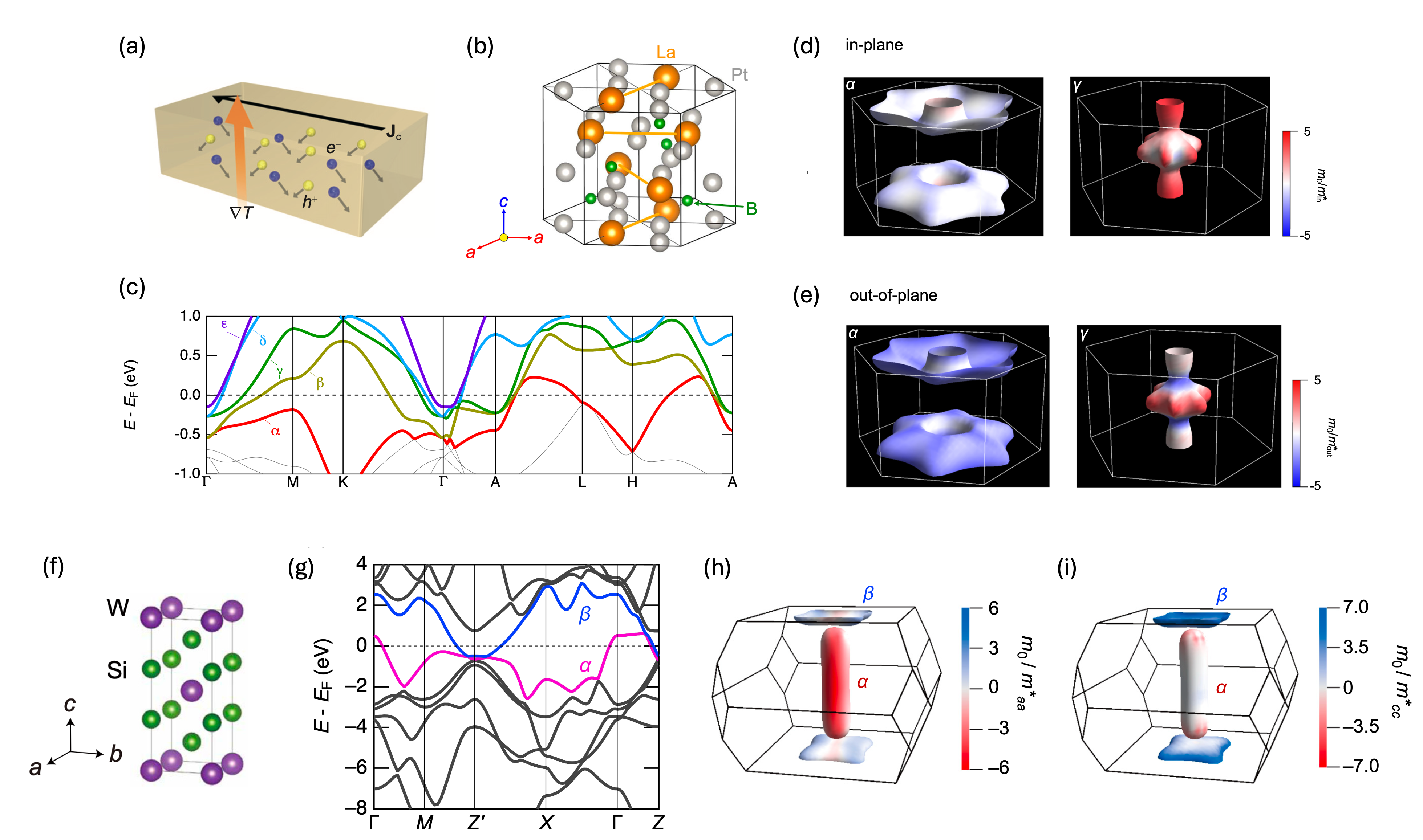}
\caption{(a) Schematic diagram of transverse TE transport using a goniopolar material \cite{uchida2022thermoelectrics}. $\nabla T$ and $J_c$ denote the temperature gradient and charge current driven by the thermoelectric effect. Reproduced from ref. 76 with permission from Elsevier, copyright 2022. (b) Crystal structure of LaPt$_2$B. (c) Electronic band structure of LaPt$_2$B. Mixed-dimensional Fermi surfaces $\alpha$ and $\gamma$, displaying the diagonal elements of the inverse mass tensor for (d) in-plane and (e) out-of-plane directions for LaPt$_2$B \cite{manako2024large}. Reproduced from ref. 75 under the terms of the Creative Commons Attribution 4.0 International License (CC BY), 2024. (f) Crystal structure of WSi$_2$. (g) Electronic band structure of WSi$_2$. Fermi surface showing (h) in-plane and (i) out-of-plane components of the inverse effective mass tensor, with blue and red colors indicating electron- and hole-like dispersions at the Fermi level for WSi$_2$, respectively \cite{ohsumi2024transverse}. Reproduced from ref. 112 under the terms of the Creative Commons Attribution 4.0 International License (CC BY), 2024.}
\end{figure*}

\begin{figure*}[t]
\centering
\includegraphics[width=0.98\textwidth]{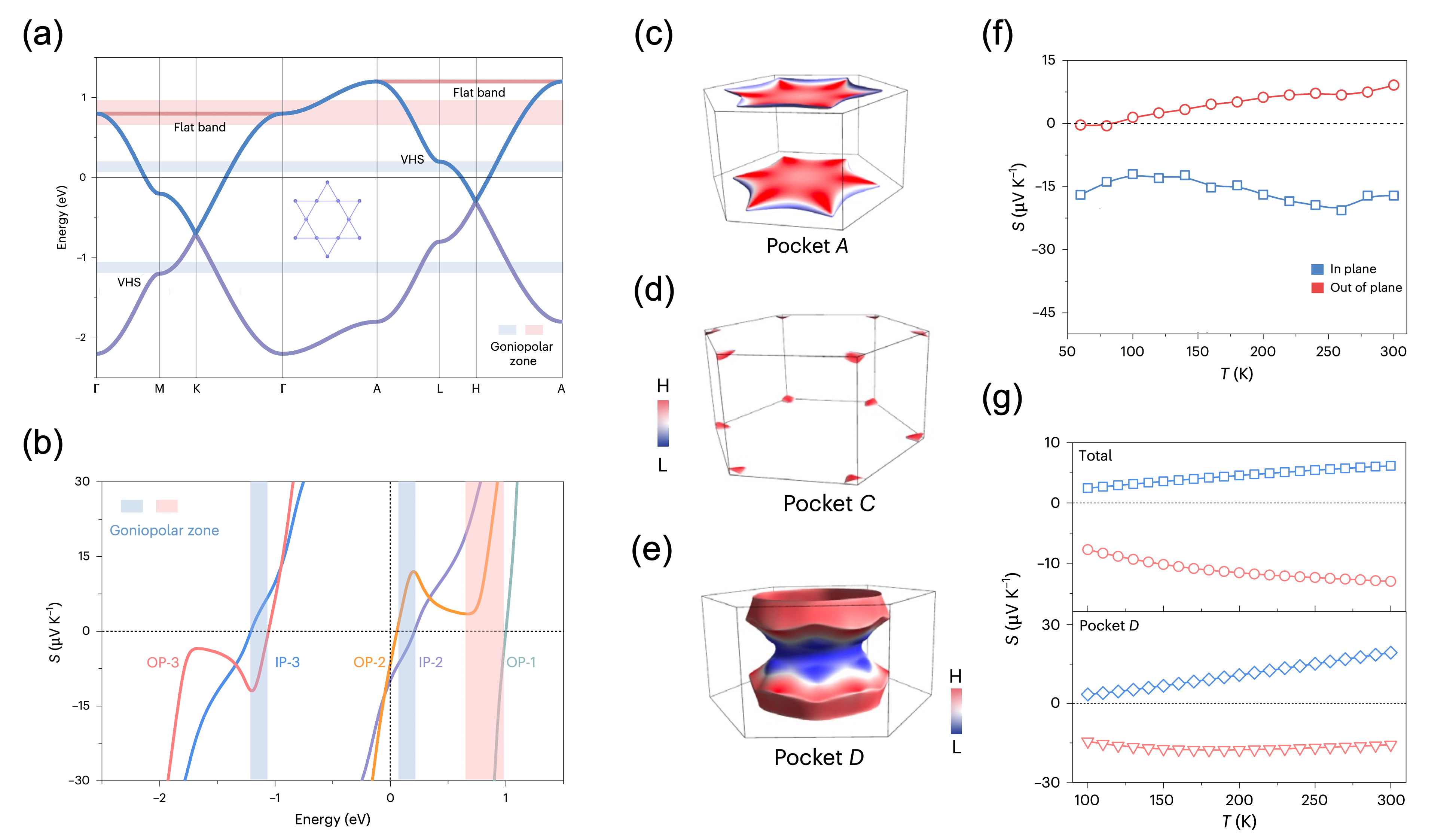}
\caption{(a) Single-orbital electronic band structure of the 3D kagome lattice, and (b) corresponding in-plane and out-of-plane Seebeck coefficients. The blue and red colors highlight the single-band and two-band goniopolar zones, respectively. (c) The hole-dominated Fermi pocket A and (d) the electron-dominated pocket C of LuCo$_6$Ge$_6$. (e) The hyperbolic pocket D of LuCr$_6$Ge$_6$. L and H denote low and high Fermi velocities, respectively. (f) Temperature-dependent in-plane and out-of-plane Seebeck coefficients of the LuCo$_6$Ge$_6$ single crystal. (g) Calculated Seebeck coefficients for all pockets (upper) and pocket D (lower) in LuCr$_6$Ge$_6$. \cite{hu2026fermiology}. Reproduced from ref. 113 under the terms of the Creative Commons Attribution 4.0 International License (CC BY), 2026.}  
\end{figure*}

\subsubsection{Bi$_{1-x}$Sb$_x$ alloys}
Bi$_{1-x}$Sb$_x$ alloys, which adopt a rhombohedral crystal structure with space group $R\overline{3}m$ [Fig. 7(a)], are of significant interest for low-temperature (T $<$ 300 K) thermoelectric cooling \cite{pan2025magneto}. The alloying of semi-metallic bismuth with antimony facilitates a transition to an inverted band insulator at $x$ $\approx 7\%$. Further increase in the concentration of Sb results in a direct gap topological insulator ($x$ $<$ 0.22), where the T band lies below the L$_s$ band [Fig. 7(b) inset]. Bi$_{88}$Sb$_{12}$ has been identified as the optimal alloy for superior thermoelectric performance among the possible compositions. The Seebeck coefficient can be enhanced mainly by applying a magnetic field until the electrons are forced into a cyclotron motion when $\mu B$ $>$ 1, where $\mu$ is the carrier mobility and B is the applied magnetic field. This phenomenon is termed as the magneto-Seebeck or magneto-thermoelectric effect. Bi$_{88}$Sb$_{12}$ possesses a sharp band dispersion [Fig. 7(b)], leading to high mobility, low resistivity and low thermal conductivity due to heavy elements. Recent investigations from Yu Pan et al. \cite{pan2025magneto} have shown that the Seebeck coefficient can be significantly enhanced with a magnetic field as low as 0.1T and a saturation trend at elevated magnetic fields [Fig. 8(b)], a consequence of electrons forced into cyclotron motion. Although this is accompanied by a slight reduction in electrical conductivity [Fig. 8(a)], the overall power factor increases substantially [Fig. 8(c)]. A concurrent decrease in electronic thermal conductivity also reduces total thermal conductivity [Fig. 8(d)], resulting in a maximum $zT$ of $\sim$ 2 and an average $zT$ of $\sim$ 1.7 at 180 K in high-quality Bi$_{88}$Sb$_{12}$ single crystals under a moderate 0.7 T field \cite{pan2025magneto} [Fig. 8(e)]. Such moderate fields can be readily achieved with permanent magnets. Under high magnetic fields, the thermoelectric figure of merit ($zT$) enhancement is also observed in Dirac semimetals like Pb$_{1-x}$Sn$_x$Se \cite{liang2013evidence}, PtSn$_4$ \cite{fu2020largely}, and Cd$_3$As$_2$ \cite{liang2017anomalous, wang2019magnetic, yue2019soft}.

\subsection{Goniopolar Thermoelectrics}
LaPt$_2$B is a a non-magnetic metal in a series of rare-earth-based chiral magnets, having a space group P6$_3$22. It has a screw axis along the $c$-axis and has layers of La-Pt and B, which stack along the $c$-axis [Fig. 9(b)]. To measure the transverse thermoelectric effect (TTE), the temperature gradient is applied 45$^o$ from the $a$-axis to the $c$-axis. The observed $S_{xy}^\prime$ reaches a maximum value of 10 $\mu\mathrm{V\,K^{-1}}$ around room temperature and the Peltier conductivity have a maximum value of 130 A m$^{-1}$ K$^{-1}$ at 15 K. LaPt$_2$B has a figure of merit of $\approx 1.5 \times 10^{-3}$ at room temperature in zero magnetic fields. The Fermi surface corresponding to five different bands has been studied to observe the multi-dimensionality nature, highlighted in Fig. 9(c). The hole-like $\alpha$-sheet has a large surface area with normal vector along the $c$-axis, contributed to positive thermopower along the $c$-axis direction. The electron-like $\gamma$- sheet Fermi surface forms a cylinder, responsible for negative thermopower along the $ab$-plane direction [Figs. 9(d) and 9(e)]. The in-plane La-Pt network and the screw-axis Pt-B connection are mainly responsible for the in-plane and $c$-axis conduction, respectively \cite{manako2024large}.

WSi$_2$ is a semimetal which crystallizes in the tetragonal crystal structure of the space group $I$4/$mmm$ [Fig. 9(f)]. The temperature gradient is applied along the 45$^o$ from the in-plane direction ($i=a$) to the out-of-plane direction ($j=c$). The thermopower is positive along the in-plane direction and negative along the out-of-plane direction, contributing to a total $S_{xy}^\prime$ of 6 $\mu\mathrm{V\,K^{-1}}$ near room temperature. The estimated figure of merit for WSi$_2$ is $\approx 1 \times 10^{-3}$ above 100K. This ADCP in WSi$_2$ originates from the mixed-dimensionality, can be understood by the coexistence of electron- and hole-like Fermi surfaces, as shown in Fig. 9(g). A quasi-two-dimensional (q-2D) hole-like Fermi surface ($\alpha$-sheet) is centered at the $\Gamma$-point, and has a cylindrical shape extended in the $c$-axis direction, dominates the in-plane conduction [Fig. 9(h)]. While, a quasi-one-dimensional (q-1D) electron-like Fermi surface ($\beta$-sheet) is centered at the $Z$-point, shows a large surface area with the normal vector along the $c$-axis direction, dominates the out-of-plane conduction [Fig. 9(i)] \cite{ohsumi2024transverse}.

Kagome lattice has a structural geometry that hosts exotic quantum states and has the potential to generate goniopolarity. Fig. 10(a) shows the single-orbital electronic band structure of 3D kagome lattice featuring in-plane and out-of-plane nearest-neighbour hoppings. It has flat bands and van Hove singularities (VHS), highlighted by the red and blue shaded areas, and their corresponding Seebeck response is shown in Fig. 10(b). Based on the opposite Seebeck-coefficient response in the out-of-plane and in-plane directions, which suggests a goinopolar nature, three 166-type kagome materials were experimentally investigated, all of which have space group P6/mmm \cite{hu2026fermiology}. Among them, LuCo$_6$Ge$_6$, has shown an exceptionally large goniopolar transverse thermopower of  18.4 $\mu\mathrm{V\,K^{-1}}$ at 300K, due to a flat-band enabled two-band mechanism. The flat band dominates the out-of-plane transport, having Seebeck coefficient increases with temperature, from 0 to 9.1 $\mu\mathrm{V\,K^{-1}}$ between 60 K to 300 K. The Fermi pocket A [see Fig. 10(c)] originating from the flat band near E$_F$, show hole-like dispersion. An additional Fermi pocket C, shown in Fig. 10 (d) from a  dispersive band, shows electron-like transport. This dominates the in-pane transport, having Seebeck coefficient of -15 $\mu\mathrm{V\,K^{-1}}$ across the temperature range [Fig. 10(f)]. Sibling compounds Lu(Ho)Cr$_6$Ge$_6$ have a VHS-driven single band goniopolar response that arises from asynchronous particle–hole asymmetry along the in-plane and out-of-plane directions. The DFT calculations reveal four anisotropic Fermi pockets (labelled as D, E, F and G), attributed to the 3d orbitals of the Cr-based kagome lattice. The hyperbolic pocket D, given in Fig. 10(e) show positive Seebeck coefficient along the in-plane directions and negative along the out-of-plane direction, unlike LuCo$_6$Ge$_6$. Either pocket D or E alone is sufficient to produce the observed goniopolar behaviour. LuCr$_6$Ge$_6$ and HoCr$_6$Ge$_6$ also demonstrate exceptional goniopolar transverse thermopower of 7.3 $\mu\mathrm{V\,K^{-1}}$ and 8.1 $\mu\mathrm{V\,K^{-1}}$ at 300 K, respectively [Fig. 10(g)]. Hence, kagome systems can host both single- and two-band mechanisms, highlighting the pivotal role of fermiology in goniopolarity \cite{hu2026fermiology}.

\begin{figure*}[t]
\centering\includegraphics[width=0.75\textwidth]{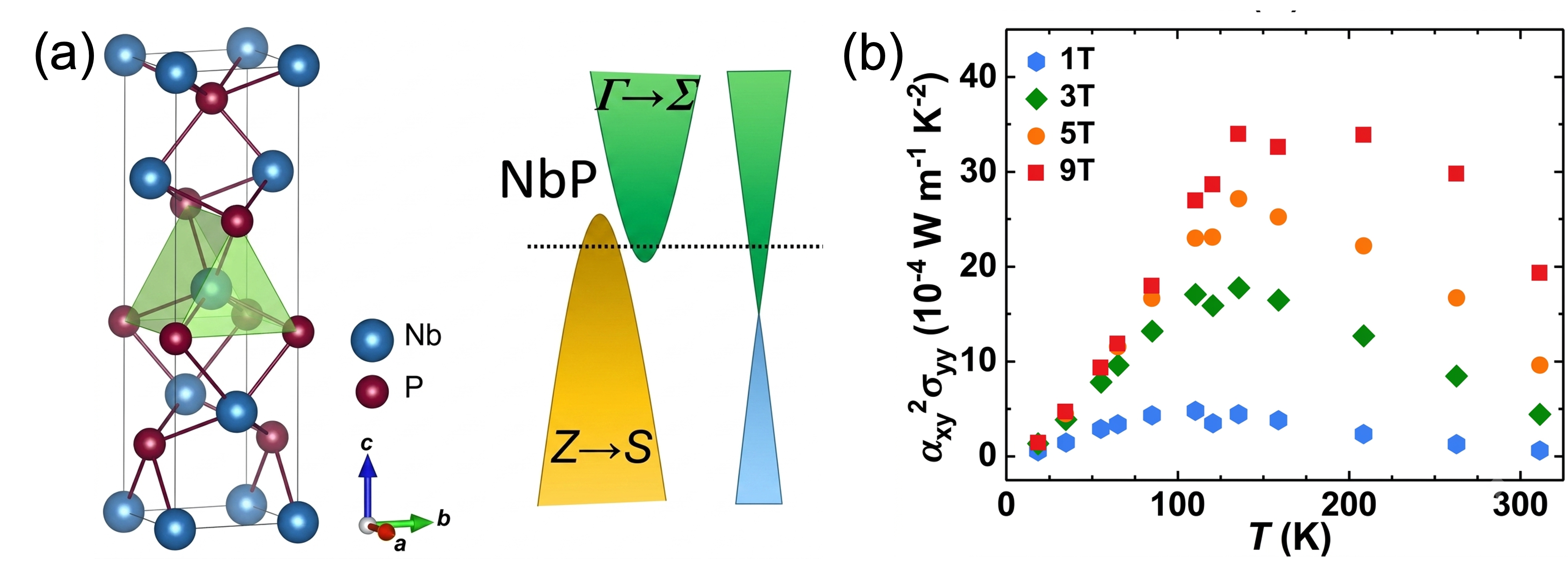}
\caption{(a) Crystal structure of NbP (left) alongside a schematic representation of its band structure. (b) Temperature-dependent $PF_{xy}$ of polycrystalline NbP measured under varying magnetic fields \cite{fu2018large}. Reproduced from ref. 119 under the terms of the Creative Commons Attribution 3.0 Unported License, 2018.}
\end{figure*}

Re$_4$Si$_7$ is a thermally robust compound, crystallizes in a monoclinic layered structure with the space group $Cm$. It has a maximum figure of merit 0.7 $\pm$ 0.15 at 980 K. $Z_GT$ values of the device based on this compound and the material are almost same, having values 0.44 and 0.42 $\pm$ 0.08 at 580K, respectively \cite{scudder2021highly}. The experimental compounds listed in the international crystal structure database, a dozen or so have been identified experimentally to exhibit ADCP, which includes Mg$_3$Sb$_2$ and Mg$_3$Bi$_2$ \cite{goto2024band}, NaSnAs \cite{ochs2021computationally}, NaSn$_2$As$_2$ \cite{wang2020anisotropic},  KMgBi \cite{ochs2024synergizing} and PdSe$_2$ \cite{nelson2023axis}. The transverse thermoelectric generation is obtained by cutting a goniopolar crystal along the direction between the one that gives the largest $n$-type thermopower and the one that gives the largest $p$-type thermopower. A temperature gradient is then applied in a direction that causes electrons and holes to move sideways. The goniopolar material can show excellent transverse thermoelectric performances, however the choice of materials is presently limited, and further material exploration is required \cite{song2025field, wang2020chemical}.

\subsection{Transverse Thermoelectrics}
In this section, we highlight transverse thermoelectricity arising from the Nernst effect in non-magnetic topological systems and then move on to the anomalous Nernst effect in magnetic topological materials.

\subsubsection{Non-centrosymmetric Weyl semimetal NbP}

\begin{figure*}[t]
\centering
\includegraphics[width=0.99\textwidth]{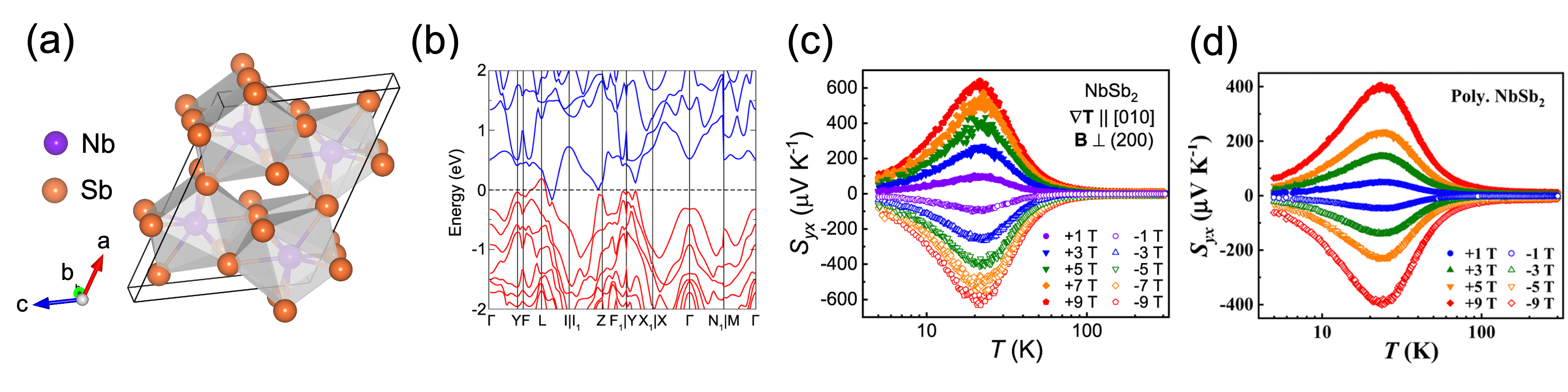}
\caption{(a) Crystal structure of NbSb$_2$. (b) Electronic band-structure of NbSb$_2$ with spin-orbit coupling (SOC). (c) Temperature dependence of Nernst thermopower ($S_{yx}$) of single-crystal NbSb$_2$ at different magnetic fields \cite{li2022colossal}. Reproduced from ref. 127 under the terms of the Creative Commons Attribution 4.0 International License (CC BY), 2022. (d) Temperature dependence of Nernst thermopower ($S_{yx}$) of polycrystalline NbSb$_2$ at different magnetic fields \cite{li2023giant}. Reproduced from ref. 128 under the terms of the Creative Commons Attribution 3.0 Unported License, 2023.}
\end{figure*}

NbP crystallises in a non-centrosymmetric tetragonal structure (I4$_1$md), isostructural to other Weyl semimetals like TaAs, TaP, and NbAs [Fig. 11(a)]. Its crystal structure consists of interconnected NbP$_6$ and PNb$_6$ trigonal prisms \cite{fu2018large, watzman2018dirac}. Electronic structure of NbP near the Fermi level exhibits a combination of features that are highly favourable for thermoelectric performance: it possesses quadratic electron and hole pockets similar to the elemental semimetal Bi, which are known to enhance the Nernst effect, as well as additional linearly dispersing electron pockets originating from its Weyl bands. This linear dispersion contributes to extremely high carrier mobility, a hallmark of topological semimetals. In contrast to small-bandgap semiconductors like PbTe, which require heavy doping to achieve high thermoelectric performance, intrinsic semi-metallic and topological electronic structure makes NbP a compelling candidate for exhibiting a significant Nernst effect and excellent transverse thermomagnetic properties \cite{fu2018large}.

The first report of the Nernst effect in single crystals of NbP by Watzman et al. has shown a very large Nernst thermopower of about $\sim$ 800 $\mu\mathrm{V\,K^{-1}}$ at 109 K and 9 T, which is of two orders of magnitude higher than the Seebeck thermopower of NbP \cite{watzman2018dirac}. Investigations on spark plasma sintered high-quality polycrystalline NbP have shown a peak thermopower of $\sim - 47 \ \mu\mathrm{V\,K^{-1}}$ (near 110 K and 9 T) and a Nernst thermopower of $\sim$ 90 $\mu\mathrm{V\,K^{-1}}$ near 136 K and 9 T. The significant thermopower difference was attributed to three main factors: first, polycrystalline nature alters the measurement direction compared to single crystals, with grain boundaries likely scattering carriers and affecting transport properties; second, extremely small Fermi surface makes its transport behaviour highly sensitive to slight stoichiometric variations, which can shift the Fermi level across samples; and third, the the uncertainity in the thermopower measurements due to single crystals' smaller size. At zero magnetic field the estimated Seebeck power factor rises with temperature and shows a peak value of $\sim 8.5 \times 10^{-4} \, \mathrm{W\,m^{-1}\,K^{-2}}$ near 100K, then declines further as temperature increases. When a magnetic field is applied, the peak shifts to higher temperatures, though its magnitude remains nearly constant. In contrast, the Nernst power factor steadily increases with temperature, reaching a peak value of $\sim 35 \times 10^{-4} \, \mathrm{W\,m^{-1}\,K^{-2}}$ at 9T, near 150K [Fig. 11(b)], which is about four times higher than the Seebeck power factor and comparable to leading thermoelectric materials like Bi$_2$Te$_3$, PbTe, CoSb$_3$, and half-Heusler compounds. Although the thermopower and power factor are high, the $z_NT$ remains low compared to leading thermoelectric materials due to high thermal conductivity. Polycrystalline samples, with enhanced grain boundary scattering, offer reduced thermal conductivity and a pathway to improve $z_NT$ through phonon engineering \cite{fu2018large}.

\subsubsection{Centrosymmetric Dirac semimetal NbSb$_2$} The topological semimetal NbSb$_2$ has a centrosymmetric monoclinic crystal structure (space group $C_{\mathrm{2}/\mathit{m}}$). Its structure features niobium atoms enclosed within Sb-atom hendecahedron. These polyhedra are interconnected face-to-face along the $b$-axis and edge-to-edge along the $c$-axis, forming atomic layers that lie parallel to the $bc$-plane [Fig. 12(a)]. Electronic structure of NbSb$_2$ with spin-orbit coupling (SOC) shows the conduction band crosses on the path from L to I and the valence band near L [Fig. 12(b)], with an about 350 $\mathrm{meV}$ overlap between the conduction and valence bands \cite{li2022colossal}. The chemical vapour transport grown single crystals show a Nernst power factor $\sim 3800 \times 10^{-4} \, \mathrm{W\,m^{-1}\,K^{-2}}$ under 5T and 25K and a $z_NT$ $\sim 71 \times 10^{-4} \, \mathrm{K^{-1}}$ under 5T at 20K. The large, unsaturated Nernst effect in NbSb$_2$ is attributed to its compensated electron-hole concentration favoured by the band structure, high Dirac-like mobilities, and a strong phonon-drag effect. The latter points to a significant phonon role in Dirac fermion transport \cite{li2022colossal}. Extended studies on polycrystalline NbSb$_2$ reveal a $S_{yx}$ $\sim$ 396 $\mu\mathrm{V\,K^{-1}}$ at 21 K and 9 T [Fig. 12(d)], which is 36\% lower than that of the single-crystalline sample $\sim$ 616 $\mu\mathrm{V\,K^{-1}}$ [Fig. 12(c)] and a giant Nernst power factor of $\sim 1269 \times 10^{-4} \, \mathrm{W\,m^{-1}\,K^{-2}}$ under 9 T at 28 K. Furthermore, a high $z_N$ $\sim 28.5 \times 10^{-4} \, \mathrm{K^{-1}}$ under 9 T at 18 K, highlighting its competitive performance. Despite similar carrier mobilities, polycrystalline NbSb$_2$ has a lower total thermopower than its single-crystal counterpart-about 57\% at its peak under a 9 T magnetic field. This deficit stems from a significantly weaker phonon-drag component, while the diffusion component remains comparable. The suppressed phonon-drag effect is attributed to grain boundary scattering, which reduces the relaxation time of the long-wavelength phonons responsible for the effect \cite{li2023giant}. A proposed refrigerator using polycrystalline NbSb$_2$ is estimated to achieve $P_{max}$ of 6.5 $\mathrm{W\,g^{-1}}$ under a 9 T field at 30 K. This cooling power is superior to conventional gas-based systems ($P_{max}$ = 0.05 $\mathrm{W\,g^{-1}}$ for He at 5 K, 0.12 $\mathrm{W\,g^{-1}}$ for H$_2$ at 26 K, and 1.01 $\mathrm{W\,g^{-1}}$ for N$_2$ at 93 K) \cite{li2023giant}.

\begin{figure*}[t]
\centering
\includegraphics[width=0.90\textwidth]{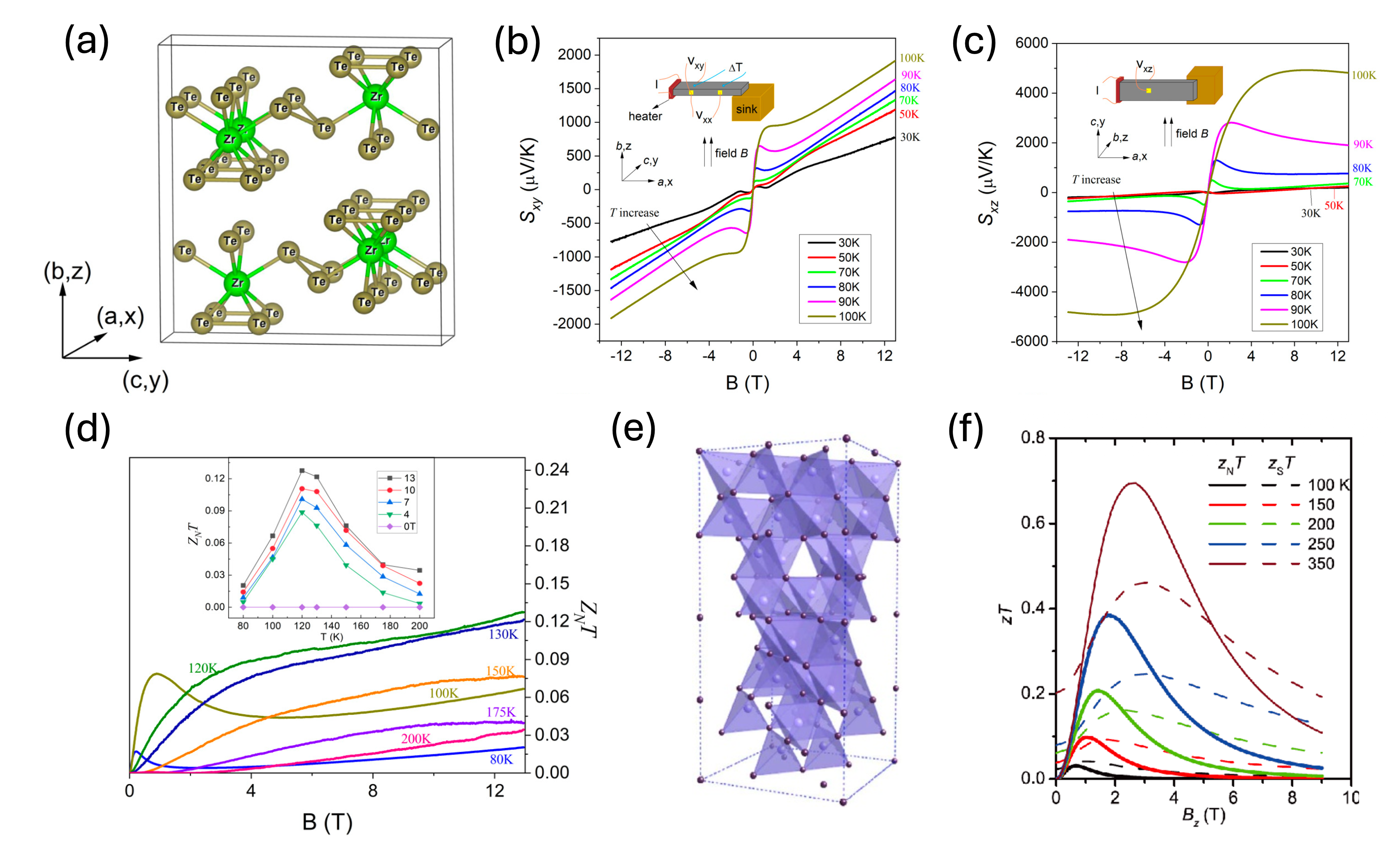}
\caption{(a) Crystal structure of ZrTe$_5$. (b,c) The in-plane ($S_{xy}$) and out-of-plane ($S_{xz}$) Nernst thermopower as a function of B under different temperatures, respectively. (d) The transverse $z_NT$ as a function of field B for ZrTe$_5$ \cite{wang2021giant}. Reproduced from ref. 129 with permission from the American Physical Society, copyright 2021. (e) Crystal structure of Cd$_3$As$_2$ having 32 unit cells \cite{neupane2014observation}. Reproduced from ref. 130 with permission from Springer Nature, copyright 2014. (f) The figure of merits $z_NT$ and $z_sT$ as a function of field B for Cd$_3$As$_2$ at selected temperatures \cite{xiang2020large}. Reproduced from ref. 131 with permission from Springer Nature, copyright 2020.}
\end{figure*}

\begin{figure*}[t]
\centering
\includegraphics[width=0.99\textwidth]{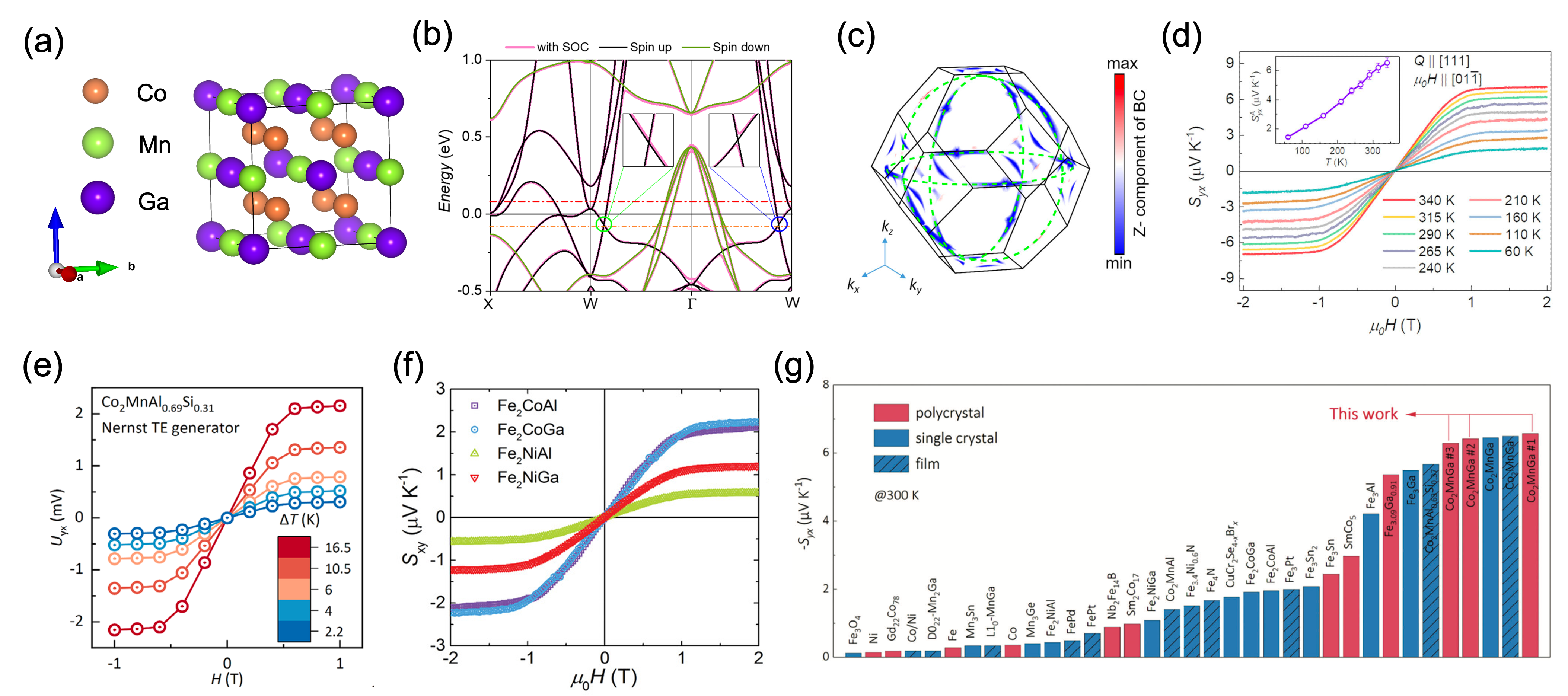}
\caption{(a) Crystal structure of Co$_2$MnGa crystallising in a cubic Heusler structure (space group $Fm\bar{3}m$). (b) Electronic band structure of Co$_2$MnGa with and without spin-orbit coupling (SOC). The regions containing nodal lines are circled and magnified. As shown in the blue circle, SOC lifts the band degeneracy, which induces a gap in the nodal line. (c) Berry curvature distribution in Brillouin zone with [$01\bar{1}$] magnetization direction. The dotted green lines indicate the nodal lines. The Berry curvature is concentrated around the gapped lines. (d) The Nernst thermopower of Co$_2$MnGa single crystals as a function of magnetic field. The inset displays the large anomalous Nernst signal at different temperatures \cite{guin2019anomalous}. Reproduced from ref. 69 under the terms of the Creative Commons Attribution 4.0 International License (CC BY), 2019. (e) Magnetic field dependence of the measured Nernst voltage for Co$_2$MnAl$_{0.69}$Si$_{0.31}$-based Nernst TE generator at different temperature differences \cite{qian2025anomalous}. Reproduced from ref. 132 with permission from John Wiley and Sons, copyright 2025. (f) Nernst thermopower versus magnetic field for various Fe$_2$YZ (Y = Co, Ni; Z = Al, Ga) compounds, measured near 340 K \cite{mende2021large}. Reproduced from ref. 133 under the terms of the Creative Commons Attribution License, 2021. (g) The $S_{xy}$ values of different magnetic materials at 300 K, plotted with Co$_2$MnGa \cite{chen2024topological}. Reproduced from ref. 134 with permission from John Wiley and Sons, copyright 2024.}
\end{figure*}

\begin{figure*}[t]
\centering
\includegraphics[width=0.95\textwidth]{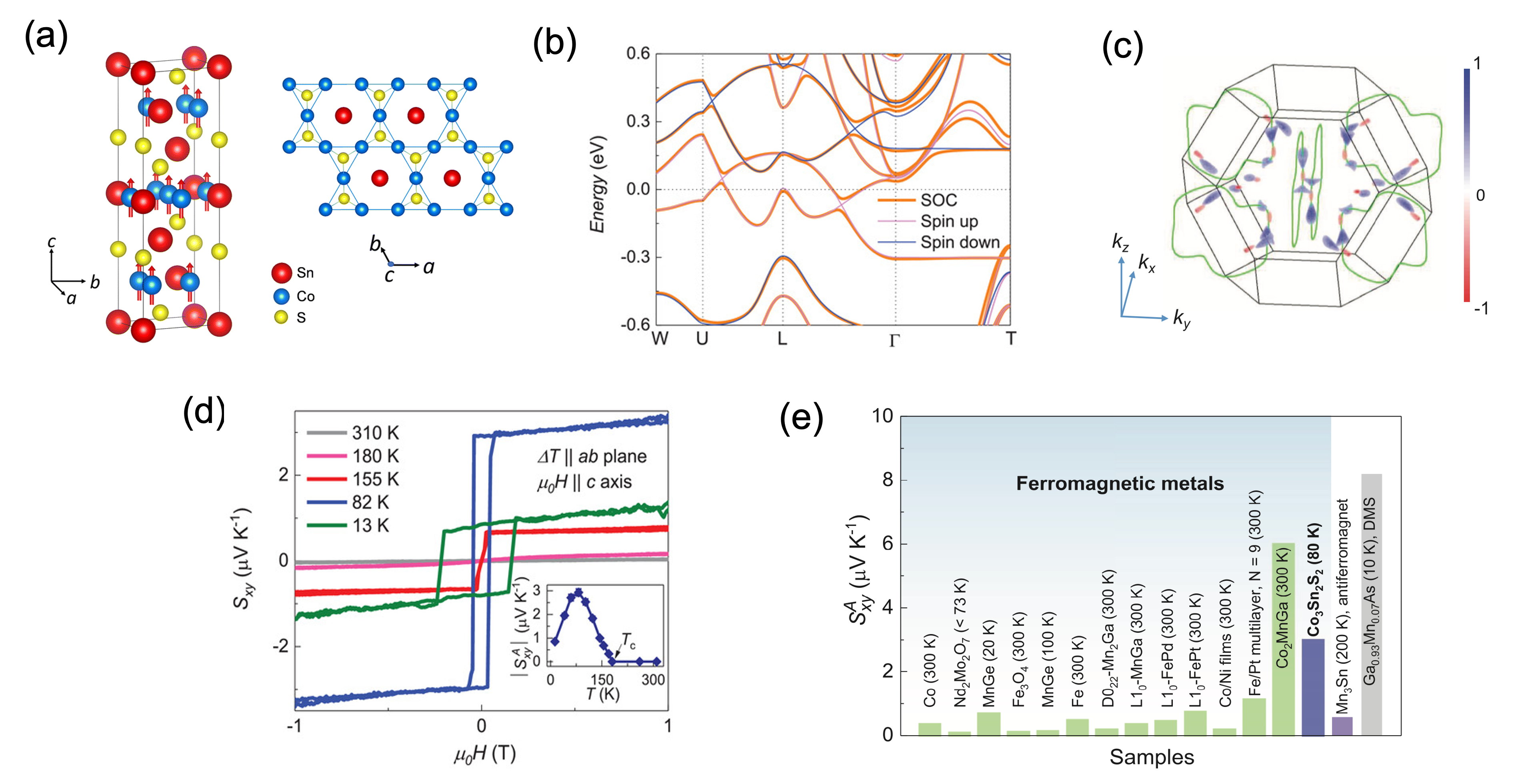}
\caption{(a) Crystal structure of Co$_3$Sn$_2$S$_2$ shown from two different viewing directions, including a view down the $c$-axis highlighting the flat hexagonal Kagome lattice. (b) Electronic band structures of Co$_3$Sn$_2$S$_2$ with and without spin-orbit coupling (SOC). (c) Temperature-modified Berry curvature distribution ($\Omega_N$) at 80 K in the Brillouin zone with the magnetization along the c-axis. (d) Magnetic field dependence of anomalous Nernst thermopower ($S_{xy}$) for a magnetised Co$_3$Sn$_2$S$_2$ sample at different temperatures \cite{guin2019zero}. (e) Absolute value of anomalous Nernst thermopowers ($S_{xy}^A$) for different materials \cite{guin2019zero}. Reproduced from ref. 71 under the terms of the Creative Commons Attribution 4.0 International License (CC BY-NC), 2019.}
\end{figure*}

\begin{figure*}[t]
\centering
\includegraphics[width=0.85\textwidth]{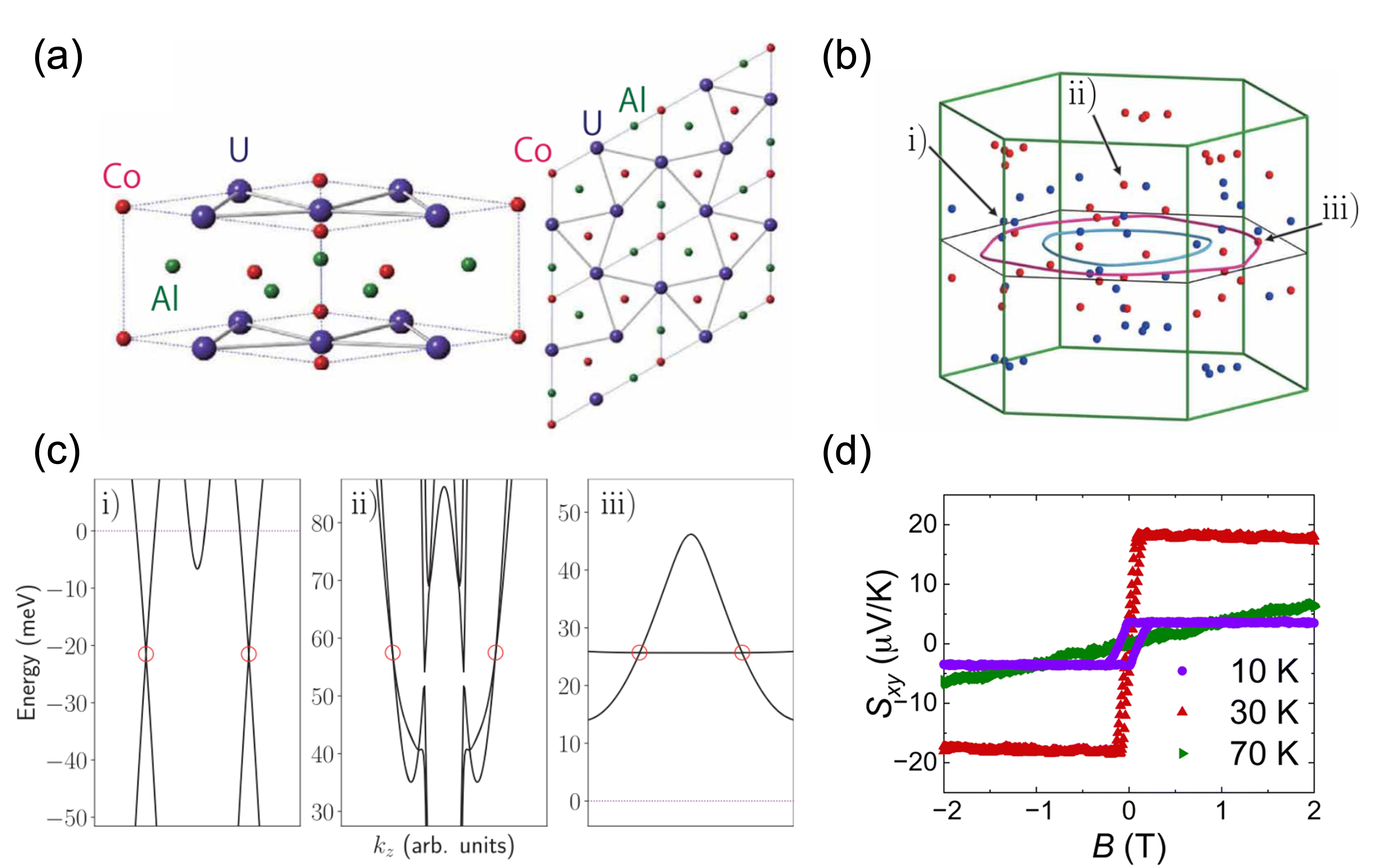}
\caption{(a) Left: The unit cell of UCoAl viewed from the [111] direction. Right: The top view of UCoAl. The solid bonds indicate the tilted kagome layer formed by uranium atoms. (b) Nodal lines and Weyl points within $\pm$60 meV of the Fermi energy in UCo$_{0.8}$Ru$_{0.2}$Al. (c) Band plots for examples of (i) type I, (ii) type II, and (iii) critically tilted Weyl points (red circles), identified with arrows in (b). (d) Anomalous Nernst thermopower of UCo$_{0.8}$Ru$_{0.2}$Al as a function of magnetic field applied along the c-axis (B‖c)\cite{asaba2021colossal}. Reproduced from ref. 52 under the terms of the Creative Commons Attribution 4.0 International License (CC BY-NC), 2021.}
\end{figure*}

\subsubsection{Centrosymmetric Dirac semimetal TaSb$_2$}
Single-crystalline topological semimetal, TaSb$_2$  \cite{wu2025dirac}, belongs to the same family as NbSb$_2$, which also shows giant Nernst transverse thermopower of 1200 $\mu\mathrm{V\,K^{-1}}$ at 25 K and 9 T, along with an exceptional transverse power factor of 8300 $\mu\mathrm{W\,cm^{-1}\,K^{-2}}$. This compound have very high mobilities, and a nearly ideal electron-hole compensation below 50 K with the charge carrier concentration of $\approx 1.3 \times 10^{20}\ \text{cm}^{-3}$ as evidenced by the comparable electron-hole Fermi pockets. The phonon-drag effect markedly amplify the Seebeck thermopower of electrons and holes, reaching a value of $S_h$ of 110.5 $\mu\mathrm{V\,K^{-1}}$ and $S_e$ of 37.9 $\mu\mathrm{V\,K^{-1}}$ at 25K. The two-carrier transport behavior leads to large $\Delta S_{xx}$, eventually contribute to large $S_{xy}$. While the polycrystalline TaSb$_2$ \cite{negi2025breakdown}, have a peak Nernst thermopower of $ \sim 37.9$ $\mu\mathrm{V\,K^{-1}}$ at 25 K under a 9 T field due to increased scattering at grain boundaries, which naturally limits the mean free path of carriers, also interrupts phonon propagation. The average value of $PF_{xx}$ displays a marginal change with magnetic field, while $PF_{xy}$ exhibits a drastic enhancement, reaching a value \(1.8 \times 10^{-4}\ \mathrm{W\ m^{-1}\ K^{-2}}\) at 9T. Importantly, the Lorenz number deviates from the Sommerfeld value between 40 K to 100 K, indicating non-Fermi liquid behavior.

\subsubsection{Centrosymmetric Dirac semimetal  ZrTe$_5$}

ZrTe$_5$ is a Dirac semimetal \cite{zheng2016transport} which crystallises in a layered orthorhombic structure (Cmcm), where ZrTe$_6$ prisms form 1D ZrTe$_3$ chains along the $a$-axis [Fig. 13(a)]. These chains are connected by additional Te atoms forming zigzag chains, creating quasi-2D layers stacked along $b$-axis via weak van der Waals interactions \cite{liang2018anomalous, wang2021giant}. A giant Nernst thermopower, reaching up to 1900 $\mu\mathrm{V\,K^{-1}}$ (in-plane) and 5000 $\mu\mathrm{V\,K^{-1}}$ (out-of-plane), has been observed at 100 K and 13 T [Figs. 13(b) and 13(c)]. This large response is widely attributed to the combined effects of strong Berry curvature and bipolar transport. Consequently, a peak transverse thermoelectric figure of merit ($z_{N}T$) $\sim$ 0.12 at 120 K and 13 T has been reported, which significantly surpasses its longitudinal counterpart [Fig. 13(d)] \cite{wang2021giant}. Cd$_3$As$_2$ is another example of centrosymmetric Dirac semimetal [Fig. 13(e)] \cite{neupane2014observation}. A $z_{N}T$ of $\sim$ 0.5 at 300 K has been realised in Cd$_3$As$_2$ single crystals under a low magnetic field of 2T, a strength attainable with commercial permanent magnets. This $z_{N}T$ has been ascribed to an enhanced Nernst effect and intrinsically low thermal conductivity [Fig. 13(f)] \cite{xiang2020large}.


\begin{figure*}[t]
\centering
\includegraphics[width=0.95\textwidth]{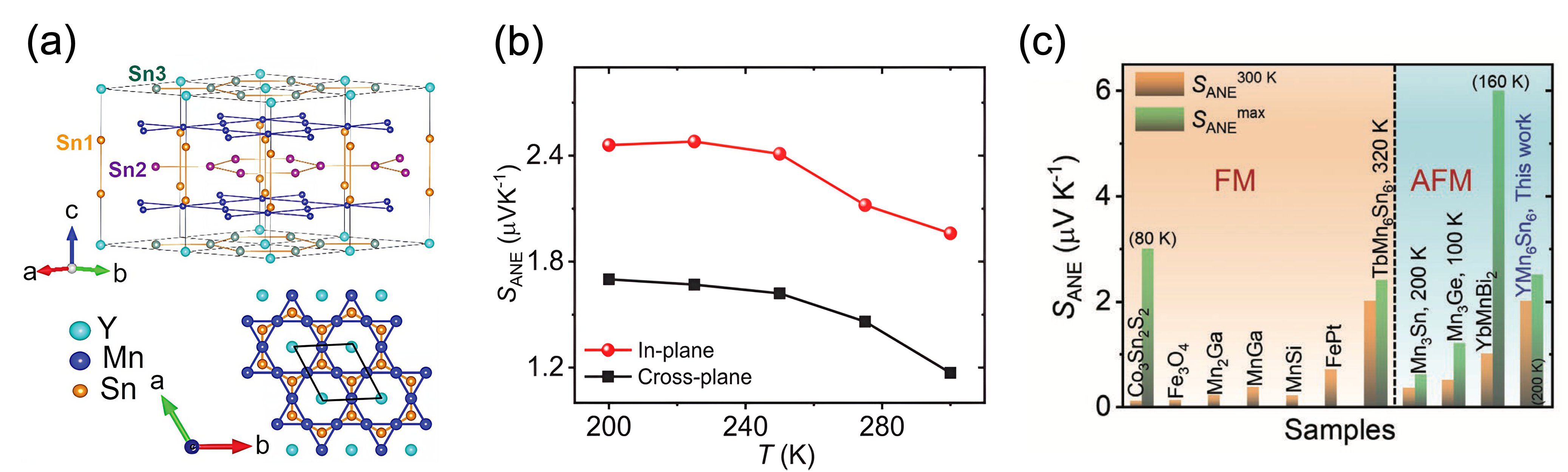}
\caption{(a) Crystal structure of hexagonal Kagome magnet YMn$_6$Sn$_6$ (P6/mmm) and (b) Temperature dependence of anomalous Nernst thermopower ($S_{ANE}$). (c) Comparison of anomalous Nernst thermopower between YMn$_6$Sn$_6$ with previously reported TE materials \cite{roychowdhury2022large}. Reproduced from ref. 137 under the terms of the Creative Commons Attribution License (CC BY), 2022.}
\end{figure*}


\subsubsection{Co and Fe-based magnetic Heusler compounds} 
Contrary to ONE, observation of a large Nernst signal in magnetic materials under low and zero magnetic fields makes the anomalous Nernst effect highly promising for transverse thermoelectrics \cite{kumar2020topological, sakai2018giant}. Our discussion of the various materials exhibiting ANE will emphasise the important breakthroughs, starting with the Heusler compounds. Co$_2$MnGa, one of the first topological Heusler magnets discovered to exhibit a large anomalous Nernst effect, possesses a Curie temperature ($T_\mathrm{C}$) of $\sim$ 690 K and a saturation magnetization of 3.7 $\mu_\mathrm{B}\,\mathrm{f.u.^{-1}}$ at 300K \cite{belopolski2019discovery}. This high $T_\mathrm{C}$, which is above room temperature, is a critical feature for thermoelectric power generation. It adopts a cubic crystal structure with a space group $Fm\bar{3}m$, which consists of four interpenetrating face-centred cubic (fcc) sublattices \cite{sakai2018giant, belopolski2019discovery, guin2019anomalous} [Fig. 14(a)]. The electronic structure of Co$_2$MnGa features three nodal lines near the $E_F$ in the absence of spin-orbit coupling [Fig. 14(b)]. SOC lifts the degeneracy of these nodal lines by breaking mirror symmetry, thereby opening a small gap. This gapping mechanism generates a substantial non-zero Berry curvature $\Omega_n(k)$, which is the origin of the material's pronounced anomalous transport properties [Fig. 14(c)] \cite{guin2019anomalous}.

Investigations on single crystals grown by the Bridgman-Stockbarger method revealed a large anomalous Nernst thermopower ($S_{xy}$) of $\sim$ 6.0 $\mu\mathrm{V\,K^{-1}}$ at 300 K, increasing to $\sim$ 8 $\mu\mathrm{V\,K^{-1}}$ at 400 K [Fig. 14(d)]. Coupled with a large anomalous Hall conductivity of $\sim$ 1600 $\Omega^{-1}\,\mathrm{cm^{-1}}$, studies on thin films exhibited anomalous signals of comparable magnitude \cite{manna2018colossal}. These values are $\sim$ 7 times more than the ANE expected from conventional ferromagnets, where the ANE thermopower is proportional to the magnetization and is given by the relation, $ |S_{xy}| = |N_{xy}| \mu_o M $, where $|N_{xy}|$ is the anomalous Nernst coefficient, ranging from 0.05 to 1 $\mu\mathrm{V\,K^{-1}T^{-1}}$ at 300 K. The observation of these large signals near room temperature is critical, considering the limited number of materials known to exhibit strong thermoelectric performance in this regime \cite{sakai2018giant}. A recent investigation on a polycrystalline Co$_2$MnGa exhibits excellent Nernst effect. Owing to the robustness of the topological effect against grain boundary scattering, polycrystalline Co$_2$MnGa demonstrates a large Nernst thermopower of approximately $\sim$ 6.5 $\mu\mathrm{V\,K^{-1}}$ at 300K, comparable to the previously reported record in its single-crystal form \cite{guin2019anomalous}. As a result, the fabricated Nernst generators achieve excellent performance, delivering an output voltage of 5.4 $\mathrm{mV}$ and power of 14.3 $\mu\mathrm{W}$, which is substantially higher than Nernst thermopiles made from conventional ferromagnets \cite{chen2024topological}.

A sister compound, Co$_2$MnAl, in the single-crystal form, was also predicted to show ANE due to presence of Weyl points near E$_F$ \cite{kubler2016weyl}. The formation of B$_2$ disorder sometime reduces the number of symmetry planes and alters the topological behaviour, reducing the ANE response, compared to its L2$_1$ ground state structure. In order to increase the atomic ordering and fine-tune the position of the Fermi level, Co$_2$MnAl is doped with Si atoms. The polycrystalline Co$_2$MnAl$_x$Si$_{(1-x)}$ has been shown to exhibit a large ANE of 4.5 $\mu\mathrm{V\,K^{-1}}$ at $x$ = 0.19 and 0.31. These samples were then annealed at 873K for three days, a temperature slightly below the L2$_1$/B$_2$ phase order–disorder transition of Co$_2$MnAl and a centimeter-sized TE Nernst generator was prepared with 11 annealed polycrystalline Co$_2$MnAl$_{0.69}$Si$_{0.31}$, having dimension 18.5 × 1.26 × 3.33 mm$^3$. With this generator an open-circuit voltage of 2.2 mV and a peak output power of 7.7 $\mu$W has been achieved at $\Delta$T of 16.5 K, as shown in Fig. 14(e) \cite{qian2025anomalous}.

A comprehensive study \cite{mende2021large} on the anomalous transverse transport in selected Fe-based Heusler single crystals (Fe$_2$YZ, Y = Co, Ni; Z = Al, Ga) revealed large intrinsic values near room temperature. These compounds, which have high Curie temperatures ($T_C$ $\sim$ 1000 K), exhibited anomalous Hall conductivities ($\sigma_{xy}^A$) of $250$--$750\,\mathrm{S\,cm^{-1}}$ and anomalous Nernst coefficients ($S_{xy}$) $\sim$ 2 $\mu\mathrm{V\,K^{-1}}$ [Fig. 14(f)]. The study proposed that anti-site disorder, contrary to its typically detrimental role in binary semiconductors, can enhance the anomalous transport properties in these complex materials. This enhancement is attributed to the introduction of additional mirror planes, which amplify the Berry curvature. This suggests that such disorder can be a viable tool for tailoring electronic and magnetic properties in ternary and quaternary systems \cite{mende2021large}. The predicted large anomalous Nernst effect in many high-$T_C$ Heusler compounds makes them promising candidates to explore for achieving high-thermoelectric performance.


\subsubsection{Ferromagnetic Kagome compounds} 

A class of materials that hosts Kagome sublattice in which atoms are arranged in a unique two-dimensional, corner-sharing triangular network. This distinct lattice geometry directly shapes their electronic structures, giving rise to flat bands and Dirac-like band crossings. These features make Kagome materials an ideal platform for investigating correlated topological phenomena, such as unconventional superconductivity and anomalous Hall effects. Notable examples of Kagome compounds include ferromagnetic systems like Co$_3$Sn$_2$S$_2$ and UCo$_{0.8}$Ru$_{0.2}$Al \cite{guin2019zero, kumar2023topological, liu2018giant}, antiferromagnets such as Mn$_3$Sn, Mn$_3$Ge, and YbMnBi$_3$, rare-earth variants including RMn$_6$Sn$_6$ (R = Tb, Gd, Tm, Lu), and the vanadium-based superconductors AV$_3$Sb$_5$ (A = K, Rb, Cs). In this section, we discuss 
ferromagnetic compounds.

\begin{figure*}[t]
\centering
\includegraphics[width=0.99\textwidth]{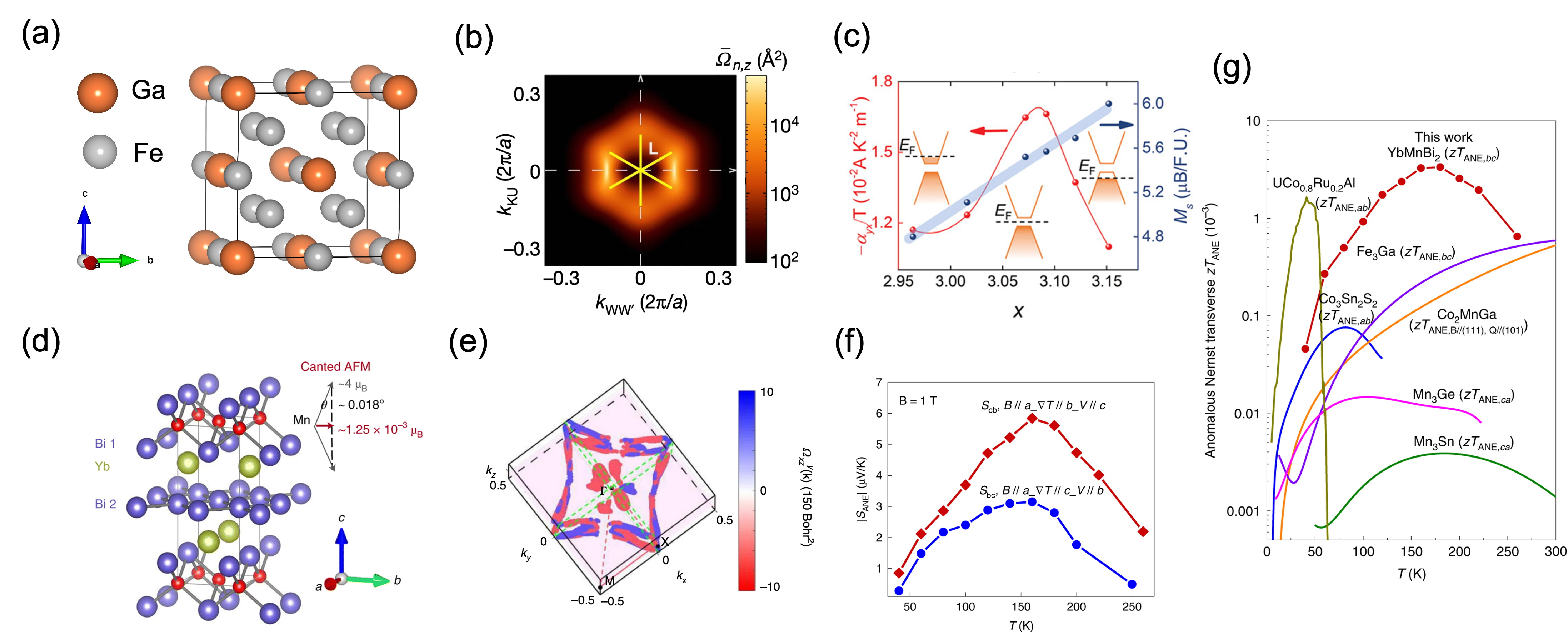}
\caption{(a) Crystal structure of Fe$_3$Ga with a fully ordered D0$_3$ configuration. (b) Contour plot of the Berry curvature $\Omega_{n,z}$ of the lower-energy band $n$, around the L point and nodal lines \cite{sakai2020iron}. Reproduced from ref. 70 with permission from Springer Nature, copyright 2020. (c) The measured thermoelectric conductivity and saturated magnetisation as a function of Fe content (x). Insets illustrate how the Fermi level shifts relative to the nodal-web electronic structure with varying x \cite{feng2022giant}. Reproduced from ref. 51 with permission from John Wiley and Sons, copyright 2022. (d) Crystal structure of YbMnBi$_2$. (e) Berry curvature distribution in the Brillouin zone calculated from first principles. (f) Temperature-dependent Nernst thermopower measured along different crystal directions. (g) Comparison of the anomalous Nernst thermoelectric figure of merit ($zT _{ANE}$) across various magnetic topological materials \cite{pan2022giant}. Reproduced from ref. 37 under the terms of the Creative Commons Attribution 4.0 International License (CC BY), 2021.}
\end{figure*}

Co$_3$Sn$_2$S$_2$, a member of the shandite family (A$_3$M$_2$X$_2$; A = Ni, Co, Rh, Pd; M = Sn, In, Pb, Tl; X = S, Se), is a ferromagnetic Weyl semimetal with a Kagome lattice with a $T_C$ = 177 K and a magnetic moment of 0.29 $\mu_\mathrm{B}\,\mathrm{Co^{-1}}$ \cite{vaqueiro2009powder}. It crystallises in a rhombohedral structure with the space group $R\overline{3}m$, where the Sn atoms occupy two distinct crystallographic sites: interlayer Sn(1) and Kagome-layer Sn(2). Sn(1) atoms exhibit trigonal antiprismatic coordination, surrounded by two Co$_3$ triangles. Sn(2) atoms form a hexagonal planar layer, while the magnetic Co atoms form a Kagome lattice within the $ab$-plane. All Co$_3$ triangles are capped by sulphur atoms positioned above and below this Kagome layer \cite{guin2019zero, liu2018giant} [Fig. 15(a)]. The material exhibits a spin-polarised electronic structure. The spin-up channel is semi-metallic, while the spin-down channel is semiconducting with a band gap of $\sim$ 0.35 eV \cite{guin2019zero, liu2018giant}. Linear band crossings are present near the Fermi level ($E_F$) in the spin-up channel along the $\Gamma-$L and L$-$U high-symmetry paths [Fig. 15(b)]. The combination of inversion and $C_3$ rotational symmetry gives rise to six protected nodal lines in the Brillouin zone, also highlighted by green lines in Fig. 15(c). Spin-orbit coupling (SOC) breaks mirror symmetry, lifting the degeneracy of these nodal lines and generating a pair of Weyl points \cite{guin2019zero} [Fig. 15(b)]. The resulting topological character is confirmed by electrical and thermal transport measurements, and by direct verification through ARPES and STM \cite{guin2019zero, liu2018giant, morali2019fermi}. Bridgman-grown single crystals exhibit a large anomalous Hall effect. Liu et al. \cite{liu2018giant} reported an anomalous Hall conductivity (AHC) of $\sim$ 1130 $\Omega^{-1}\,\mathrm{cm^{-1}}$ and an anomalous Hall angle (AHA) of above 20\%. Furthermore, a significant zero-field Nernst effect is observed in thermal transport measurements, reaching $S_{xy}$ $\sim$ 3 $\mu\mathrm{V\,K^{-1}}$ at 80 K. This, along with a hysteretic loop in the field-dependent Nernst thermopower, confirms the material's hard-magnetic nature \cite{guin2019zero} [Fig. 15(d)].

UCo$_{0.8}$Ru$_{0.2}$Al, a ferromagnetic material with a distorted Kagome lattice and strong uniaxial anisotropy, having saturation magnetization of $3.7\,\mu_\mathrm{B}\,\mathrm{U^{-1}}$ and critical temperature of 56 K [Fig. 16(a)]. This compound possesses a colossal ANE thermopower of 23 $\mu\mathrm{V\,K^{-1}}$ at T $=$ 40 K, approximately four to five times higher than the commonly reported values [Fig. 16(d)]. The reason to this colossal value is the large SOC due to uranium atom and the strong electronic correlation, which shifts multiple band crossing points near the $E_F$. The partial density of states shows a narrow band of uranium f-states dominating the electronic states close to $E_F$ \cite{asaba2021colossal}. There are 148 Weyl points within $\pm$60 $\mathrm{meV}$ of the Fermi level, including both type I and type II Weyl nodes [Figs. 16(b) and 16(c)]. Also, two nodal lines close to $E_F$ within the $\sigma_z$ mirror plane are observed. These large number of Weyl nodes and nodal line contribute to the large Berry curvature and gigantic Nernst response.

The RMn$_6$Sn$_6$ family (R = rare earth) also hosts an Mn Kagome lattice in a hexagonal structure with $P$6/$mmm$ space group [Fig. 17(a)]. In YMn$_6$Sn$_6$, this lattice gives rise to a helical antiferromagnetic state with a high Neel temperature of $\sim$ 340 K \cite{roychowdhury2022large, ghimire2020competing}. The variation in the in-plane electrical resistivity indicates that YMn$_6$Sn$_6$  is metallic, and the positive sign of the longitudinal Seebeck coefficient confirms that $p$-type carriers dominate in the system. This high magnetic ordering temperature allows this compound to exhibit a substantially large ANE $\sim$ 2.4 $\mu\mathrm{V\,K^{-1}}$ at 200K and $\sim 2.0$ $\mu\mathrm{V\,K^{-1}}$ at room temperature [Fig. 17(b)]. The presence of a Dirac point and a flat band near the Fermi level, leads to large Berry curvature-driven response signifies the family's potential for transverse thermoelectric applications \cite{li2021dirac}.

\begin{figure*}[t]
\centering
\includegraphics[width=0.95\textwidth]{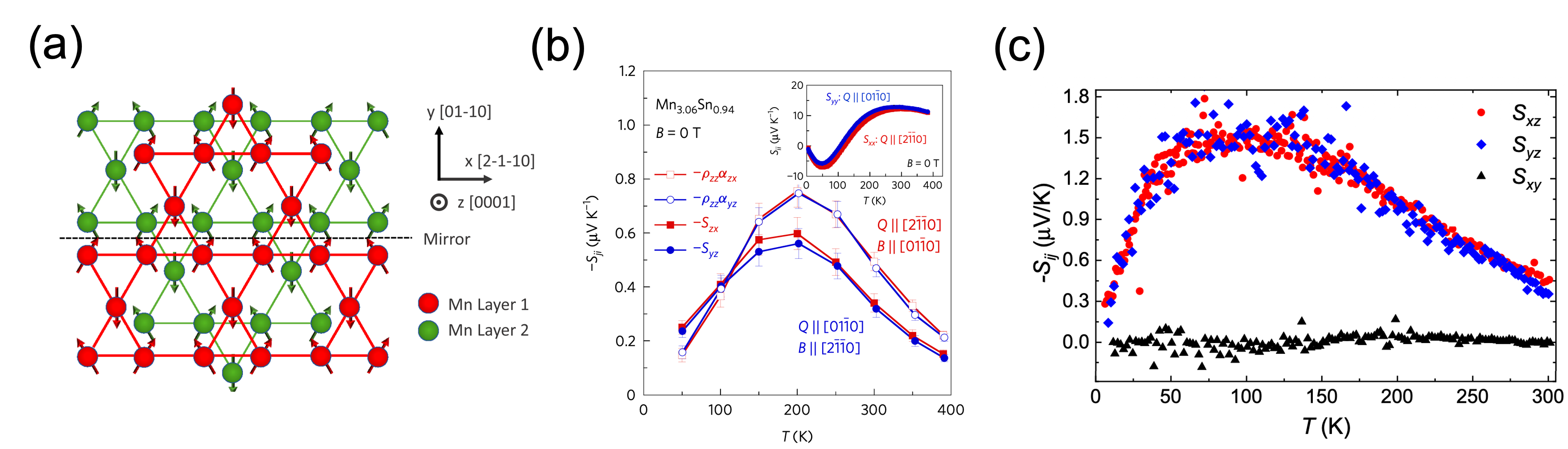}
\caption{(a) Mn atoms and magnetic structure of Mn$_3$Sn or Mn$_3$Ge. Sn or Ge atoms are in the center of each Mn hexagon (not sketched here) \cite{wuttke2019berry}. Reproduced from ref. 146 with permission from the American Physical Society, copyright 2019. (b) T dependence of the zero-field remnant Nernst signal $S_{ji}$ for Mn$_{3.06}$Sn$_{0.94}$. Insets: Seebeck coefficients $S_{xx}$, $S_{yy}$ versus T measured under zero field \cite{ikhlas2017large}. Reproduced from ref. 147 with permission from Springer Nature, copyright 2017. (c) T dependence of Nernst signal for Mn$_3$Ge \cite{wuttke2019berry}. Reproduced from ref. 146 with permission from the American Physical Society, copyright 2019.}
\end{figure*}

\subsubsection{Fe$_3$X ferromagnets} 
While the large anomalous Nernst effect (ANE) linked to extended Berry curvature has been observed in various ferromagnets, the common bcc ferromagnet $\alpha$-Fe does not exhibit a strong ANE. However, doping iron with 25\% aluminium or germanium to form Fe$_3$X compounds (X = Al, Ge) has been predicted theoretically to reveal that a "nodal web" structure (around the high-symmetry point L) formed by interconnected flat nodal lines, where valence and conduction bands touch, generates strong Berry curvature under spin-orbit coupling to show large ANE [Fig. 18(b)] \cite{sakai2020iron}. These Fe$_3$X materials crystallise in derivative bcc structures, exhibiting a spectrum of structural order from the fully ordered [Fig. 18(a)] D0$_3$ ($Fm\bar{3}m$) and partially ordered B$_2$ ($Pm\bar{3}m$) to the fully disordered A2 ($I m \overline{3} m$) structure, depending on the atomic disorder, a variable that is key to tuning their properties \cite{sakai2020iron, feng2022giant}. Investigations on Czochralski method-grown single crystals exhibiting D0$_3$ ordering by Sakai et al, have shown anomalous Nernst thermopower of $\sim$ 6 $\mu\mathrm{V\,K^{-1}}$ and $\sim$ 4 $\mu\mathrm{V\,K^{-1}}$ for Fe$_3$Ga and Fe$_3$Al, respectively, at 300 K, which is 20-fold higher than the $\alpha$-Fe ($\sim$ 0.3 $\mu\mathrm{V\,K^{-1}}$). Such anomalous Nernst thermopower couldn’t be achieved in disordered phases \cite{sakai2020iron}. Thin films of Fe$_3$Ga and Fe$_3$Al exhibit a large ANE of $\sim$ 4 $\mu\mathrm{V\,K^{-1}}$ for Fe$_3$Ga and $\sim$ 2 $\mu\mathrm{V\,K^{-1}}$ for Fe$_3$Al at room temperature under an in-plane temperature gradient. They also exhibit an ANE at zero field, for an out-of-plane temperature gradient because the films have in-plane magnetization with coercivities ($B_c$) of approximately 40 Oe and 20 Oe, respectively. Given the need for easy compositional tuning and economic viability in devices, researchers have extended these studies to polycrystalline samples. Z. Feng et al. \cite{feng2022giant} investigated the compositional range Fe$_x$Ga$_{4-x}$ (2.95 $< x <$ 3.15), finding a large ANE of $\sim$ 5.4 $\mu\mathrm{V\,K^{-1}}$ at 300 K in the polycrystalline compound Fe$_{3.09}$Ga$_{0.91}$. The magnitude of the ANE is governed by the position of the nodal web relative to the Fermi level ($E_F$). Theoretical studies indicate that increasing the Fe content ($x$) raises the energy of this nodal web [Fig. 18(c)]. The observed peak in Berry curvature and conductivity at $x$ = 3.09 signifies the optimal condition where $E_F$ lies within the energy gap of the nodal web, maximizing the effect [Fig. 18(c)]. Further increase in $x$ pushes $E_F$ below the nodal web, degrading performance.

\subsubsection{Canted antiferromagnet} 
Manganese-based canted antiferromagnets, such as YbMnBi$_2$, are promising materials due to their unique electronic properties \cite{borisenko2019time, soh2019magnetic, pan2022giant}. They possess topological electronic structures and highly dispersive bands that result in low electrical resistivity. YbMnBi$_2$ has a tetragonal crystal structure $(P4/nmm)$. This structure consists of magnetic MnBi$_4$ layers and Bi square nets \cite{pan2022giant} [Fig. 18(d)], which are predicted to host Weyl fermions. Below 290 K, Mn spins in the compound align antiferromagnetically, but stack ferromagnetically along the $c$-axis, causing band splitting. A slight canting of these spins leads to a small net magnetisation ($\sim$ $1.25 \times 10^{-3} \, \mu_\mathrm{B} \, \mathrm{f.u.}^{-1}$). When this canting is combined with the strong spin-orbit coupling of bismuth atoms, it breaks time-reversal symmetry. This breaking of symmetry generates a large Berry curvature by transforming nodal lines in the electronic structure into pairs of Weyl nodes [Fig. 18(e)]. The sharp dispersion of p(x/y) orbital of Bi will contribute of high conductivity \cite{soh2019magnetic, pan2022giant}. Measurements show peak Nernst thermopower values of $\sim$ 6 $\mu\mathrm{V\,K^{-1}}$ and $\sim$ 3 $\mu\mathrm{V\,K^{-1}}$ at 160 K for crystal orientations $cb$ and $bc$, respectively [Fig. 18(f)]. Herein, $cb$ indicates the measurement of the Nernst voltage along the $c$ axis and the temperature gradient along the $b$ axis, and vice versa for $bc$, when the magnetic field B is always along the $a$ axis. Furthermore, it achieves a high ANE conductivity of $\sim$ 10 A m$^{-1}$ K$^{-1}$  in the $bc$ configuration. Thus, YbMnBi$_2$ shows a anomalous Nernst thermoelectric figure of merit of $3.25 \times 10^{-3}$ at 160K in the $bc$ configuration \cite{pan2022giant, wen2025ultrahigh}. These values are comparable to those observed in ferromagnetic topological materials \cite{pan2022giant} [Fig. 18(g)]. These findings demonstrate that canted antiferromagnets like YbMnBi$_2$ are strong candidates for applications in transverse thermoelectrics. Similarly, SmMnBi$_2$ has a Neel temperature of 235 K, with spins largely along the easy $c$-axis but having a small canting towards the $ab$-plane. SmMnBi$_2$ has maximum ANE of 1.8 $\mu\mathrm{V\,K^{-1}}$, and 1.6 $\mu\mathrm{V\,K^{-1}}$ at 300K. Tang et. al \cite{tang2024unconventional} showed that the unconventional AHE (UAHE) and large ANE with unique field dependence are observed both above and below the T$_N$. The UAHE is associated with spin canting, implying that both the vector spin chirality and the SOC effect contribute to the Berry curvature. He showed that the giant ANE is probably dominated by some extrinsic mechanisms, because it violates the universal relation of $|\alpha_{xy}^A/\sigma_{xy}^A|$ $\sim k_B$/e, valid for several topological materials.

\begin{figure*}[t]
\centering
\includegraphics[width=0.99\textwidth]{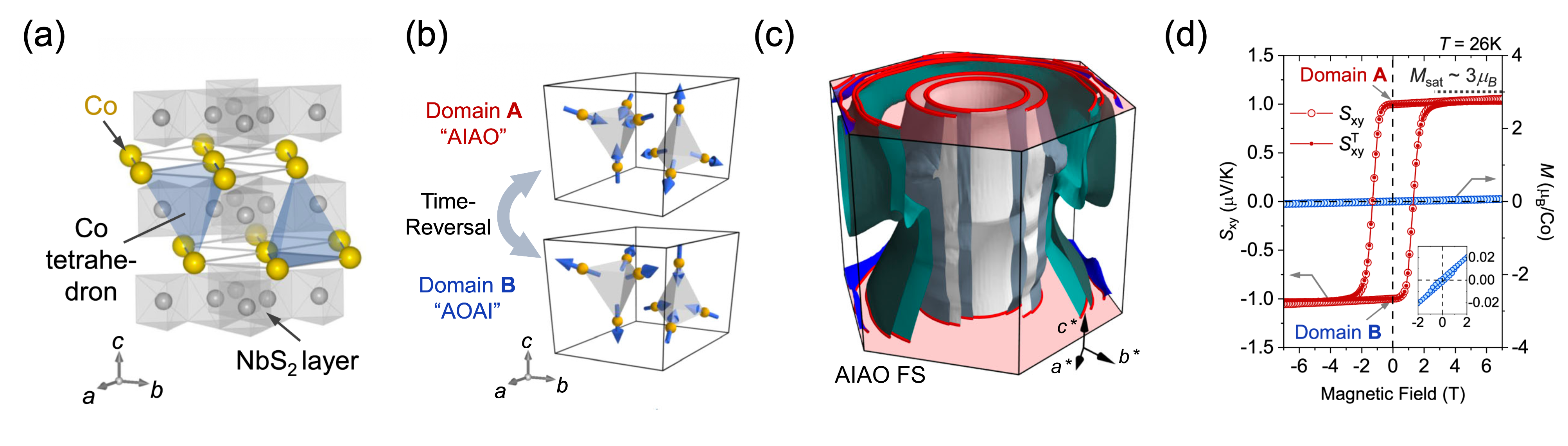}
\caption{(a) Layered CoNb$_3$S$_6$ with chiral lattice, (b) Non-coplanar all-in-all-out (AIAO) magnetic order of cobalt ions (yellow spheres). Two magnetic domains are related by time reversal symmetry. (c) Fermi surface and first Brillouin zone in the AIAO compensated antiferromagnetic state. The red shading indicates nodal planes at $k_z = \pm \pi/c$, protected by chiral P6$_3$22 symmetry in PM, and slightly gapped in AIAO. Red lines are cross-sections of Fermi surface sheets with $k_z = \pm \pi/c$. (d) Isotherms of the total Nernst effect $S_{xy}(B)$ before subtraction of the linear part related to the classical Lorentz force on moving charge carriers, and topological Nernst effect $S^T_{xy}(B)$ after subtraction (left axis). The right axis shows the magnetization M at T = 26 K. Inset: expanded view of M in the low-B regime, with hysteresis. \cite{khanh2025gapped}. Reproduced from ref. 152 under the terms of the Creative Commons Attribution 4.0 International License (CC BY-NC-ND), 2025.}
\end{figure*}

\subsubsection{Antiferromagnetic Kagome compounds} 
The intrinsic anomalous Hall effect (AHE) requires a non-zero Berry curvature, which is present in ferromagnets due to broken time-reversal symmetry but vanishes in collinear antiferromagnets. In non-collinear antiferromagnets, however, the unique spin arrangement generates the necessary Berry curvature for a large AHE \cite{ikhlas2017large}, a phenomenon that has been observed in materials such as Mn$_3$Sn and Mn$_3$Ge crystallising in Ni$_3$Sn-type hexagonal structures with P6$_3$/mmc space group. In this layered structure, the hexagonal unit cell consists of two layers of Mn triangles stacked along the $c$-axis [Fig. 19(a)]. In each layer, the Mn atoms form a Kagome lattice, with the Sn or Ge atoms sitting at the centre of the hexagons \cite{ikhlas2017large, nayak2016large}. Despite possessing only, a tiny net magnetic moment (on the order of 10$^{-2}$ to 10$^{-3}$ $\mu_\mathrm{B}\,\mathrm{Mn^{-1}}$) due to their orthorhombic symmetry, these chiral antiferromagnets exhibit a remarkably large AHE. The Nernst coefficient takes a value of $\sim 0.35$ $\mu\mathrm{V\,K^{-1}}$ at room temperature at zero magnetic field and reaches at $\sim 0.6$ $\mu\mathrm{V\,K^{-1}}$ at 200K for Mn$_3$Sn [Fig. 19(b)]. The magnetization increases linearly beyond the coercive field of 100 G also, as sublattice moments cant towards the field direction, while the Nernst signal saturates. This shows that the contribution from normal Nernst effect and the conventional ANE is negligible. Mn$_3$Ge \cite{wuttke2019berry, nayak2016large} has a large Nernst signal of around $ 0.4 - 1.5$ $\mu\mathrm{V\,K^{-1}}$, depending on the temperature [Fig. 19(c)]. These antiferromagnetic materials have Weyl points near the Fermi level that contribute to having a net non-zero Berry curvature \cite{guo2017large, yang2017topological}. 
Mn$_5$Si$_3$ has two inequivalent Mn lattice sites, say Mn$_1$ and Mn$_2$. It exhibits collinear antiferromagnetic (AFM) ordering between 60 K and 100 K, where Mn$_1$ sites have zero magnetic moment, and two-thirds of Mn$_2$ sites display oppositely oriented moments on adjacent sites in the orthorhombic structure. Below 60 K, a noncollinear and noncoplanar AFM ordering exists, with both Mn$_1$ and two-thirds of Mn$_2$ sites acquire magnetic moments \cite{ surgers2024anomalous}. Epitaxial thin films of Mn$_5$Si$_3$ grown on Si, exhibit altermagnetic ordering, have a zero-field $S_{xy}$ of 0.26 $\mu\mathrm{V\,K^{-1}}$ and an ANE conductivity of 0.22 \( \mathrm{A\,m^{-1}\,K^{-1}}\). These films possess a small net magnetization ranging from $20$--$60\,\mathrm{m}\mu_\mathrm{B}\,\mathrm{f.u.^{-1}}$ \cite{badura2025observation}.

\subsubsection{Chiral antiferromagnet} 
CoNb$_3$S$_6$ has a hexagonal chiral structure with the space group P6$_3$22, in which the intercalation of Co ions breaks the mirror planes and inversion symmetry of NbS$_2$ in the paramagnetic state [Fig. 20(a)]. It exhibits time-reversal symmetry and a $6_3$ screw axis in this space group, enforcing nodal plane degeneracy at $k_z = \pm \pi/c$. At T = 28 K, CoNb$_3$S$_6$ undergoes a magnetic phase transition to a twisted all-in-all-out (AIAO) antiferromagnetic state with the magnetic space group P32' [Fig. 20(b)]. It has a small net magnetization of \( 2 \times 10^{-3}\,\mu_\mathrm{B}\,\mathrm{f.u.^{-1}} \) in the AIAO order. The magnetic unit cell is four times larger than the paramagnetic cell. The magnetic order in the AIAO state breaks both time reversal symmetry and $6_3$ screw symmetry, lifting the degeneracy of Kramers pairs even at  $k_z = \pm \pi/c$ [Fig. 20(c)]. The non-symmorphic spin-space group symmetry in the AIAO order prevents a uniform band hybridization across the Brillouin zone, and suppresses contributions to $z$-component of emergent magnetic field when $k_z \neq \pm \pi/c$. The scalar spin chirality $\chi$ from AIAO order and the nodal plane give rise to a large ANE. At a temperature just below the transition to the AIAO state, T = 26 K, it exhibits a $S_{xy}$ of $1$ $\mu\mathrm{V\,K^{-1}}$ [Fig. 20(d)] \cite{khanh2025gapped}. This work highlights how spin topology and band topology together enable strong anomalous transport in compensated antiferromagnets.

\subsection{Effect of Bipolar conduction}

Semimetals having comparable carrier concentration of holes and electrons, have an additional component in  electronic thermal conductivity due to bipolar diffusion ($\kappa_{bi}$). The thermal energy at the hotter end create electron-hole pairs in zero or narrow band gap materials. These pair drift towards the colder end and recombine, releasing the energy as heat. The bipolar thermal conductivity can be expressed as \cite{price1955cxxxv, goldsmid1956thermal}

\[
\kappa_{bi} = \frac{\sigma_n \sigma_p}{\sigma_n + \sigma_p} (S_p - S_n)^2 T
\]

and the total electronic thermal conductivity is defined as,

\[ \kappa_e = \kappa_n + \kappa_p + \kappa_{bi} \]

Where, $\sigma_i$ and $S_i$  $(i=n,p)$ are the $n$- or $p$-type electrical conductivity and Seebeck coefficient for electrons and holes, respectively. In  longitudinal effect, the electrons and holes  move simultaneously in the same direction, often having significant bipolar contribution to detriment the thermoelectric performance. Band engineering has been used to increase the band gap, which effectively reduce the bipolar diffusion \cite{wang2020bi8se9}. Increasing the majority carrier concentration (either holes or electrons) to suppress the $(S_p - S_n)$ contribution and mitigate the bipolar effect is employed \cite{qu2024synergistic}. Semimetals having linear dispersive bands accompanied with a heavy regular band also behave as heavily doped and have very low bipolar conductivity \cite{pan2021thermoelectric}. In the transverse mode, the bipolar effect still remain intact, but acts distinctly. The charge carriers are differentiated by their sign and velocity, and carry both heat and charge in the transverse direction. The bipolar effect, which was a performance limitation in traditional longitudinal thermoelectricity, can act as a performance enhancer in transverse thermoelectricity \cite{chen2021leveraging}.

\subsubsection{Supressing the bipolar effect in longitudinal thermoelectrics}

The bipolar diffusion has been the primary bottleneck limiting the performance of traditional narrow-bandgap longitudinal thermoelectric materials. It is intrinsic to narrow-band-gap materials, and its magnitude is higher in semimetals. In the case of longitudinal thermoelectric effects, the temperature gradient drives both electrons and holes in the same direction, producing concurrent effects. First, charge compensation between electrons and holes reduces the total Seebeck coefficient ($S_{xx}$) and the Seebeck conductivity ($\alpha_{xx}$), thereby lowering the power factor ($PF = S\alpha$ = $S^2\sigma$). Second, the partial electron and hole currents induced by the temperature gradient generate additional Peltier heat flow ($\kappa_{bi}$), thereby increasing the total thermal conductivity \cite{chen2021leveraging}.

Hence, to suppress bipolar diffusion, one effective strategy has been to widen the bandgap through doping or alloying. For instance, in Se alloying, Se substitution for Te in $n$-type Bi$_2$Te$_3$ lowers the formation energy of donor-type Te$_{Bi}$/Se$_{Bi}$ defects, thus increasing electron concentration and improving thermopower. At low Se content, alloying also widens the band gap and reduces lattice thermal conductivity \cite{greenaway1965band, lou2026recent}. In $p$-type Bi$_2$Te$_3$, Sb plays a complementary role; here, Sb on Bi sites lowers the formation energy of acceptor-type Sb$_{Te}$/Bi$_{Te}$ antisite defects and increases hole concentration, while excess Te can suppress defect formation when the carrier concentration becomes too high and leads to widening of the band gap. This could be attributed to the larger bandgap of Sb$_2$Te$_3$ (0.23 eV) than that of Bi$_2$Te$_3$ (0.13 eV) \cite{zheng2014high, lou2026recent}. Thus, Sb and Se alloying provide analogous defect and band-gap engineering in $p$- and $n$-type Bi$_2$Te$_3$, respectively. However, such strategies in SnTe and Pb$_{0.6}$Sn$_{0.4}$Te-based TCI's could be detrimental for near room temperature performance \cite{wang2025advances}. Generally, physical perturbations or dopants such as Na or K are employed to break the crystal symmetry, thereby widening the vanishingly low bulk band gap. These dopants also increase $p$-type carrier concentration and shift the high-$zT$ window to higher temperatures. The resulting loss of mirror symmetry transforms the material into a topologically trivial regime \cite{roychowdhury2015tailoring, roychowdhury2016effect, wang2025advances}. Similarly, In doping enhances the Seebeck coefficient via resonant states by breaking crystal symmetry. In contrast, theoretical predictions suggest SnTe antisite defects can achieve comparable or better thermoelectric enhancement while fully preserving the nontrivial topology\cite{muzaffar2020antisite}, warranting extensive experimental investigations. To further suppress high temperatures, bipolar diffusion energy-filtering approaches have been used in TCIs through strategies such as ordered grain-boundary reconstruction\cite{deng2024ordered}, dislocation arrays\cite{meng2017grain}, and composites \cite{lin2020expression} to effectively scatter low-energy carriers while allowing  the high-energy carriers to transfer more heat, which in turn leads to enhanced S. Similarly, tetradymites near RT performance has been enhanced by such interfacial engineering\cite{kim2024selective}. Beyond such interfacial engineering, topological surface states offer an intrinsic energy-filtering channel. For example, in Bi$_{2-x}$Sb$_x$Te$_{3-y}$Se$_y$ films, isolated metallic Dirac states yield $S= -$ 44 $\mu\mathrm{V\,K^{-1}}$, while gap-opened states reach $-$ 212 $\mu\mathrm{V\,K^{-1}}$, governed by strongly energy-dependent relaxation time beyond the Born approximation \cite{matsushita2017thermoelectric}.

\subsubsection{Exploiting the bipolar effect to enhance transverse thermoelectric performance}

The bipolar effect usually decreases the Seebeck coefficient in longitudinal thermoelectrics. However, optimal performance in transverse thermoelectrics favors a strong bipolar effect. In the transverse case, the electrons and holes are deflected in opposite directions in the presence of magnetic field or Berry curvature because they carry opposite charges. Consequently, their transverse currents add constructively rather than canceling each other.

The ideal band structure for maximum transverse effect consists of slightly overlapping conduction and valence bands near their edges. The electronic band structure of the cubic semimetal Mg$_2$Pb satisfies the mentioned criteria, with $\Gamma_8$ valence band and $X_7$ conduction band contributes to charge transport \cite{chen2021leveraging}. The transverse current usually have two components, the charge current driven by the thermal Hall effect and the Nernst effect. To maximize the Nernst effect, we must reduce the thermal Hall contribution, which vanishes under perfect electron-hole compensation. The longitudinal thermopower (S$_{xx}$) is small becuase of charge compensation. At low temperatures (T $<$ 30K), it is dominated by holes, whereas at relatively high temperaturess (T $>$ 30K), it is dominated by electrons. The contributions from electrons and holes become identical at 30 K, where the transverse thermopower (S$_{xy}$) reaches its maximum. When the partial conductivities of electrons and holes become nearly identical at $\sim$ 30K, the bipolar effect is maximized. At 30 K and 10 T, a giant transverse thermoelectric power factor of 400 $\mathrm{\mu W\,cm^{-1}\,K^{-2}}$ is achieved, reaching a three order of magnitude enhancement over the longitudinal configuration. In a single band Nernst response, the transverse thermopower peaks at low fields ($\mu B \approx 1$), and decays toward zero as the magnetic field enters the high-field limit ($\mu B \gg 1$). When both electrons and holes are present, the bipolar component does not vanish. In the high-field limit, the bipolar transverse thermopower scales linearly with $B$, while the conductivity is inversely proportional to $B^2$. This causes the transverse power factor to saturate at a certain field. Any deviation from the optimal band structure, including highly asymmetric conduction and valence bands in InSb \cite{nakamura1997transport}, an overly overlapped band gap in Bi-Sb \cite{jandl1994thermogalvanomagnetic} and a high residual carrier concentration in Pb$_{1-x}$Sn$_x$Te \cite{jovovic2008low}, leads to a decrease in transverse power factor.

A narrow band gap semiconductor, Ag$_{2-\delta}Te$, can achieve a peak transverse thermopower of 625 $\mu\mathrm{V\,K^{-1}}$ at 75 K under a 14 T magnetic field due to an enhanced bipolar effect. At this temperature, the partial conductivities of electrons and holes are balanced, resulting in a zero longitudinal thermopower. Above 75 K, thermal excitation increases the electron concentration, thereby suppressing bipolar conduction. Around 200 K, minority holes are thermally activated across the band gap, causing S$_{xy}$ to increase again up to room temperature. The highest Nernst power factor, PF$_N$, achieved is 5.32 $\mathrm{\mu W\,cm^{-1}\,K^{-2}}$ at 300 K under 12 T. At 300 K, the maximum $Z_N$ value reaches $4.7 \times 10^{-4} \, \mathrm{\,K^{-1}}$ under 14 T, corresponding to a peak $z_NT$ of 0.15. Materials such as NbSb$_2$ \cite{li2022colossal}, TaSb$_2$ \cite{wu2025dirac} and ZrTe$_5$ \cite{neupane2014observation}, which exhibits large Nernst response, as discussed previously in the paper, are attributed to compensated electron-hole concentrations and bipolar transport, along with other non-trivial topological properties.

\subsection{Near Room-temperature Thermoelectrics}

\subsubsection{Texture engineering and lattice planification in longitudinal thermoelectrics}

Enhancing the in-plane texture, specifically, promoting alignment along the ($00l$) planes, has been recognised as an effective strategy to maximise TE performance in textured polycrystalline materials for $n$-type conduction. $n$-type polycrystals are more vulnerable to texture loss than $p$-type polycrystals; hence, maximising in-plane ($00l$) alignment is essential for high thermoelectric performance. Floating-zone growth provides highly oriented Bi$_2$Te$_3$-based polycrystals with large columnar grains and near-single-crystal ($00l$) texture \cite{zhai2018tunable, lou2026recent}, but weak van der Waals bonding between quintuple layers leads to easy cleavage; thermomechanical deformation (extrusion, rolling, hot forging) can recrystallise and reorient grains to restore texture and improve in-plane transport\cite{hu2015tuning, hu2018synergistic}. Complementarily, trace Cu in $n$-type Bi$_2$(Te,Se)$_3$ enables lattice planification, where Cu fills Bi vacancies to reduce defect scattering and modulate bands, while interstitial Cu bonds with the matrix to create additional electron pathways. These multiple Cu occupations boost carrier mobility and electrical performance, yielding a room-temperature $zT$ of $\sim$ 1.3 and $zT_{ave}$ of $\sim$ 1.2 between 300–523 K. Further, the performance was found to be uniform across kilogram-scale ingots \cite{liu2025lattice}.  In a parallel study, Ag-doped $n$-type Bi$_2$(Te,Se)$_3$, where trace Ag intercalates into the van der Waals gaps, bonds with Te$_1$, suppresses Te$_1$ vacancies, and acts as an $n$-type dopant to boost carrier concentration and mobility while lowering lattice thermal conductivity. The optimised doped with 0.3\% Ag reaches $zT$ $\sim$ 1.1 at room temperature, while 0.4\% Ag doping leads to $zT_{ave}$ $\sim$ 1.1 over 300–523 K. Further full-scale devices deliver a cooling $\Delta T_{max}$ $\sim$ 68.3 K at 300 K, and 84.8 K at 343 K, and a power-generation efficiency $\eta_{max} \sim$ 6.0 \% at $\Delta T$ = 240K \cite{peng2025lattice}. Together, these studies establish lattice planification as an effective design principle for n-type Bi$_2$Te$_3$-based thermoelectrics. Cu is optimal for band-structure-driven mobility, while Ag is effective for intercalation-driven defect passivation and thermal-conductivity reduction.

\subsubsection{Topological phase transition and Dirac cone engineering in longitudinal thermoelectrics}

Beyond Bi$_2$Te$_3$, Mg$_3$Sb$_{2-x}$Bi$_x$ solid solutions are promising for near-room-temperature applications, offering improved mechanical stability and intrinsically low thermal conductivity. While $n$-type Mg$_3$Sb$_{2-x}$Bi$_x$ benefits from a high valley degeneracy ($N_V$ = 6) at the conduction band minimum, its $p$-type counterpart suffers from low $N_V$ at the valence band maximum, resulting in inferior power factor and $zT$. A topological electronic transition offers a route to overcome this, where tuning the Bi content ($x$) reduces the bandgap and strengthens spin–orbit coupling, driving a transition from a trivial semiconductor to a topological semimetal. At the optimal composition Mg$_3$Sb$_{0.5}$Bi$_{1.5}$, a positive bandgap of 0.07 eV and a high $N_V$ of 6 are simultaneously achieved. This yields a synergy of high carrier concentration, mobility, and effective mass (1.4 $m_0$). The Ag-doped Mg$_3$Sb$_{0.5}$Bi$_{1.5}$ films achieves power factors of 3.54 and 6.93 $\mathrm{\mu W\,cm^{-1}\,K^{-2}}$ at 300 and 575 K, respectively \cite{xie2024topological}. A complementary strategy for topological band engineering is demonstrated in the Bi$_4$Te$_3$, a member of the [Bi$_2$]$_m$[Bi$_2$Te$_3$]$_n$ homologous series. Increasing the Bi$_2$ bilayer ratio in Bi$_4$Te$_3$ ($m$=3, $n$=3) sharpens the Dirac cone and increases the carrier mobility by twofold to 117 $\mathrm{cm^2\,V^{-1}\,s^{-1}}$ from 58 $\mathrm{cm^2\,V^{-1}\,s^{-1}}$ in BiTe ($m$=1, $n$=2). This is attributed to strengthened $\sigma$-bonding interlayer coupling via orbitals, which reduces the carrier effective mass from 3.35 $m_e$ to 0.06 $m_e$. Simultaneously, the abundant Bi$_2$ bilayer raises the Te vacancy formation energy, thereby suppressing electron-carrier concentration and enhancing the Seebeck coefficient from $-$34 to $-$ 63 $\mu\mathrm{V\,K^{-1}}$. The room-temperature $zT$ increases from 0.05 to 0.11. This demonstrates that Dirac cone sharpening via natural heterostructure engineering is a distinct topological knob for simultaneously improving mobility and the Seebeck coefficient \cite{guo2026sharpened}.

\subsubsection{Current bottlenecks and design strategies in Transverse thermoelectric materials}

Transverse thermoelectrics offer major architectural advantages over longitudinal thermoelectrics, such as single-material operation, decoupled charge and heat flow, and simpler contacts. However, to date, high-performance transverse thermoelectric materials generally achieve peak performance well below room temperature and often require large magnetic fields that are unattainable without low-temperature superconducting magnets, thereby necessitating cryogenic conditions. For practical deployment and low-grade heat recovery, materials with competent transverse thermoelectric performance are essential. Yet only a few are known to exhibit considerable performance near room temperature, and these remain far from reaching the performance of their counterparts observed at cryogenic temperatures. Furthermore, materials with large transverse power factors exhibit high longitudinal thermal conductivity, leading to a trade-off between power factor and thermal conductivity that, in turn, limits their $zT$ to well below that of longitudinal thermoelectrics. Hence, materials with low thermal conductivity and strong bipolar effects near room temperatures are necessary. Moreover, only a handful of magnetic topological materials combine a high magnetic ordering temperature, strong spin–orbit coupling, and nonzero Berry curvature near $E_F$, and good transport properties. High-$T_c$ Heusler compounds such as Co$_2$MnGa ($T_c$ $\sim$ 690K) and Fe$_2$YZ ($T_c$ $\sim$ 1000K) are promising, but their ANE values at 300 K are still limited. Goniopolar materials are elusive, with only about a dozen compounds experimentally identified as exhibiting axis-dependent conduction polarity. Hence, the two essential avenues to explore alongside the quest for new materials are: (i) Engineering the known room temperature transverse thermoelectric materials to achieve higher performance, and (ii) shifting the transverse thermopower of low temperature materials towards room temperature \cite{feng2024transverse}.

High-performance transverse thermoelectrics depend on engineering Berry curvature, which acts as an effective magnetic field and controls the anomalous Nernst conductivity $\alpha_{xy}^A$. Large Berry curvature near $E_F$ arises in topological magnets with Weyl points, and nodal lines. For instance, in Co$_2$MnGa, the combination of large Berry curvature and nodal-line semi-metallic states provides strongly enhanced Berry curvature near $E_F$, enabling macroscopic responses that exceed empirical scaling laws based on magnetisation alone \cite{guin2019anomalous, sakai2025berry}. In electron-doped Co$_2$MnGa, it is further illustrated by achieving a giant anomalous Nernst power factor of $\sim$ 47.8 $\mathrm{\mu W\,m^{-1}\,K^{-2}}$ at room temperature through synergistic tuning of the entropy-density-weighted Berry curvature and the Fermi surface\cite{chen2025design}. As exemplified by Chen et al.\cite{chen2025design} large Berry curvature alone is insufficient, the anomalous Nernst conductivity is proportional to the energy derivative of the anomalous Hall conductivity, $\alpha_{xy}^{A} \propto \left. \frac{\partial \sigma_{xy}^{A}(\varepsilon)}{\partial \varepsilon} \right|_{\varepsilon=E_F}$. This makes fermi level engineering as an equally important design strategy. Often, maximum  occurs when $E_F$ is shifted away from the Berry-curvature hotspot, where the slope of $\sigma_{xy}^A$ is steepest. In Co$_2$MnGa, the maximum $\sigma_{xy}^A$  was theoretically predicted slightly above intrinsic $E_F$, making electron doping a viable route, as experimentally demonstrated by Chen et al.\cite{chen2025design}.

Beyond these, the mixed-dimensionality of Fermi surfaces as demonstrated in LaPt$_2$B offers a zero-field route with competitive performance at 300 K, achieving a transverse $zT$ $\sim 1.5 \times 10^{-3}$, which is $\sim$ 3 times larger than that of full-Heusler topological magnets\cite{manako2024large}. In TbPtBi at 300 K and 13.5 T, a transverse thermopower (S$_{yx}$) of $\sim 214$ $\mu\mathrm{V\,K^{-1}}$ and power factor $\sim 35$ $\mathrm{\mu W\,cm^{-1}\,K^{-2}}$ were observed, which are attributed to intrinsic Berry curvature of its Weyl nodes and the high carrier mobility of its linear bands, while the bipolar effect above 79 K further enhances the transverse response through thermally excited electron–hole pairs \cite{wang2023large}. This shows that narrow-gap magnetic semimetals can achieve competitive room-temperature transverse performance by optimising Berry curvature and bipolar transport. Low-temperature mechanisms such as phonon drag and dimensionality reduction, while not directly applicable at room temperature, highlight the importance of anisotropic Fermi surfaces and sharp density of states \cite{li2025thermomagnetic, wang2026observation}, which may inspire new room-temperature materials when combined with high-$T_c$ magnetic order or robust carrier compensation.

\subsection{Device research and application}

Historically, longitudinal thermoelectrics have been the primary focus, and remain the only architecture utilised in practice, with significant progress in materials performance and device integration. These advancements have been extensively documented in the recent review by Chen et. al \cite{shi2025toward}. Recent developments in the entropy-engineered tetradymite-based modules have further demonstrated the promising efficiencies and power output for the practical deployment of thermoelectric generators \cite{ahmad2024largely}. In addition to being a leading choice for power generation, tetradymite compounds have also attracted considerable interest for Peltier-effect cooling, alongside Bi$_{1-x}$Sb$_x$ alloys \cite{mao2021thermoelectric, qian2025thermoelectric}. Moreover, recent advances in the magneto-Seebeck effect in Bi$_{1-x}$Sb$_x$ alloys have achieved a $zT$ of $\sim$ 1.5 over the 120-225 K temperature range, with moderate magnetic fields achievable with permanent magnets, highlighting significant potential for this material system \cite{qian2025thermoelectric, liu2011thermoelectric}.

\begin{figure*}[t]
\centering
\includegraphics[width=0.95\textwidth]{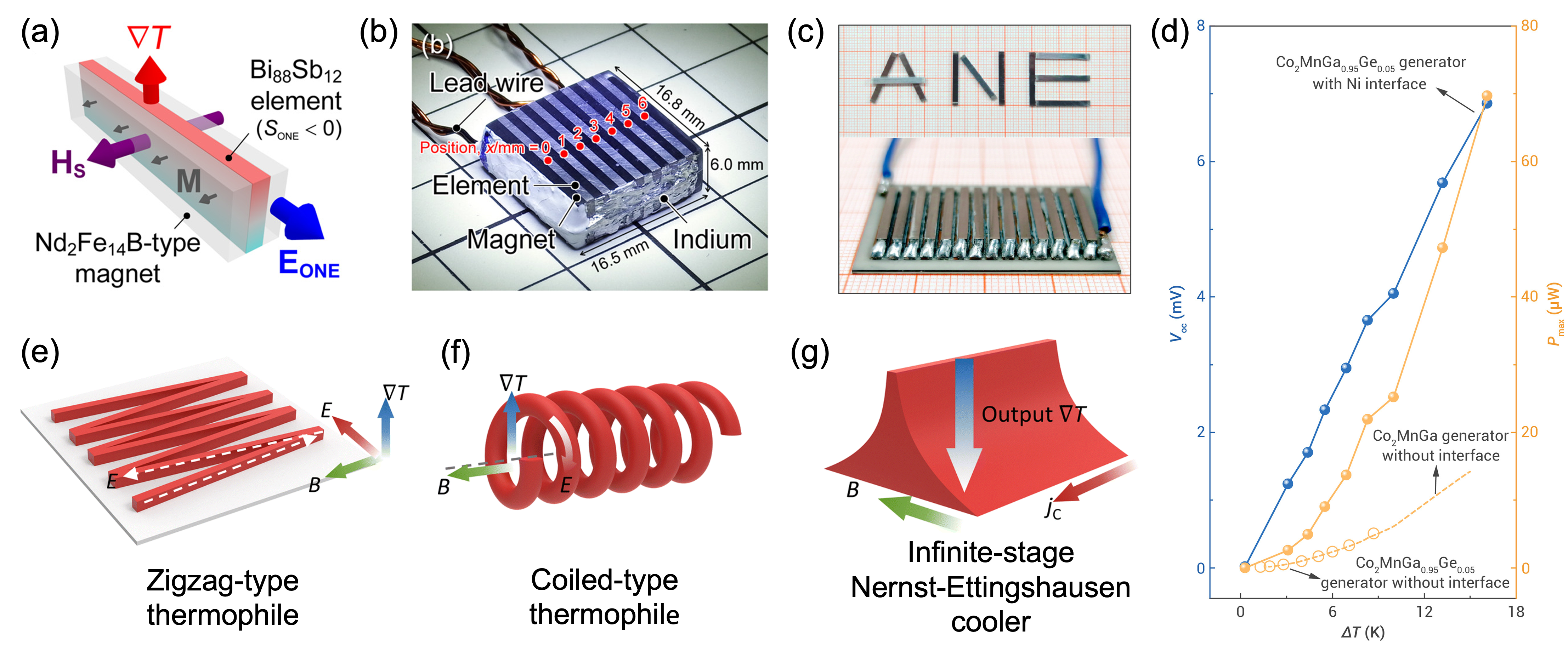}
\caption{(a) Schematic of ONE-based transverse thermoelectric module operating without an external magnetic field, and (b) an optical image of the corresponding assembled module \cite{murata2024zero} . Reproduced from ref. 193 with permission from AIP Publishing, copyright 2024. (c) Optical image Co$_2$MnGa based ANE generator with a Ni reactive wetting interlayer. (d) Open-circuit voltage (V$_{OC}$) and maximum output power P$_{max}$ versus temperature difference $\Delta$T for electron doped Co$_2$MnGa based Nernst generators: with and without a reactive Ni layer, and a previously reported Co$_2$MnGa Nernst generator without an interface \cite{chen2025design}. Reproduced from ref. 185 with permission under Creative Commons CC BY license, 2025. Schematic of (e) a zigzag‑type Nernst thermopile using tilted bicrystals, (f) a helical-type device that accumulates voltage along the helical transverse material under a radial temperature gradient and (g) an infinite-stage Nernst–Ettingshausen cooler \cite{lou2026recent}. Reproduced from ref. 165 with permission from the Royal Society of Chemistry, copyright 2026.}
\end{figure*}

Despite this progress, the traditional longitudinal architecture’s parallel thermal and electric fields require complex $\Pi$-shaped p-n junctions with many interfaces, leading to high thermal and electrical resistance and to tightly coupled thermal and electrical conductivities, presenting a formidable hurdle in translating the high performance of individual material legs into an equally efficient complete device. In contrast, transverse thermoelectrics decouple these fields, with the output voltage following a simple scaling law without increasing thermal bypass, and the gate-type design reduces electrical connections, lowering both resistances. Devices based on ordinary and anomalous transverse effects have been demonstrated, establishing a new paradigm. Hybrid gate-type architectures that combine magnetic, conventional, and anomalous transverse effects via tilted multilayers and alternating materials \cite{uchida2024hybrid, hirai2024hybridizing, ando2025multifunctional} further enable efficient conversion across diverse conditions, greatly enhancing output and application potential.

Recent demonstrations have validated two distinct device paradigms without the bulky external magnets. Murata et. al. \cite{murata2024zero} implemented an ordinary‑Nernst‑effect (ONE) module by alternately stacking semi-metallic Bi$_{88}$Sb$_{12}$ slabs and Nd$_2$Fe$_{14}$B‑type permanent magnets [Figs. 21(a) and 21(b)]. The stray field ($\sim$ 0.43 T) from the embedded permanent magnets drives the ordinary Nernst effect in Bi$_{88}$Sb$_{12}$, yielding a power output of 13.2 $\mu W$ at $\Delta$T = 120 K (average 260 K). Although the fill factor ($\sim$ 40 \%) and internal resistance still leave room for improvement, the work establishes that ordinary‑Nernst‑effect modules can operate robustly at zero external field, fulfilling the symmetry requirements for pure transverse generation. In parallel, Chen et. al. \cite{chen2025design} demonstrated a new benchmark in ANE-based device output, based on an electron-doped Co$_2$MnGa-based device [Fig. 21(c)]. The authors constructed a Ni‑interfacial reactive wetting layer via one‑step spark plasma sintering, reducing the interfacial resistivity to below 0.5 $\mu \Omega$ cm$^2$ and significantly reducing the internal resistance. The resulting anomalous Nernst generator delivered an output power of 69.7 $\mu W$ at a temperature difference of only 16.1K [Fig. 21(d)]. The study further demonstrated the intrinsic structural advantage of transverse generators, output voltage and power scale nearly linearly with the length‑to‑thickness ratio of the thermoelectric leg, without increasing thermal bypass.

Artificially tilted multilayers (ATMLs) of SmCo$_5$/Bi$_{0.2}$Sb$_{1.8}$Te$_3$ \cite{ando2025multifunctional} were shown to hybridise the off-diagonal Seebeck effect with the anomalous Ettingshausen effect, achieving a magnetically tunable $zT$ that far surpasses the record ANE alone. A parallel nonmaterial strategy uses high‑aspect‑ratio tilted bicrystals of Mn$_3$Sn arranged in a zig-zag pattern [Fig. 21(e)], where the large transverse magnetisation during domain reversal generates a contact‑free “hermaphroditic” Nernst thermopile whose voltage scales linearly with the number of elements \cite{li2021monomaterial}. Further, to better utilise the magnetic field, coiled or helical thermopiles [Fig. 21(f)] are explored, in which a tubular heat source can be used \cite{sakuraba2016potential, yang2017scalable}. For cooling, the foundational concept of the exponential cascade shape \cite{harman1963theory} [Fig. 21(g)], which maximises the temperature difference by increasing the cross-sectional area along the heat flow, was experimentally validated by Harman et al. \cite{harman1964oriented} in 1964 using oriented single-crystal bismuth. Their device achieved a record cooling from 302 K to 201 K, demonstrating the critical role of shape ratio. More recently, this classic theory has been revisited by Polash et al. \cite{polash2021infinite} to address non-ideal assumptions. They have derived a fully temperature-dependent numerical model that shows that Harman's original shaping function overestimated Joule heating. The corrected infinite-stage geometry, which accounts for the strong temperature dependence of electrical resistivity, yields higher coefficients of performance (COP) and heat flow density, particularly at large temperature differences.

In addition, a purely structural route to high-performance transverse thermoelectrics has been demonstrated in the single crystals of Re$_4$Si$_7$, a layered semiconductor exhibiting Goniopolarity. The authors have demonstrated that materials performance ($z_{xy}T$ $\sim$ 0.7 at 980 K) translates into comparable device performance \cite{scudder2021highly}, highlighting the importance of scarce Goniopolar materials and their further exploration. Together, these advances solidify the gate‑type transverse architecture as a viable, high‑performance alternative to conventional longitudinal modules, with the combination of embedded permanent magnets and low‑resistance interfaces opening a clear path toward practical thermal energy harvesting and solid‑state cooling.

Beyond power generation and cooling, anomalous Nernst devices have been extended to heat-flux sensing and passive energy harvesting. Chiba et al. \cite{chiba2022single} demonstrated a junction‑free, single‑material thermopile by alternately coating a zigzag SmCo film with black ink; the structure generates continuous voltage from solar heating during the day and through radiative cooling to the cold sky at night. Tanaka et al. \cite{tanaka2023roll} minimized the Seebeck-induced offset in ANE sensors by matching the Seebeck coefficients of Fe$_{79}$Ga$_{21}$ and Ni$_{10}$Cu$_{90}$ electrodes, achieving a flexible, roll‑to‑roll printed sensor that directly measures perpendicular heat flux with 0.23 $\mu$VW$^{-1}$m$^2$ sensitivity and stable performance under bending. Odagiri et al. \cite{odagiri2024bipolar} identified amorphous GdCo alloys as a promising heat flux sensor, offering bipolar transverse thermopower (over 1 $\mu$V K$^{-1}$) and low thermal conductivity ($<$ 5 Wm$^{-1}$K$^{-1}$), yielding a material sensitivity up to 0.23 $\mu$mA$^{-1}$, comparable to Co$_2$MnGa and enabling single‑material thermopiles with effectively doubled output \cite{odagiri2024bipolar}.

\begin{figure*}[t]
\centering
\includegraphics[width=0.95\textwidth]{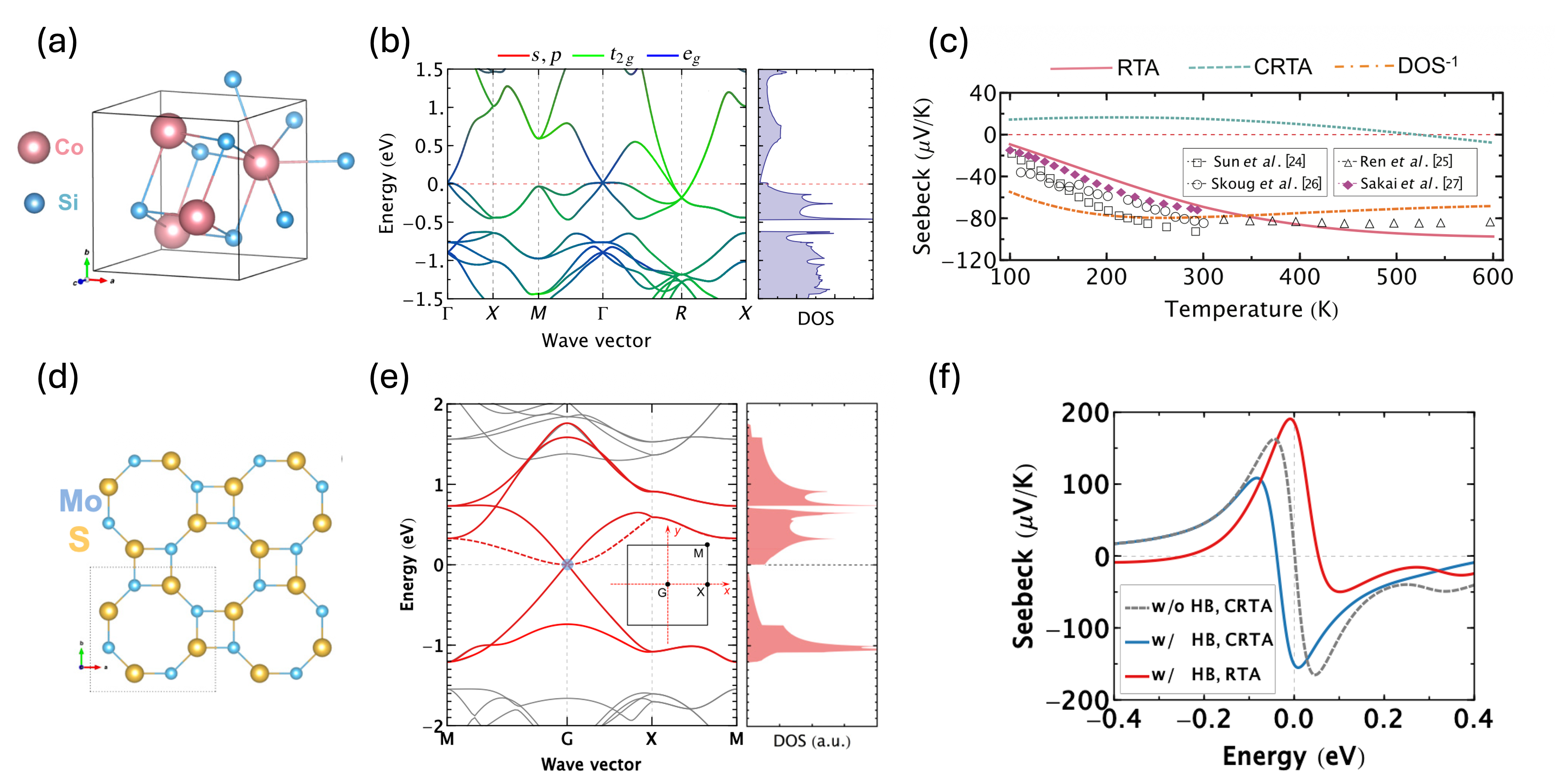}
\caption{(a) Crystal structure, (b) electronic band structure and DOS, and (c) Calculated temperature-dependent S under RTA, CRTA, and DOS$^{-1}$ approximation in comparison with experimental measurements of B20-type CoSi \cite{xia2019high}. Reproduced from ref. 206 with permission from the American Physical Society, copyright 2019. Similary, the (d) Crystal structure, (e) electronic band structure and DOS, and (f) Calculated $S$ at 300 K as a function of chemical potential using band structures with (solid blue lines) and without (dashed gray lines) the heavy band (HB) under the CRTA of monolayer MoS$_2$. The red line traces the values calculated with the HB under the RTA \cite{xia2019leveraging}. Reproduced from ref. 207 with permission from the American Physical Society, copyright 2019.}
\end{figure*}

Further, for hotspot thermal management in electronic systems, materials with high thermal conductivity and a large power factor are crucial \cite{mao2021thermoelectric}, alongside a compact on-chip energy-harvesting design. De Sousa et. al. demonstrated a compact transverse thermoelectric device ($<$ 0.17 mm$^2$) that leverages the anomalous Nernst and spin Seebeck effects to convert heat into electricity using a CoTb-based heterostructure. The device achieves high output and field-free operation within realistic microelectronic environments. The harvested energy can be stored in capacitors with no degradation over hundreds of cycles \cite{de2025miniaturized}. Similarly, Razeghi et al. observed a giant anomalous Ettingshausen coefficient of 3.1 mV in nanoscale Co$_2$MnGa magnetic Weyl semimetal devices at room temperature, and $\sim$ 57 \% of this contribution has been attributed to the intrinsic topological properties, particularly the Berry curvature. By combining this with a geometrically induced Peltier effect in nano ribbons, the cooling performance is boosted by 170\%, demonstrating a promising on-chip spot-cooling approach \cite{razeghi2025giant}.

\section{\textbf{Tuning thermoelectricity with electron-phonon interactions}}

To achieve good TE performance, both $S$ and $\sigma$ should be high, as they together determine the power factor. $\sigma$  for semi-metals are generally high due to the increased density of states or the presence of linear band dispersion near the Fermi level, but $S$ is often negligible due to mutually counterbalancing contribution from electrons and holes. This is not the case in doped semiconductors, which possess an inherent band gap that effectively separates the contributions of holes from those of electrons. The $S$ is dependent on the derivative of transport distribution function $\sigma(E) = e \tau(E)N(E) \langle v^2(E) \rangle$, given as \cite{mott2012electronic}

\[ S = -\frac{\pi^2 k_B^2 T}{3e} \left[\frac{N'(E)}{N(E)} + \frac{\langle v^2(E) \rangle'}{\langle v^2(E) \rangle} + \frac{\tau'(E)}{\tau(E)} \right] \]

in which $N(E)$, $v^2(E)$ and $\tau(E)$ is the energy-dependent DOS, the squared group velocity and carrier lifetime, respectively. The first two terms are dependent on the electronic band structure, while the third term is determined by the scattering of the carriers. The scattering term is dominant in materials having strong energy filtering due to electron-phonon interactions (EPIs), such as CoSi \cite{xia2019high}, so-MoS$_2$ (monolayer $MoS_2$ allotrope constructed of repeating square-octagon pairs in a square lattice] \cite{xia2019leveraging} and strained GeH \cite{li2024electron}. These compounds have high $S$ values due to the asymmetry in the bands around the Fermi level, having a massless Dirac band accompanied with a heavy band either in valence or conduction bands, facilitates strong scattering of the low-energy electrons by phonons into the heavy band, resulting in eﬀective energy filtering. Therefore, the derivative $\tau'(E)$ is expected to be most pronounced with Dirac bands in the background assuming $1/\tau(E ) \propto$ DOS.

\begin{figure*}[t]
\centering
\includegraphics[width=0.80\textwidth]{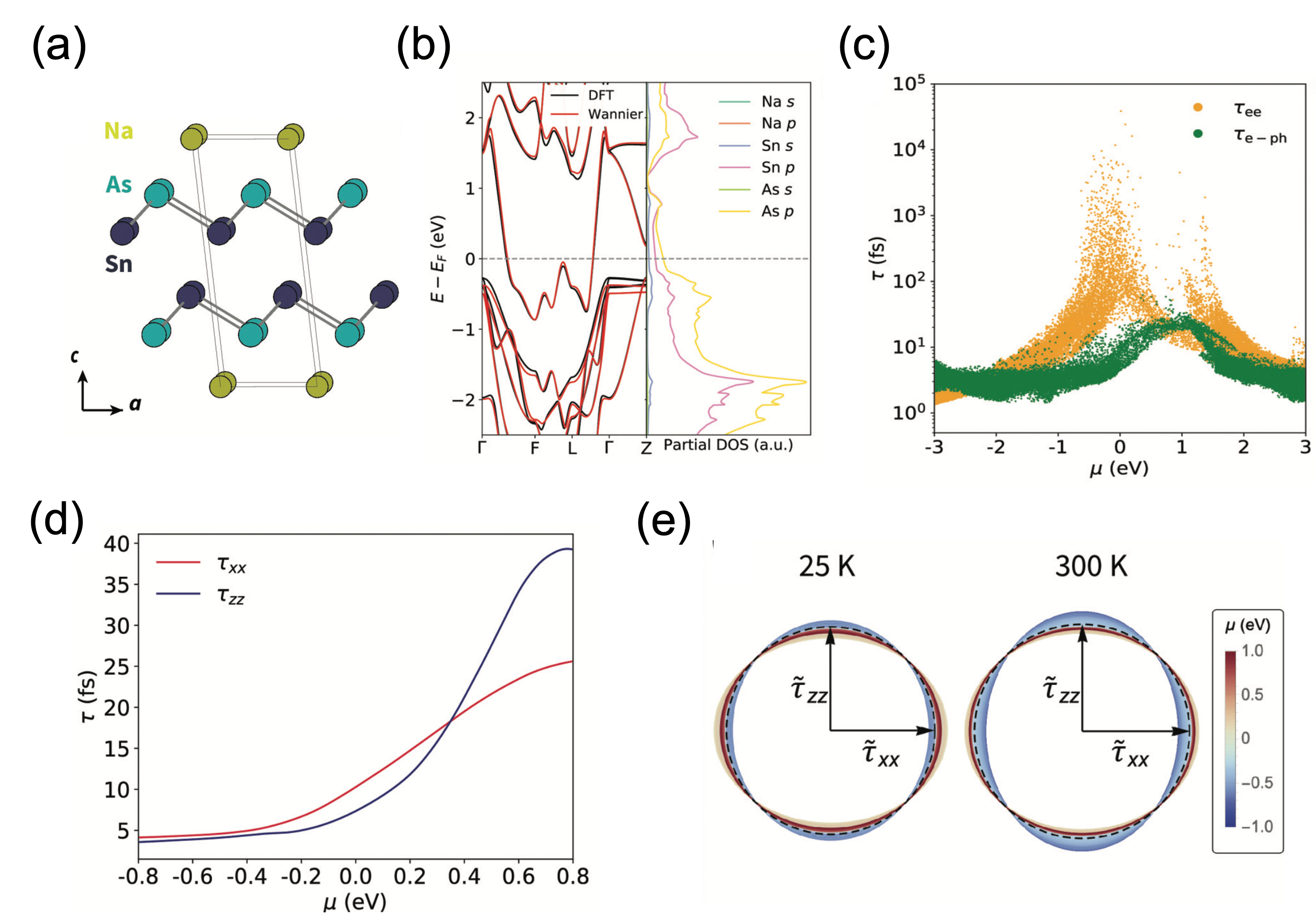}
\caption{(a) Crystal structure, (b) electronic band structure and DOS of NaSn$_2$As$_2$. (c) Comparison of e-ph and e-e scattering lifetimes, and (d) axis dependent e-ph scattering lifetimes as a function of chemical potentials for NaSn$_2$As$_2$. (e) Normalized $\tilde\tau_{xx}$ and $\tilde\tau_{zz}$ at various electron chemical potentials calculated at 25 and 300 K for NaSn$_2$As$_2$ \cite{wang2020anisotropic}. The black dashed circle is the isotropic reference as a guide. Reproduced from ref. 122 with permission from the American Physical Society, copyright 2020.}
\end{figure*}

B20-type CoSi \cite{xia2019high} belongs to space group P2$_1$3, having a cubic structure [Fig 22(a)]. The electronic structure has a massless Dirac  point at the $\Gamma$ point, and a heavy band just below the $E_F$. Due to this, the DOS have a quadratic increase followed by a sharp step, as shown in Fig. 22(b). $S$ calculated by taking into account the scattering term, under the constant relaxation time approximation (CRTA) yields a positive $S$ value. However, experimental observation \cite{sun2013effects, ren2005effects, skoug2009high, sakai2007filling} shows a negative $S$ value as shown in Fig. 22(c). Under the RTA, negative $S$ values are obtained that agree with experimental observations. Thus, accurate momentum- and energy-resolved treatment of EPI is important to study $S$. This negative $S$ under RTA is due to the asymmetric behavior of $\sigma(E)$ around the Fermi energy. A long electron lifetimes dropping quickly near the Fermi level is observed due to abrupt change of DOS, leads to preferential scattering of low-energy electrons in the massless bands, thus creating a large gradient in $\tau(E)$. In addition, the hole lifetimes are short due to larger hole DOS near the Fermi level, allowing electrons to dominate transport.

Similar effects are more dominant in 2D materials due to quantum confinement \cite{sun2012atomically}. The monolayer MoS$_2$ \cite{xia2019leveraging}  bears  P4 symmetry (\#75), has a band structure with parabolic heavy band above the Fermi level [Figs. 22(d) and 22(e)]. The Seebeck coefficient, calculated under the CRTA with (blue line)  and without (grey dashed line) this heavy band (HB), varies nearly symmetrically around the Dirac point when no EPI was taken into account in the calculations. However, the HB provides an excess electronic DOS, and shifts  $S$ to the negative side at the Dirac point (blue line) as shown in Fig. 22(f). After carefully treating and the inclusion of the the EPI under the RTA, the maximum value of $S$ undergoes a sign reversal from -153 to 186 $\mu\mathrm{V\,K^{-1}}$ at the Dirac point, signifying the strong effect of the EPI on $S$ caused by the asymmetry of the DOS and the ensuing carrier filtering mechanism. 

In order to find the optimal position of HB with respect to the Fermi level to maximize S, the band was shifted manually by $\pm 2 eV$, $\pm 1 eV$ and $0 eV$ around the $E_F$. Shifting the HB down suppresses $\sigma$ for holes and also decreases $S$ due to undesirable hole scattering, while upward shift suppresses the $PF$ due to insufficient filtering of electrons. Hence, we conclude that the crossing of heavy bands and massless Dirac bands at the Fermi level is a favorable type of band structure for combining high $S$ and $\sigma$. This analysis suggest that tunability of the Fermi level and band asymmetry that may be achieved in a material through doping, alloying or band engineering could provide a control over the sign and magnitude of the Seebeck coefficient through the relative modulation of the electron-phonon interaction strength and charge carrier filtering mechanism  for the electronic bands across the Fermi level.

These studies show that the energy filtering effect derived from the intrinsic EPI leads to exceptionally high $S$ in semi-metals. Hence, engineering the EPI in semimetallic compounds will be efficient in improving TE performances. Like, the 4-fold $PF$ enhancement is observed in hydrogenated germanene (GeH) with the 8\% biaxial strain \cite{li2024electron}. On application of biaxial strain the GeH goes through a topological phase transition (TPT) from a semiconducting state of bandgap 1.03 eV to a semi-metallic state. It has the desired asymmetric band-crossing, where the light valence band and the conduction band merge into a linear band, crossing with the heavy valence band at the Dirac point. This approach opens an avenue for attaining ultrahigh power factors via engineering the EPIs in semimetals. Furthermore, the strength of EPI can induce topological phase transitions in materials  such as BiTl(S$_{1-\delta}$Se$_\delta$)$_2$ \cite{antonius2016temperature}, BiTeI \cite{brousseau2020temperature}, SnTe \cite{querales2020towards}. Such EPI induced electronic phases are expected to have important implications on their thermoelectric transport properties that remain to be explored in future studies.

EPI has also been shown to affect the thermopower of goniopolar materials. NaSn$_2$As$_2$ was one of the first goniopolar materials reported to exhibit axis-dependent conduction polarity (ADCP) originating from the Fermi surface geometry of a single band \cite{he2019fermi}. It crystallizes in the $R\bar{3}m$ space group and possesses a layered structure, as shown in Fig. 23(a). A single electron band crosses the Fermi energy, giving rise to a hyperboloidal single-band Fermi surface [Fig. 23(b)]. When only the band topology was considered, the predicted in-plane thermopower was $-2$ $\mu\mathrm{V\,K^{-1}}$, while the experimentally measured value ranged from $-2$ to $-6$ $\mu\mathrm{V\,K^{-1}}$ for different samples. Similarly, the predicted out-of-plane thermopower was 21 $\mu\mathrm{V\,K^{-1}}$, whereas the measured value at 300 K was between 8 and 10 $\mu\mathrm{V\,K^{-1}}$. To understand this discrepancy, both electron–electron ($\tau_{ee}$) and electron–phonon ($\tau_{e-ph}$) scattering processes were included, which are determined from the imaginary part of the electron self-energy \cite{wang2020anisotropic}. The electron–electron scattering dominates in the cross-plane ($\Gamma$-$Z$) direction, whereas electron–phonon scattering is dominant along the in-plane directions ($\Gamma$-L and $\Gamma$-F). Figure 23(c) shows that electron–phonon interactions dominate near the Fermi energy in NaSn$_2$As$_2$. The axis-dependent lifetime tensor [$\tau_{e-ph}$]$_{ij}$ was calculated. The axis-resolved $\tau_{xx}$ and $\tau_{zz}$ as functions of electron chemical potential at 300 K are shown in Fig. 23(d). Both $\tau_{xx}$ and $\tau_{zz}$ exhibit a positive slope near $E_F$, with an anisotropy ratio $\tau_{zz}/\tau_{xx} < 2$. The variation of scattering-time anisotropy with chemical potential is smaller at lower temperatures [Fig. 22(e)], suggesting that the axis-dependent conduction polarity in NaSn$_2$As$_2$ primarily originates from its Fermi surface topology. However, anisotropic lifetimes still contribute substantially to the overall thermopower. The lifetime tensor contribution to both $S_{xx}$ and $S_{zz}$ are negative. When electron–phonon interactions are included, the magnitude of $S_{xx}$ increases, while that of $S_{zz}$ decreases, bringing the theoretical predictions closer to the experimental results. Therefore, studying EPIs are essential in determining the Seebeck coefficient of goniopolar material.

\section{Connection between topology and thermoelectricity}

While evaluating longitudinal and transverse thermoelectric effects in various materials, we clarify the relative contributions of topological band properties versus conventional transport mechanisms to their overall thermoelectric performance. Here, we present three representative classes of topological materials to illustrate thermoelectric transport that are largely influenced by topology-independent (i.e., conventional), topology-dependent or shared properties of the materials.

\subsection{Topology-independent factors}

Some material properties are not directly related to topology, but can still enhance the TE performance. For example, the presence of heavy elements and structural anisotropy in Bi$_2$Te$_3$ and Sb$_2$Te$_3$ contributes to their low thermal conductivity. Copper doping can further improve their TE performance by regulating the Te vacancies \cite{han2011formation, liu2011thermoelectric}. Although their topological behaviour is induced by the strong SOC of heavy atoms such as Bi, Sb and Te, their TE performance is not directly related to topology. Similarly, the metavalent bonding in the strong topological insulator, TlBiSe$_2$ \cite{maria2023metavalent} and the dual topological material, BiTe\cite{samanta2018localized} causes bonding anharmonicity, thereby suppressing the lattice thermal conductivity. Here also, the connection between topology, metavalent bonding, and thermoelectric performance is indirect. Band inversion, a common feature of topological compounds, can lead to multiple band extrema, or band convergence. This increases the density of states near the Fermi energy, thereby enhancing the Seebeck coefficient. Prominent materials exhibiting band convergence include Bi$_2$Te$_3$ \cite{cook2007minerals} and SnTe \cite{tanaka2012experimental}, whose good TE performance arises primarily from conventional mechanisms rather than their topological properties.

\subsection{Topology-dependent factors}

It has been observed that topological properties can help improve the TE properties. The presence of robust topological surface states with linear dispersion can enhance the carrier mobility, particularly in nanoscale materials. The Bi$_{1.5}$Sb$_{0.5}$Te$_{1.7}$Se$_{1.3}$ nanowires have very high TE performance than their bulk counterparts, attributed to enhanced surface-dominated electrical conductivity \cite{hsiung2014surface}. The Dirac and Weyl points in topological semimetals lead to linear band crossings near the Fermi energy in the bulk, which can enhance electron mobility by reducing the effective mass in materials like Cd$_3$As$_2$\cite{xiang2020large} and NbP \cite{watzman2018dirac}. The presence of SOC can gap the nodal-lines in few semimetals, generating large Berry curvature near the gapped crossings. This mechanism contributes to the large anomalous Nernst effect observed in Co$_2$MnGa \cite{guin2019anomalous}, Co$_3$Sn$_2$S$_2$ \cite{guin2019zero} and YMn$_6$Sn$_6$ \cite{li2021dirac}.

\subsection{Shared effects of conventional and topology-dependent factors}

The synergy between the topology and conventional factors has enhanced in the TE properties of various compounds. In [Bi$_2$]$_m$[Bi$_2$Te$_3$]$_n$, increasing the ratio of Bi$_2$ bilayer sharpens the Dirac-cone which decreases the effective mass, and increases the carrier mobility \cite{guo2026sharpened}. In addition, bilayers raise the Te vacancy formation energy, which decreases the electron concentration, and increases the Seebeck coefficient. This suggests that natural heterostructure engineering can simultaneously improve both mobility and the Seebeck coefficient. Similarly, varying the Bi content in Mg$_3$Sb$_{2-x}$Bi$_x$ induces a transition from a trivial semiconductor to a topological semimetal. At $x=1.5$, Mg$_3$Sb$_{0.5}$Bi$_{1.5}$ retains a small bandgap with increased valence band degeneracy. This improves the TE power factor which is attributed to high hole density, enhanced carrier mobility and a moderate Seebeck coefficient \cite{xie2024topological}. The high Nernst response in the Weyl semimetal, NbP is due to nearly compensated electron and hole pockets, along with high Dirac-like mobilities \cite{watzman2018dirac}. Similarly, the presence of compensated carriers and strong phonon drag in Dirac semimetals such as NbSb$_2$ \cite{li2022colossal} and TaSb$_2$ \cite{wu2025dirac} is also responsible for their large Nernst thermopower. In ZrTe$_5$, the large Nernst response is attributed to the combined effects of non-zero Berry curvature and bipolar transport \cite{wang2021giant}. These examples indicate that conventional factors and topology can work together to enhance thermoelectric performance.

\section{Summary}

\begin{table*}[htbp]
\caption{A table summarizing all the thermoelectric materials (experimentally reported) that are discussed in this Review, with references. It lists the operating temperature at which each material shows its maximum thermoelectric performance, along with the reported maximum thermopower, power factor, and figure of merit at that temperature. Pink highlights materials based on the Seebeck effect, blue highlights goniopolar materials, grey highlights materials exhibiting the ordinary Nernst effect, and orange highlights materials exhibiting the anomalous Nernst effect. We note that for thermoelectrics which are either topological, magnetic or both, an internal analogous magnetic field due to the Berry curvature or remanent magnetization is present. Blank boxes with a single dash indicate quantities that are not explicitly reported for these materials.}
\label{tab:comparison}
\centering
\renewcommand{\arraystretch}{1.2}

\begin{tabular*}{\textwidth}{@{\extracolsep{\fill}}lccccc>{\columncolor{orange!30}}c}
\hline
\hline
\textbf{Material} &
\textbf{Temperature} &
\textbf{External Magnetic Field} & \textbf{Thermopower} & \textbf{Power Factor } & \textbf{Figure of Merit }\\ 
\hline

\multicolumn{6}{c}{Longitudinal Thermoelectrics} \\[2.5pt]
\hline

\yellowrow
Bi$_2$Te$_{2.85}$Se$_{0.15}$ \cite{wang2012enhanced} & 340 K & Absent & - & - & 1.1 \\[1.7pt]

\yellowrow
Bi$_{0.5}$Sb$_{1.5}$Te$_3$ \cite{ii2015kim} & 320 K & Absent & - & - & 1.86 & \\[1.7pt]

\yellowrow
TlBiSe$_2$ \cite{maria2023metavalent} & 715 K & Absent & 172 $\mu\mathrm{V\,K^{-1}}$ & $7.4 \times 10^{-4} \, \mathrm{W\,m^{-1}\,K^{-2}}$ & 0.6 \\[1.7pt]

\yellowrow
TlBiSeS \cite{maria2023metavalent}  & 715 K & Absent & 178 $\mu\mathrm{V\,K^{-1}}$ & $6.9 \times 10^{-4} \, \mathrm{W\,m^{-1}\,K^{-2}}$ & 0.8  \\[1.7pt]

\yellowrow
BiSe \cite{samanta2018localized}  & 425 K & Absent & 98 $\mu\mathrm{V\,K^{-1}}$ & $10.5 \times 10^{-4} \, \mathrm{W\,m^{-1}\,K^{-2}}$ & 0.36 \\[1.7pt]

\yellowrow
Bi$_{88}$Sb$_{12}$ \cite{pan2025magneto} & 180 K & 0.7 T & 190 $\mu\mathrm{V\,K^{-1}}$ & $305 \times 10^{-4} \, \mathrm{W\,m^{-1}\,K^{-2}}$ & 1.7  \\[1.7pt]

\hline
\multicolumn{6}{c}{Goniopolar Thermoelectrics} \\[2.5pt]
\hline

\bluerow
LaPt$_2$B \cite{manako2024large} & 300 K & Absent & 10 $\mu\mathrm{V\,K^{-1}}$ & $1.7 \times 10^{-4} \, \mathrm{W\,m^{-1}\,K^{-2}}$ & $1.5 \times 10^{-3}$  \\[1.7pt]

\bluerow
WSi$_2$ \cite{ohsumi2024transverse} & 100 K & Absent & 6 $\mu\mathrm{V\,K^{-1}}$ & - & $1 \times 10^{-3}$  \\[1.7pt]

\bluerow
Re$_4$Si$_7$ \cite{scudder2021highly} & 580 K & Absent & 270 $\mu\mathrm{V\,K^{-1}}$ & - & 0.44 \\[1.7pt]

\bluerow
LuCo$_6$Ge$_6$ \cite{hu2026fermiology} & 300 K & Absent & 18.4 $\mu\mathrm{V\,K^{-1}}$ & $8.2 \times 10^{-4} \, \mathrm{W\,m^{-1}\,K^{-2}}$ &  $1.15 \times 10^{-2}$ \\[1.7pt]

\bluerow
LuCr$_6$Ge$_6$ \cite{hu2026fermiology} & 300 K & Absent & 7.3 $\mu\mathrm{V\,K^{-1}}$ & $2.1 \times 10^{-4} \, \mathrm{W\,m^{-1}\,K^{-2}}$ &  $1.7 \times 10^{-3}$ \\[1.7pt]

\bluerow
HoCr$_6$Ge$_6$ \cite{hu2026fermiology} & 300 K & Absent & 8.1 $\mu\mathrm{V\,K^{-1}}$ & $1.1 \times 10^{-4} \, \mathrm{W\,m^{-1}\,K^{-2}}$  &  $1.7 \times 10^{-3}$ \\[1.7pt]

\hline
\multicolumn{6}{c}{Transverse Thermoelectrics} \\[2.5pt]
\hline

\grayrow
Single crystal NbP \cite{watzman2018dirac} & 109 K & 9 T & 800 $\mu\mathrm{V\,K^{-1}}$ & - & - \\[1.7pt]

\grayrow
Polycrystalline NbP \cite{fu2018large} & 136 K & 9 T & 90 $\mu\mathrm{V\,K^{-1}}$ & $35 \times 10^{-4} \, \mathrm{W\,m^{-1}\,K^{-2}}$ & - \\[1.7pt]

\grayrow
Single crystal NbSb$_2$ \cite{li2022colossal} & 28 K & 9 T & 616 $\mu\mathrm{V\,K^{-1}}$ & $3100 \times 10^{-4} \, \mathrm{W\,m^{-1}\,K^{-2}}$ & 0.14 \\[1.7pt]

\grayrow
Polycrystalline NbSb$_2$ \cite{li2023giant} & 21 K & 9 T & 396 $\mu\mathrm{V\,K^{-1}}$ & $1269 \times 10^{-4} \, \mathrm{W\,m^{-1}\,K^{-2}}$ & 0.05 \\[1.7pt]

\grayrow
Single crystal TaSb$_2$ \cite{wu2025dirac} & 25 K & 9 T & 1200 $\mu\mathrm{V\,K^{-1}}$ & $8300 \times 10^{-4} \, \mathrm{W\,m^{-1}\,K^{-2}}$ & - \\[1.7pt]

\grayrow
Polycrystalline TaSb$_2$ \cite{negi2025breakdown} & 25 K & 9 T & 37.9 $\mu\mathrm{V\,K^{-1}}$ & $1800 \times 10^{-4} \, \mathrm{W\,m^{-1}\,K^{-2}}$ & - \\[1.7pt]

\grayrow
ZrTe$_5$ \cite{wang2021giant} & 100 K & 13 T & 1900 $\mu\mathrm{V\,K^{-1}}$ (in-plane) & $37 \times 10^{-4} \, \mathrm{W\,m^{-1}\,K^{-2}}$ & 0.07   \\ [1.7pt]

\grayrow
Cd$_3$As$_2$ \cite{xiang2020large} & 300 K & 2 T & 100 $\mu\mathrm{V\,K^{-1}}$ & - & 0.5   \\ [1.7pt]

\redrow
Single crystal Co$_2$MnGa \cite{guin2019anomalous} & 300 K & Absent & 6 $\mu\mathrm{V\,K^{-1}}$ & - & - \\ [1.7pt]

\redrow
Polycrystalline Co$_2$MnGa \cite{guin2019anomalous} & 300 K & Absent & 6.5 $\mu\mathrm{V\,K^{-1}}$ & - & - \\ [1.7pt]


\redrow
Fe$_2$YZ (Y=Co, Z= Al, Ga) \cite{mende2021large} & 340 K & 1 T & 2 $\mu\mathrm{V\,K^{-1}}$ & - & - \\ [1.7pt]

\redrow
Co$_3$Sn$_2$S$_2$ \cite{guin2019zero} & 80 K & Absent & 3 $\mu\mathrm{V\,K^{-1}}$ & - & - \\ [1.7pt]

\redrow
UCo$_{0.8}$Ru$_{0.2}$Al \cite{asaba2021colossal} & 40 K & 0.1 T & 23 $\mu\mathrm{V\,K^{-1}}$ & - & - \\ [1.7pt]

\redrow
YMn$_6$Sn$_6$ \cite{roychowdhury2022large} & 200 K & Absent & 2.4 $\mu\mathrm{V\,K^{-1}}$ & - & - \\ [1.7pt]

\redrow
Fe$_3$Ga/Al \cite{sakai2020iron} & 300 K & Absent & 6/4 $\mu\mathrm{V\,K^{-1}}$ & - & - \\ [1.7pt]

\redrow
Thin film Fe$_3$Ga/Al \cite{sakai2020iron} & 300 K & Absent & 4/2 $\mu\mathrm{V\,K^{-1}}$ & - & - \\ [1.7pt]

\redrow
Fe$_{3.09}$Ga$_{0.91}$ \cite{feng2022giant} & 300 K & Absent & 5.4 $\mu\mathrm{V\,K^{-1}}$ & - & - \\ [1.7pt]

\redrow
YbMnBi$_2$ (bc configuration) \cite{pan2022giant} & 160 K & Absent & 3 $\mu\mathrm{V\,K^{-1}}$ & - & $3.25 \times 10^{-3}$ & \\ [1.7pt]

\redrow
YbMnBi$_2$ (cb configuration) \cite{pan2022giant} & 160 K & Absent & 6 $\mu\mathrm{V\,K^{-1}}$ & - & - \\ [1.7pt]

\redrow
SmMnBi$_2$ \cite{tang2024unconventional} & 300 K & Absent & 1.6 $\mu\mathrm{V\,K^{-1}}$ & - & - \\ [1.7pt]

\redrow
Mn$_3$Sn \cite{ikhlas2017large} & 300 K & Absent & 0.35 $\mu\mathrm{V\,K^{-1}}$ & - & - \\ [1.7pt]

\redrow
Mn$_3$Ge \cite{wuttke2019berry} & 125 K & Absent & 1.5 $\mu\mathrm{V\,K^{-1}}$ & - & - \\ [1.7pt]

\redrow
Thin-film Mn$_5$Si$_3$ \cite{badura2025observation} & 58 K & Absent & 0.26 $\mu\mathrm{V\,K^{-1}}$ & - & - \\ [1.7pt]

\redrow
CoNb$_3$S$_6$ \cite{khanh2025gapped} & 26 K & Absent & 1 $\mu\mathrm{V\,K^{-1}}$ & - & - \\ [1.7pt]

\hline
\hline
\end{tabular*}
\end{table*}

Materials having a high thermoelectric (TE) figure of merit are difficult to engineer due to several conflicting requirements. The Seebeck coefficient and electrical conductivity show opposite trends with effective mass and carrier concentrations. The Wiedemann-Franz law establishes a linear relationship between electronic thermal conductivity and electrical conductivity.

Interestingly, materials exhibiting non-trivial electronic topology often support good TE performance. These materials commonly have heavy atoms and possess small band gaps. The presence of heavy atoms lowers the phonon frequencies, leading to reduced lattice thermal conductivity. Heavy atoms also give rise to strong spin-orbit coupling (SOC). In topological insulators (TIs), strong SOC can induce band inversion, leading to multiple band extrema present near the Fermi level, also known as band convergence. This enhances the density of states near the Fermi level and can increase the Seebeck coefficient. TIs have an insulating bulk band gap, but have conducting surface states. These states exhibit linear dispersion and can contribute to high electrical conductivity at the surface, which becomes dominant in nanoscale devices or samples having a high surface-to-volume ratio. The TE properties of TIs depend strongly on density of states close to the Fermi level. The presence of a flat band around the VBM or CBM increase the DOS, and eventually the Seebeck coefficient also increases.

In semimetals, linearly dispersing bands are present near the Fermi energy, resulting in small effective masses and high electrical conductivity. The carrier concentrations of holes and electrons in semimetals are comparable. The Seebeck coefficients from both carriers cancel each other, as they move in the same direction under a temperature gradient, reducing the longitudinal thermopower. Non-trivial band topology in semimetals can generate significant Berry curvature near the Fermi level. Semimetals have Dirac points, Weyl points, or nodal lines protected by crystal symmetries. The presence of spin–orbit coupling lifts band degeneracies along different directions in momentum space in few nodal semimetals, except at certain high-symmetry points protected by certain symmetries. These points form a Dirac crossing if time-reversal and inversion symmetries, along with certain rotational symmetries, are present in the system. In contrast, Weyl points require the breaking of either time-reversal or inversion symmetry. The change in the mirror symmetry representation after the inclusion of SOC, allows the hybridization of bands, and opens a gap in mirror-symmetric nodal-line semimetals. This generates a large Berry curvature, which acts analogously to a magnetic field in momentum space and accumulate the opposite charge carriers in the lateral direction. This increases the transverse Nernst thermopower and can eventually lead to a large transverse Nernst figure of merit. TE performance is not solely related to non-trivial topological properties, but also depends strongly on conventional factors such as structural anisotropy, carrier concentration, doping, and the bipolar effect. In studying various materials for longitudinal and transverse thermoelectric effects, we have emphasized both topologically nontrivial and conventional factors that contribute to their excellent thermoelectric performance.

In this review article, we provide a brief overview of different topological quantum materials that are relevant to thermoelectric energy conversion and technology. We started our discussions with the prototypical strong topological insulators Bi$_2$Te$_3$ and Sb$_2$Te$_3$,  which belong to the tetradymite family of compounds and were long realized as good thermoelectric materials before coming to be known as topological insulators. These materials contribute to the thermoelectric transport via a longitudinal mechanism, wherein the applied temperature gradient and the induced electrical voltage   appear in the same spatial direction in the materials. The resulting transport quantity is known as the Seebeck coefficient (e.g., $S_{xx}$ along x-direction). Combing $S_{xx}$, electrical conductivity ($\sigma_{xx}$) and the thermal conductivity ($\kappa_{xx}$) occurring in the same spatial crystallographic direction, we define the thermoelectric figure of merit ($zT$) that quantifies the efficiency of thermoelectric energy conversion. Therefore, in longitudinal thermoelectrics, charge and heat transports take place in the same direction.   Most of the conventional high-performance thermoelectric materials belong to this category regardless of whether they are topologically trivial or non-trivial and hence, these materials are termed as longitudinal thermoelectrics. A few more prominent examples in this class include SnTe-based family of topological crystalline insulators Pb$_{1-x}$Sn$_x$Te, Bi-based weak topological insulator such as BiSe, and Bi$_{1-x}$Sb$_x$ based alloys of topological materials. On the other hand, there are a class of topological quantum materials which lack band gaps in the bulk and exhibit semimetallic electronic structures. Examples include Dirac semimetals, Weyl, nodal line semimetals. The longitudinal thermoelectric response of these materials are usually very small owing to the presence of both types of charge carriers which lead to very small Seebeck coefficient ($S_{xx}$).

Interestingly, some of these topological semimetallic compounds can develop an electrical voltage in the transverse direction with respect to an applied temperature gradient ($\nabla_xT$) when an external magnetic field ($B_z$) is applied perpendicular to the temperate gradient, resulting in a transport coefficient known as the Nernst coefficient ($S_{xy}$). This effect is known as the ordinary Nernst effect (ONE). Examples of compounds that shows ONE  include  Weyl semimetal NbP, Dirac semimetals NbSb$_2$ and ZrTe$_5$. In topological semimetals which exhibit non-zero and large Berry curvature, large Nernst effect is observed even in the absence of an external magnetic field. Such phenomenon is known as the anomalous Nernst effect (ANE). Here, Berry curvature acts as a pseudo-magnetic field that induce the transverse voltage. Magnetic topological materials in the Heusler family such as  Co$_2$MnGa, Fe$_2$YZ (Y= Co, Ni; Z = Al, Ga), ferromagnetic Kagome compound  Co$_3$Sn$_2$S$_2$, antiferromagnetic Kagome compounds Mn$_3$X (X = Sn, Ge), ferromagnetic Fe$_3$X (X=Al, Ge), canted antiferromagnet YbMnBi$_2$, chiral antiferromagnet CoNb$_3$S$_6$ are some of the prominent examples that exhibit ANE. Combining $S_{xy}$, electrical conductivity ($\sigma_{yy}$) and thermal conductivity ($\kappa_{xx}$) in the transverse directions, we can define a thermoelectric figure of merit ($z_NT$) which depend on orthogonal charge and heat transports. Therefore, although the longitudinal thermoelectric response and $zT$ in such semimetals can be small due to compensated electron and hole carriers, the transverse thermoelectric response and hence, $z_NT$ can be quite high. As an example, Dirac semimetal Cd$_3$As$_2$ exhibits $z_NT$ = 0.5 at 300K under an applied magnetic field of 2T. Since the charge and heat transports in such transverse thermoelectrics are decoupled, it is more advantageous to tune the heat and transport properties independently than in longitudinal thermoelectrics where heat and charge transport are coupled through the Wiedemann-Franz law.

Apart from the above two classes of thermoelectricity, there is another class of materials which can exhibit both $p$- and $n$-type charge carriers simultaneously but along different crystallographic directions in the same material.   This can happen due to the presence of a single band which has a positive  curvature along a direction in the Brillouin zone and a negative curvature along another direction. It may also result from  multiple bands in which conduction electrons have a higher mobility in one direction and holes have higher mobility along another. Materials exhibiting such axis-dependent conduction polarity is termed as goniopolar materials. Goniopolarity can also result from the presence of mixed-dimensionality of the Fermi surface. Therefore, due to the presence of anisotropy in the crystal, electronic structures, and Fermi surfaces, transport coefficients vary both in magnitudes (Seebeck coefficient, electrical, and thermal conductivity) and sign (Seebeck coefficient) along different crystallographic directions. Consequently, the measurement of these properties can be done at a certain crystallographic angle to maximize the transverse response of the Seebeck coefficient ($S_{xy}^{\prime}$) and the resulting thermoelectric figure of merit. Goniopolar materials exhibit transverse thermoelectricity  through a combination of their longitudinal thermoelectric responses occurring along different directions in the crystal structures, which is different from the transverse thermoelectricity shown by topological semimetals that  utilize (ordinary or anomalous) Nernst effect. Some examples of gonioplar materials include LaPt$_2$B, Re$_4$Si$_7$, WSi$_2$ and KMgBi. It is worth mentioning that some of the well known longitudinal thermoelectrics (such as Mg$_3$Sb$_2$ and Mg$_3$Bi$_2$) have also been shown to exhibit goniopolar thermoelectricity.

A summary of all the experimentally reported thermoelectric materials discussed in this review article is provided in Table I. The longitudinal Seebeck coefficient is relatively high and requires no external magnetic field. However, Seebeck based modules need alternating $p$-type and $n$-type legs connected electrically in series, which makes the device geometry complex. Transverse devices based on the Nernst effect require a single material, avoiding the need for alternating $p-n$ legs. The ordinary Nernst effect can produce a thermopower comparable to the Seebeck coefficient, but require a strong external magnetic field. This limits the practical application of ONE-based devices because maintaining such strong fields is difficult and costly. The anomalous Nernst effect, observed in materials with intrinsic remanent magnetization or non-zero Berry curvature, can produce a transverse response in the absence of an external magnetic field. But, the ANE thermopower is generally much smaller than the Seebeck coefficient. The Goniopolar materials can exhibit both $p$-type and $n$-type responses within the same material, reducing device complexity and eliminating the need for a magnetic field. Their real-world application is also currently limited due to the small number of materials known to have axis-dependent conduction polarity. Thus, transverse thermoelectric effects offer advantages in terms of device geometry and eliminate the need of external magnetic fields, but achieving a large transverse thermopower in a material for practical applications remains a major challenge.

Finally, we have provided a discussion how electron-phonon interaction engineering can be extremely useful to enhance the thermoelectric responses of topological quantum materials. By studying CoSi and monolayer MoS$_2$, we showed that including electron-phonon interactions in the calculation of the Seebeck coefficient is important in materials with strong energy filtering. Thermopower can be enhanced in materials by modifying the band topology such that a heavy band crosses a Dirac band at the Fermi energy, resulting in an asymmetry in density of states at the Fermi energy. This asymmetry  enables electron energy filtering through preferential scattering of electron/hole carriers into the heavy band, which is facilitated by strong electron-phonon interaction. We further show that a high $PF$ is obtained when the heavy-band and Dirac-band crossings coincide exactly at the Fermi energy, that
could be achieved through doping. The energy-filtering effect enhanced the power factor by a factor of four in hydrogenated germanene (GeH), as it transitions from a semiconducting phase to a semi-metallic state upon application of strain. This allows us to think beyond conventional mechanisms such as doping, alloying to  enhance the thermoelectric energy conversion efficiency in quantum topological materials.

\section{\textbf{Future outlook}}
\subsection{Harnessing advantages in topological thermoelectrics}
By converging the principles from quantum materials, topology, and thermoelectrics, these exotic compounds enable functionalities that extend beyond the limits of conventional thermoelectric materials. Their unconventional electronic structures, characterized by linear band dispersion, spin-momentum locking, and protection against backscattering, provide viable routes toward simultaneously enhancing the Seebeck coefficient, Nernst coefficient, electrical, or Hall conductivity while suppressing lattice thermal conductivity, thereby improving the thermoelectric figure of merit ($zT$). In topological materials, gapless metallic surface or chiral states dominate charge transport and remain robust against disorder, allowing high electrical conductivity without a proportional increase in thermal conductivity. This decoupling of electrical and thermal transport is particularly pronounced in low-dimensional and nanostructured systems, where surface states significantly enhance the thermoelectric response. Furthermore, the presence of nontrivial Berry curvature in Dirac and Weyl systems gives rise to large anomalous Nernst and thermomagnetic responses, enabling transverse thermoelectric conversion even in the absence of external magnetic fields and opening new device architectures, which is inaccessible to conventional longitudinal thermoelectrics. In addition, many topological materials intrinsically possess low lattice thermal conductivity due to heavy constituent elements, strong spin-orbit coupling, low energy phonon modes. In many cases the layered or low-dimensional crystal structures, which further suppress phonon-mediated heat transport while preserving high carrier mobility. 

\subsection{\textbf{Utilization towards quantum science and technology}}
Beyond energy conversion, topological thermoelectrics provide a unique interface with quantum science. This can be possible by enabling thermoelectric control of quantum states, including thermal manipulation of spin-polarized surface channels, modulation of quantum oscillations in Dirac and Weyl systems, and control of Majorana modes in hybrid topological insulator-superconductor structures. The strong spin-orbit coupling and spin-momentum-locked transport in topological systems also useful for spin-caloritronics, wherein thermal gradients generate spin currents for applications in low-power spintronic logic and magnetic memory devices.

Moreover, thermoelectric and Nernst measurements serve as sensitive probes of quantum anomalies, such as the chiral anomaly, quantum Hall effects, and topological phase transitions. Therefore, thermoelectric transport can be a useful characterization technique in quantum materials research. From the device application perspective, these properties position topological thermoelectrics as promising candidates for high-efficiency waste heat recovery in industrial and automotive systems, on-chip energy harvesting for microelectronics, flexible and wearable power generators, highly sensitive thermal detectors. In hard magnetic topological semimetals, large anomalous Nernst effect facilitate zero-field thermomagnetic generators, compact transverse cooling devices, and high-resolution magnetic and thermal gradient sensors. 

The confluence of thermoelectrics and quantum transport in topologically non-trivial compounds could lead to quantum thermoelectric hybrid devices, such as the Majorana-coupled quantum heat engine based on topological superconductor junctions, which have been proposed theoretically to generate thermoelectric power from nanoscale heat gradients \cite{mishra2024majorana}. Moreover, recent experimental realisation of quantum heat engines in superconducting circuits \cite{uusnakki2026initial} suggests that similar thermodynamic cycles could be achieved in topological platforms, thus highlighting topological thermoelectrics as a promising platform for next-generation quantum energy and information technologies.


\section*{Author Contributions}
K.P. conceived the idea and designed the outlines of this article. All authors contributed to the writing and editing. The overall project was supervised by S.N.G and K.P.

\section*{Conflict of interest}

The authors declare no conflict of interest.

\section*{Acknowledgments}
J.D. is grateful for the financial support provided by IIT Kanpur through the institute fellowship. A.K.B.M. thanks BITS Pilani for a Ph.D. fellowship. S.N.G. thanks Core Research Grant (CRG) (CRG/2022/004125), Anusandhan National Research Foundation (ANRF), Govt. of India and Max Planck Society, Germany for Max Planck Partner Group grant. K.P. acknowledges financial support from Indian Institute of Technology Kanpur through the Initiation Grant and the Anusandhan National Research Foundation for an ARG grant (ANRF/ARG/2025/002328/ENS).

\bibliography{rsc}

@article{kumar2020topological,
  title={Topological quantum materials from the viewpoint of chemistry},
  author={Kumar, Nitesh and Guin, Satya N and Manna, Kaustuv and Shekhar, Chandra and Felser, Claudia},
  journal={Chemical Reviews},
  volume={121},
  number={5},
  pages={2780--2815},
  year={2020},
  publisher={ACS Publications}
}

@article{yan2022high,
  title={High-performance thermoelectrics and challenges for practical devices},
  author={Yan, Qingyu and Kanatzidis, Mercouri G},
  journal={Nature materials},
  volume={21},
  number={5},
  pages={503--513},
  year={2022},
  publisher={Nature Publishing Group UK London}
}

@article{pal2015thermoelectric,
  title={Thermoelectric properties of materials with nontrivial electronic topology},
  author={Pal, Koushik and Anand, Shashwat and Waghmare, Umesh V},
  journal={Journal of Materials Chemistry C},
  volume={3},
  number={46},
  pages={12130--12139},
  year={2015},
  publisher={Royal Society of Chemistry}
}

@article{samanta2018localized,
  title={Localized vibrations of Bi bilayer leading to ultralow lattice thermal conductivity and high thermoelectric performance in weak topological insulator n-type BiSe},
  author={Samanta, Manisha and Pal, Koushik and Pal, Provas and Waghmare, Umesh V and Biswas, Kanishka},
  journal={Journal of the American Chemical Society},
  volume={140},
  number={17},
  pages={5866--5872},
  year={2018},
  publisher={ACS Publications}
}

@article{samanta2020intrinsically,
  title={Intrinsically low thermal conductivity and high carrier mobility in dual topological quantum material, n-type BiTe},
  author={Samanta, Manisha and Pal, Koushik and Waghmare, Umesh V and Biswas, Kanishka},
  journal={Angewandte Chemie International Edition},
  volume={59},
  number={12},
  pages={4822--4829},
  year={2020},
  publisher={Wiley Online Library}
}

@article{guin2019anomalous,
  title={Anomalous Nernst effect beyond the magnetization scaling relation in the ferromagnetic Heusler compound Co2MnGa},
  author={Guin, Satya N and Manna, Kaustuv and Noky, Jonathan and Watzman, Sarah J and Fu, Chenguang and Kumar, Nitesh and Schnelle, Walter and Shekhar, Chandra and Sun, Yan and Gooth, Johannes and others},
  journal={NPG Asia Materials},
  volume={11},
  number={1},
  pages={16},
  year={2019},
  publisher={Nature Publishing Group UK London}
}

@article{sakai2020iron,
  title={Iron-based binary ferromagnets for transverse thermoelectric conversion},
  author={Sakai, Akito and Minami, Susumu and Koretsune, Takashi and Chen, Taishi and Higo, Tomoya and Wang, Yangming and Nomoto, Takuya and Hirayama, Motoaki and Miwa, Shinji and Nishio-Hamane, Daisuke and others},
  journal={Nature},
  volume={581},
  number={7806},
  pages={53--57},
  year={2020},
  publisher={Nature Publishing Group UK London}
}

@article{guin2019zero,
  title={Zero-field Nernst effect in a ferromagnetic kagome-lattice Weyl-semimetal Co3Sn2S2},
  author={Guin, Satya N and Vir, Praveen and Zhang, Yang and Kumar, Nitesh and Watzman, Sarah J and Fu, Chenguang and Liu, Enke and Manna, Kaustuv and Schnelle, Walter and Gooth, Johannes and others},
  journal={Advanced Materials},
  volume={31},
  number={25},
  pages={1806622},
  year={2019},
  publisher={Wiley Online Library}
}

@article{pan2022giant,
  title={Giant anomalous Nernst signal in the antiferromagnet YbMnBi2},
  author={Pan, Yu and Le, Congcong and He, Bin and Watzman, Sarah J and Yao, Mengyu and Gooth, Johannes and Heremans, Joseph P and Sun, Yan and Felser, Claudia},
  journal={Nature materials},
  volume={21},
  number={2},
  pages={203--209},
  year={2022},
  publisher={Nature Publishing Group UK London}
}

@article{kumar2023topological,
  title={Topological quantum magnets for transverse thermoelectric energy conversion},
  author={Kumar, Anil and Guin, Satya N},
  journal={Materials Chemistry Frontiers},
  volume={7},
  number={19},
  pages={4202--4214},
  year={2023},
  publisher={Royal Society of Chemistry}
}

@article{heremans2008enhancement,
  title={Enhancement of thermoelectric efficiency in PbTe by distortion of the electronic density of states},
  author={Heremans, Joseph P and Jovovic, Vladimir and Toberer, Eric S and Saramat, Ali and Kurosaki, Ken and Charoenphakdee, Anek and Yamanaka, Shinsuke and Snyder, G Jeffrey},
  journal={Science},
  volume={321},
  number={5888},
  pages={554--557},
  year={2008},
  publisher={American Association for the Advancement of Science}
}

@article{xiao2006berry,
  title={Berry-phase effect in anomalous thermoelectric transport},
  author={Xiao, Di and Yao, Yugui and Fang, Zhong and Niu, Qian},
  journal={Physical review letters},
  volume={97},
  number={2},
  pages={026603},
  year={2006},
  publisher={APS}
}

@article{noky2018characterization,
  title={Characterization of topological band structures away from the Fermi level by the anomalous Nernst effect},
  author={Noky, Jonathan and Gooth, Johannes and Felser, Claudia and Sun, Yan},
  journal={Physical Review B},
  volume={98},
  number={24},
  pages={241106},
  year={2018},
  publisher={APS}
}

@article{liu2018giant,
  title={Giant anomalous Hall effect in a ferromagnetic kagome-lattice semimetal},
  author={Liu, Enke and Sun, Yan and Kumar, Nitesh and Muechler, Lukas and Sun, Aili and Jiao, Lin and Yang, Shuo-Ying and Liu, Defa and Liang, Aiji and Xu, Qiunan and others},
  journal={Nature physics},
  volume={14},
  number={11},
  pages={1125--1131},
  year={2018},
  publisher={Nature Publishing Group UK London}
}

@article{sakai2018giant,
  title={Giant anomalous Nernst effect and quantum-critical scaling in a ferromagnetic semimetal},
  author={Sakai, Akito and Mizuta, Yo Pierre and Nugroho, Agustinus Agung and Sihombing, Rombang and Koretsune, Takashi and Suzuki, Michi-To and Takemori, Nayuta and Ishii, Rieko and Nishio-Hamane, Daisuke and Arita, Ryotaro and others},
  journal={Nature Physics},
  volume={14},
  number={11},
  pages={1119--1124},
  year={2018},
  publisher={Nature Publishing Group UK London}
}

@article{badura2025observation,
  title={Observation of the anomalous Nernst effect in altermagnetic candidate Mn5Si3},
  author={Badura, Anton{\'\i}n and Campos, Warlley H and Bharadwaj, Venkata K and Kounta, Isma{\"\i}la and Michez, Lisa and Petit, Matthieu and Rial, Javier and Leivisk{\"a}, Miina and Baltz, Vincent and Krizek, Filip and others},
  journal={Nature Communications},
  volume={16},
  number={1},
  pages={7111},
  year={2025},
  publisher={Nature Publishing Group UK London}
}

@article{surgers2024anomalous,
  title={Anomalous Nernst effect in the noncollinear antiferromagnet Mn5Si3},
  author={S{\"u}rgers, Christoph and Fischer, Gerda and Campos, Warlley H and Hellenes, Anna Birk and {\v{S}}mejkal, Libor and Sinova, Jairo and Merz, Michael and Wolf, Thomas and Wernsdorfer, Wolfgang},
  journal={Communications Materials},
  volume={5},
  number={1},
  pages={176},
  year={2024},
  publisher={Nature Publishing Group UK London}
}

@article{asaba2021colossal,
  title={Colossal anomalous Nernst effect in a correlated noncentrosymmetric kagome ferromagnet},
  author={Asaba, Tomoya and Ivanov, Vsevolod and Thomas, Sean Michael and Savrasov, SY and Thompson, Joe David and Bauer, Eric Dietzgen and Ronning, Filip},
  journal={Science Advances},
  volume={7},
  number={13},
  pages={eabf1467},
  year={2021},
  publisher={American Association for the Advancement of Science}
}

@article{khanh2025gapped,
  title={Gapped nodal planes and large topological Nernst effect in the chiral lattice antiferromagnet CoNb3S6},
  author={Khanh, Nguyen Duy and Minami, Susumu and Hirschmann, Moritz M and Nomoto, Takuya and Jiang, Ming-Chun and Yamada, Rinsuke and Heinsdorf, Niclas and Yamaguchi, Daiki and Hayashi, Yudai and Okamura, Yoshihiro and others},
  journal={Nature Communications},
  volume={16},
  number={1},
  pages={2654},
  year={2025},
  publisher={Nature Publishing Group UK London}
}

@article{tang2024unconventional,
  title={Unconventional anomalous Hall effect and large anomalous Nernst effect in antiferromagnet SmMnBi2},
  author={Tang, Kaixin and Yang, Ye and Shen, Jianchang and Shi, Mengzhu and Zhang, Nan and Li, Houpu and Li, Hongyu and Liu, Zhengtai and Shen, Dawei and Wang, Rui and others},
  journal={Communications Materials},
  volume={5},
  number={1},
  pages={89},
  year={2024},
  publisher={Nature Publishing Group UK London}
}

@article{wen2025ultrahigh,
  title={Ultrahigh anomalous Nernst thermopower and thermal Hall angle in YbMnBi2},
  author={Wen, Jiamin and Manna, Kaustuv and Vu, Dung and Bej, Subhadeep and Pan, Yu and Felser, Claudia and Skinner, Brian and Heremans, Joseph P},
  journal={Communications Materials},
  volume={6},
  number={1},
  pages={138},
  year={2025},
  publisher={Nature Publishing Group UK London}
}

@article{watzman2018dirac,
  title={Dirac dispersion generates unusually large Nernst effect in Weyl semimetals},
  author={Watzman, Sarah J and McCormick, Timothy M and Shekhar, Chandra and Wu, Shu-Chun and Sun, Yan and Prakash, Arati and Felser, Claudia and Trivedi, Nandini and Heremans, Joseph P},
  journal={Physical Review B},
  volume={97},
  number={16},
  pages={161404},
  year={2018},
  publisher={APS}
}

@article{ohsumi2024transverse,
  title={Transverse Thermoelectric Conversion in the Mixed-Dimensional Semimetal WSi 2},
  author={Ohsumi, Shoya and Sato, Yoshiki J and Okazaki, Ryuji},
  journal={PRX Energy},
  volume={3},
  number={4},
  pages={043007},
  year={2024},
  publisher={APS}
}

@article{liang2017anomalous,
  title={Anomalous Nernst effect in the dirac semimetal Cd 3 As 2},
  author={Liang, Tian and Lin, Jingjing and Gibson, Quinn and Gao, Tong and Hirschberger, Max and Liu, Minhao and Cava, Robert Joseph and Ong, Nai Phuan},
  journal={Physical review letters},
  volume={118},
  number={13},
  pages={136601},
  year={2017},
  publisher={APS}
}

@article{ikhlas2017large,
  title={Large anomalous Nernst effect at room temperature in a chiral antiferromagnet},
  author={Ikhlas, Muhammad and Tomita, Takahiro and Koretsune, Takashi and Suzuki, Michi-To and Nishio-Hamane, Daisuke and Arita, Ryotaro and Otani, Yoshichika and Nakatsuji, Satoru},
  journal={Nature Physics},
  volume={13},
  number={11},
  pages={1085--1090},
  year={2017},
  publisher={Nature Publishing Group UK London}
}

@article{wuttke2019berry,
  title={Berry curvature unravelled by the anomalous Nernst effect in Mn 3 Ge},
  author={Wuttke, Christoph and Caglieris, Federico and Sykora, Steffen and Scaravaggi, Francesco and Wolter, Anja UB and Manna, Kaustuv and S{\"u}ss, Vicky and Shekhar, Chandra and Felser, Claudia and B{\"u}chner, Bernd and others},
  journal={Physical Review B},
  volume={100},
  number={8},
  pages={085111},
  year={2019},
  publisher={APS}
}

@article{manako2024large,
  title={Large transverse thermoelectric effect induced by the mixed-dimensionality of Fermi surfaces},
  author={Manako, Hikari and Ohsumi, Shoya and Sato, Yoshiki J and Okazaki, R and Aoki, D},
  journal={Nature communications},
  volume={15},
  number={1},
  pages={3907},
  year={2024},
  publisher={Nature Publishing Group UK London}
}

@article{roychowdhury2022large,
  title={Large room temperature anomalous transverse thermoelectric effect in Kagome antiferromagnet YMn6Sn6},
  author={Roychowdhury, Subhajit and Ochs, Andrew M and Guin, Satya N and Samanta, Kartik and Noky, Jonathan and Shekhar, Chandra and Vergniory, Maia G and Goldberger, Joshua E and Felser, Claudia},
  journal={Advanced Materials},
  volume={34},
  number={40},
  pages={2201350},
  year={2022},
  publisher={Wiley Online Library}
}

@article{wu2025dirac,
  title={Dirac Fermion and Phonon Coupling Toward Giant Nernst Thermoelectric Performance in Topological Semimetal TaSb2},
  author={Wu, Shuai and Pan, Yu and Wang, Xincan and Xie, Tian and Zheng, Sikang and Liu, Yanhong and Li, Wenhao and Zhou, Zizhen and Ruan, Binbin and Lu, Xu and others},
  journal={Advanced Materials},
  pages={e08627},
  volume={37},
  number={43},
  year={2025},
  publisher={Wiley Online Library}
}

@article{scudder2021highly,
  title={Highly efficient transverse thermoelectric devices with Re 4 Si 7 crystals},
  author={Scudder, Michael R and He, Bin and Wang, Yaxian and Rai, Akash and Cahill, David G and Windl, Wolfgang and Heremans, Joseph P and Goldberger, Joshua E},
  journal={Energy \& Environmental Science},
  volume={14},
  number={7},
  pages={4009--4017},
  year={2021},
  publisher={Royal Society of Chemistry}
}

@article{li2021dirac,
  title={Dirac cone, flat band and saddle point in kagome magnet YMn6Sn6},
  author={Li, Man and Wang, Qi and Wang, Guangwei and Yuan, Zhihong and Song, Wenhua and Lou, Rui and Liu, Zhengtai and Huang, Yaobo and Liu, Zhonghao and Lei, Hechang and others},
  journal={Nature communications},
  volume={12},
  number={1},
  pages={3129},
  year={2021},
  publisher={Nature Publishing Group UK London}
}

@article{goto2024band,
  title={Band anisotropy generates axis-dependent conduction polarity of Mg3Sb2 and Mg3Bi2},
  author={Goto, Yosuke and Usui, Hidetomo and Murata, Masayuki and Goldberger, Joshua E and Heremans, Joseph P and Lee, Chul-Ho},
  journal={Chemistry of Materials},
  volume={36},
  number={4},
  pages={2018--2026},
  year={2024},
  publisher={ACS Publications}
}

@article{heremans2017tetradymites,
  title={Tetradymites as thermoelectrics and topological insulators},
  author={Heremans, Joseph P and Cava, Robert J and Samarth, Nitin},
  journal={Nature Reviews Materials},
  volume={2},
  number={10},
  pages={1--21},
  year={2017},
  publisher={Nature Publishing Group}
}

@article{samanta2020layered,
  title={Layered materials with 2D connectivity for thermoelectric energy conversion},
  author={Samanta, Manisha and Ghosh, Tanmoy and Chandra, Sushmita and Biswas, Kanishka},
  journal={Journal of Materials Chemistry A},
  volume={8},
  number={25},
  pages={12226--12261},
  year={2020},
  publisher={Royal Society of Chemistry}
}

@article{cook2007minerals,
  title={Minerals of the system Bi--Te--Se--S related to the tetradymite archetype: review of classification and compositional variation},
  author={Cook, Nigel J and Ciobanu, Cristiana L and Wagner, Thomas and Stanley, Christopher J},
  journal={The Canadian Mineralogist},
  volume={45},
  number={4},
  pages={665--708},
  year={2007},
  publisher={Mineralogical Association of Canada}
}

@article{bradlyn2017topological,
  title={Topological quantum chemistry},
  author={Bradlyn, Barry and Elcoro, Luis and Cano, Jennifer and Vergniory, Maia G and Wang, Zhijun and Felser, Claudia and Aroyo, Mois I and Bernevig, B Andrei},
  journal={Nature},
  volume={547},
  number={7663},
  pages={298--305},
  year={2017},
  publisher={Nature Publishing Group UK London}
}

@article{zhang2009topological,
  title={Topological insulators in Bi2Se3, Bi2Te3 and Sb2Te3 with a single Dirac cone on the surface},
  author={Zhang, Haijun and Liu, Chao-Xing and Qi, Xiao-Liang and Dai, Xi and Fang, Zhong and Zhang, Shou-Cheng},
  journal={Nature physics},
  volume={5},
  number={6},
  pages={438--442},
  year={2009},
  publisher={Nature Publishing Group UK London}
}

@article{pan2019synergistic,
  title={Synergistic modulation of mobility and thermal conductivity in (Bi, Sb) 2 Te 3 towards high thermoelectric performance},
  author={Pan, Yu and Qiu, Yang and Witting, Ian and Zhang, Liguo and Fu, Chenguang and Li, Jing-Wei and Huang, Yi and Sun, Fu-Hua and He, Jiaqing and Snyder, G Jeffrey and others},
  journal={Energy \& Environmental Science},
  volume={12},
  number={2},
  pages={624--630},
  year={2019},
  publisher={Royal Society of Chemistry}
}

@article{han2011formation,
  title={Formation of Cu nanoparticles in layered Bi 2 Te 3 and their effect on ZT enhancement},
  author={Han, Mi-Kyung and Ahn, Kyunghan and Kim, HeeJin and Rhyee, Jong-Soo and Kim, Sung-Jin},
  journal={Journal of Materials Chemistry},
  volume={21},
  number={30},
  pages={11365--11370},
  year={2011},
  publisher={Royal Society of Chemistry}
}

@article{liu2011thermoelectric,
  title={Thermoelectric property studies on Cu-doped n-type CuxBi2Te2. 7Se0. 3 nanocomposites},
  author={Liu, Wei-Shu and Zhang, Qinyong and Lan, Yucheng and Chen, Shuo and Yan, Xiao and Zhang, Qian and Wang, Hui and Wang, Dezhi and Chen, Gang and Ren, Zhifeng},
  journal={Advanced Energy Materials},
  volume={1},
  number={4},
  pages={577--587},
  year={2011},
  publisher={Wiley Online Library}
}

@article{ii2015kim,
  title={Kim, KH Lee, HA Mun, HS Kim, SW Hwang, JW Roh, DJ Yang, WH Shin, XS Li, YH Lee, GJ Snyder and SW Kim},
  author={II, S},
  journal={Science},
  volume={348},
  pages={109},
  year={2015}
}

@article{li2025plasticity,
  title={Plasticity of Bi2Te3-family thermoelectric crystals},
  author={Li, Ze and Deng, Tingting and Qiu, Pengfei and Ming, Chen and Gao, Zhiqiang and Chen, Lidong and Shi, Xun},
  journal={Nature Communications},
  volume={16},
  number={1},
  pages={5190},
  year={2025},
  publisher={Nature Publishing Group UK London}
}

@article{kumar2025facile,
  title={A facile one-pot recipe for topological insulator Bi 2 Se 3 with thermoelectric properties},
  author={Kumar, Anil and Dhar, Rittika and Biswas, Shuva and Guin, Satya N},
  journal={Chemical Communications},
  volume={61},
  number={40},
  pages={7301--7304},
  year={2025},
  publisher={Royal Society of Chemistry}
}

@article{hong2015enhanced,
  title={Enhanced thermoelectric performance of ultrathin Bi2Se3 nanosheets through thickness control},
  author={Hong, Min and Chen, Zhi-Gang and Yang, Lei and Han, Guang and Zou, Jin},
  journal={Advanced Electronic Materials},
  volume={1},
  number={6},
  pages={1500025},
  year={2015},
  publisher={Wiley Online Library}
}

@article{tanaka2012experimental,
  title={Experimental realization of a topological crystalline insulator in SnTe},
  author={Tanaka, Y and Ren, Zhi and Sato, T and Nakayama, K and Souma, S and Takahashi, T and Segawa, Kouji and Ando, Yoichi},
  journal={Nature Physics},
  volume={8},
  number={11},
  pages={800--803},
  year={2012},
  publisher={Nature Publishing Group UK London}
}

@article{dziawa2012topological,
  title={Topological crystalline insulator states in Pb1- x Sn x Se},
  author={Dziawa, P and Kowalski, BJ and Dybko, K and Buczko, R and Szczerbakow, A and Szot, M and {\L}usakowska, E and Balasubramanian, T and Wojek, Bastian M and Berntsen, MH and others},
  journal={Nature materials},
  volume={11},
  number={12},
  pages={1023--1027},
  year={2012},
  publisher={Nature Publishing Group UK London}
}

@article{hsieh2012topological,
  title={Topological crystalline insulators in the SnTe material class},
  author={Hsieh, Timothy H and Lin, Hsin and Liu, Junwei and Duan, Wenhui and Bansil, Arun and Fu, Liang},
  journal={Nature communications},
  volume={3},
  number={1},
  pages={982},
  year={2012},
  publisher={Nature Publishing Group UK London}
}

@article{xu2012observation,
  title={Observation of a topological crystalline insulator phase and topological phase transition in Pb1- x Sn x Te},
  author={Xu, Su-Yang and Liu, Chang and Alidoust, N and Neupane, M and Qian, D and Belopolski, I and Denlinger, JD and Wang, YJ and Lin, H and Wray, LA a and others},
  journal={Nature communications},
  volume={3},
  number={1},
  pages={1192},
  year={2012},
  publisher={Nature Publishing Group UK London}
}

@article{liu2013two,
  title={Two types of surface states in topological crystalline insulators},
  author={Liu, Junwei and Duan, Wenhui and Fu, Liang},
  journal={Physical Review B—Condensed Matter and Materials Physics},
  volume={88},
  number={24},
  pages={241303},
  year={2013},
  publisher={APS}
}

@article{tan2014high,
  title={High thermoelectric performance of p-type SnTe via a synergistic band engineering and nanostructuring approach},
  author={Tan, Gangjian and Zhao, Li-Dong and Shi, Fengyuan and Doak, Jeff W and Lo, Shih-Han and Sun, Hui and Wolverton, Chris and Dravid, Vinayak P and Uher, Ctirad and Kanatzidis, Mercouri G},
  journal={Journal of the American Chemical Society},
  volume={136},
  number={19},
  pages={7006--7017},
  year={2014},
  publisher={ACS Publications}
}

@article{banik2014lead,
  title={Lead-free thermoelectrics: promising thermoelectric performance in p-type SnTe 1- x Se x system},
  author={Banik, Ananya and Biswas, Kanishka},
  journal={Journal of Materials Chemistry A},
  volume={2},
  number={25},
  pages={9620--9625},
  year={2014},
  publisher={Royal Society of Chemistry}
}

@article{wang2025advances,
  title={Advances and Challenges in SnTe-Based Thermoelectrics},
  author={Wang, Lijun and Moshwan, Raza and Yuan, Ningyi and Chen, Zhi-Gang and Shi, Xiao-Lei},
  journal={Advanced Materials},
  volume={37},
  number={10},
  pages={2418280},
  year={2025},
  publisher={Wiley Online Library}
}

@article{maria2023metavalent,
  title={Metavalent bonding-mediated dual 6s2 lone pair expression leads to intrinsic lattice shearing in n-type TlBiSe2},
  author={Maria, Ivy and Arora, Raagya and Dutta, Moinak and Roychowdhury, Subhajit and Waghmare, Umesh V and Biswas, Kanishka},
  journal={Journal of the American Chemical Society},
  volume={145},
  number={16},
  pages={9292--9303},
  year={2023},
  publisher={ACS Publications}
}

@article{rajaji2023pressure,
  title={Pressure-induced topological crystalline insulating phase in TlBiSe 2: Experiments and theory},
  author={Rajaji, V and Arora, Raagya and Joseph, B and Roychowdhury, Subhajit and Waghmare, Umesh V and Biswas, Kanishka and Narayana, Chandrabhas},
  journal={Physical Review B},
  volume={107},
  number={20},
  pages={205139},
  year={2023},
  publisher={APS}
}

@article{kuroda2010experimental,
  title={Experimental Realization of a Three-Dimensional Topological Insulator<? format?> Phase in Ternary Chalcogenide TlBiSe 2},
  author={Kuroda, K and Ye, M and Kimura, A and Eremeev, SV and Krasovskii, EE and Chulkov, EV and Ueda, Y and Miyamoto, K and Okuda, T and Shimada, K and others},
  journal={Physical review letters},
  volume={105},
  number={14},
  pages={146801},
  year={2010},
  publisher={APS}
}

@article{sato2010direct,
  title={Direct Evidence for the Dirac-Cone Topological Surface<? format?> States in the Ternary Chalcogenide TlBiSe 2},
  author={Sato, Takafumi and Segawa, Kouji and Guo, Hua and Sugawara, Katsuaki and Souma, Seigo and Takahashi, <? format?> Takashi and Ando, Yoichi},
  journal={Physical review letters},
  volume={105},
  number={13},
  pages={136802},
  year={2010},
  publisher={APS}
}

@article{das2024enhanced,
  title={Enhanced Topological Surface State Mediated Transport Elevates Thermoelectric Performance in SnSb2Te4 Quantum Material},
  author={Das, Subarna and Sarkar, Debattam and Biswas, Kanishka},
  journal={Chemistry of Materials},
  volume={36},
  number={7},
  pages={3359--3368},
  year={2024},
  publisher={ACS Publications}
}

@article{pan2025magneto,
  title={A magneto-thermoelectric with a high figure of merit in topological insulator Bi88Sb12},
  author={Pan, Yu and He, Bin and Feng, Xiaolong and Li, Fan and Chen, Dong and Burkhardt, Ulrich and Felser, Claudia},
  journal={Nature Materials},
  volume={24},
  number={1},
  pages={76--82},
  year={2025},
  publisher={Nature Publishing Group UK London}
}

@article{liang2013evidence,
  title={Evidence for massive bulk Dirac fermions in Pb1- x Sn x Se from Nernst and thermopower experiments},
  author={Liang, Tian and Gibson, Quinn and Xiong, Jun and Hirschberger, Max and Koduvayur, Sunanda P and Cava, Robert Joseph and Ong, Nai Phuan},
  journal={Nature communications},
  volume={4},
  number={1},
  pages={2696},
  year={2013},
  publisher={Nature Publishing Group UK London}
}

@article{fu2020largely,
  title={Largely suppressed magneto-thermal conductivity and enhanced magneto-thermoelectric properties in PtSn4},
  author={Fu, Chenguang and Guin, Satya N and Scaffidi, Thomas and Sun, Yan and Saha, Rana and Watzman, Sarah J and Srivastava, Abhay K and Li, Guowei and Schnelle, Walter and Parkin, Stuart SP and others},
  journal={Research},
  pages={4643507},
  volume={2020},
  year={2020},
  publisher={AAAS}
}

@article{wang2019magnetic,
  title={Magnetic field-enhanced thermoelectric performance in Dirac semimetal Cd3As2 crystals with different carrier concentrations},
  author={Wang, Honghui and Luo, Xigang and Peng, Kunling and Sun, Zeliang and Shi, Mengzhu and Ma, Donghui and Wang, Naizhou and Wu, Tao and Ying, Jianjun and Wang, Zhengfei and others},
  journal={Advanced Functional Materials},
  volume={29},
  number={37},
  pages={1902437},
  year={2019},
  publisher={Wiley Online Library}
}

@article{yue2019soft,
  title={Soft phonons and ultralow lattice thermal conductivity in the Dirac semimetal Cd 3 As 2},
  author={Yue, Shengying and Chorsi, Hamid T and Goyal, Manik and Schumann, Timo and Yang, Runqing and Xu, Tashi and Deng, Bowen and Stemmer, Susanne and Schuller, Jon A and Liao, Bolin},
  journal={Physical Review Research},
  volume={1},
  number={3},
  pages={033101},
  year={2019},
  publisher={APS}
}

@article{fu2018large,
  title={Large Nernst power factor over a broad temperature range in polycrystalline Weyl semimetal NbP},
  author={Fu, Chenguang and Guin, Satya N and Watzman, Sarah J and Li, Guowei and Liu, Enke and Kumar, Nitesh and S{\"u}$\beta$, Vicky and Schnelle, Walter and Auffermann, Gudrun and Shekhar, Chandra and others},
  journal={Energy \& Environmental Science},
  volume={11},
  number={10},
  pages={2813--2820},
  year={2018},
  publisher={Royal Society of Chemistry}
}

@article{li2022colossal,
  title={Colossal Nernst power factor in topological semimetal NbSb2},
  author={Li, Peng and Qiu, Pengfei and Xu, Qing and Luo, Jun and Xiong, Yifei and Xiao, Jie and Aryal, Niraj and Li, Qiang and Chen, Lidong and Shi, Xun},
  journal={Nature Communications},
  volume={13},
  number={1},
  pages={7612},
  year={2022},
  publisher={Nature Publishing Group UK London}
}

@article{li2023giant,
  title={A giant Nernst power factor and figure-of-merit in polycrystalline NbSb 2 for Ettingshausen refrigeration},
  author={Li, Peng and Qiu, Pengfei and Xiao, Jie and Deng, Tingting and Chen, Lidong and Shi, Xun},
  journal={Energy \& Environmental Science},
  volume={16},
  number={9},
  pages={3753--3759},
  year={2023},
  publisher={Royal Society of Chemistry}
}

@article{soh2019magnetic,
  title={Magnetic structure and excitations of the topological semimetal YbMnBi 2},
  author={Soh, Jian-Rui and Jacobsen, Henrik and Ouladdiaf, Bachir and Ivanov, Alexandre and Piovano, Andrea and Tejsner, Tim and Feng, Zili and Wang, Hongyuan and Su, Hao and Guo, Yanfeng and others},
  journal={Physical Review B},
  volume={100},
  number={14},
  pages={144431},
  year={2019},
  publisher={APS}
}

@article{borisenko2019time,
  title={Time-reversal symmetry breaking type-II Weyl state in YbMnBi2},
  author={Borisenko, Sergey and Evtushinsky, Daniil and Gibson, Quinn and Yaresko, Alexander and Koepernik, Klaus and Kim, Timur and Ali, Mazhar and van den Brink, Jeroen and Hoesch, Moritz and Fedorov, Alexander and others},
  journal={Nature communications},
  volume={10},
  number={1},
  pages={3424},
  year={2019},
  publisher={Nature Publishing Group UK London}
}

@article{liang2018anomalous,
  title={Anomalous hall effect in ZrTe5},
  author={Liang, Tian and Lin, Jingjing and Gibson, Quinn and Kushwaha, Satya and Liu, Minhao and Wang, Wudi and Xiong, Hongyu and Sobota, Jonathan A and Hashimoto, Makoto and Kirchmann, Patrick S and others},
  journal={Nature Physics},
  volume={14},
  number={5},
  pages={451--455},
  year={2018},
  publisher={Nature Publishing Group UK London}
}

@article{wang2021giant,
  title={Giant Nernst effect and field-enhanced transversal z NT in ZrTe 5},
  author={Wang, Peipei and Cho, Chang-woo and Tang, Fangdong and Wang, Peng and Zhang, Wenjie and He, Mingquan and Gu, Genda and Wu, Xiaosong and Shao, Yonghong and Zhang, Liyuan},
  journal={Physical Review B},
  volume={103},
  number={4},
  pages={045203},
  year={2021},
  publisher={APS}
}

@article{belopolski2019discovery,
  title={Discovery of topological Weyl fermion lines and drumhead surface states in a room temperature magnet},
  author={Belopolski, Ilya and Manna, Kaustuv and Sanchez, Daniel S and Chang, Guoqing and Ernst, Benedikt and Yin, Jiaxin and Zhang, Songtian S and Cochran, Tyler and Shumiya, Nana and Zheng, Hao and others},
  journal={Science},
  volume={365},
  number={6459},
  pages={1278--1281},
  year={2019},
  publisher={American Association for the Advancement of Science}
}

@article{chen2024topological,
  title={Topological Heusler Magnets-Driven High-Performance Transverse Nernst Thermoelectric Generators},
  author={Chen, Mengzhao and Wang, Jialu and Liu, Kai and Fan, Wusheng and Sun, Yan and Felser, Claudia and Zhu, Tiejun and Fu, Chenguang},
  journal={Advanced Energy Materials},
  volume={14},
  number={21},
  pages={2400411},
  year={2024},
  publisher={Wiley Online Library}
}

@article{mende2021large,
  title={Large Anomalous Hall and Nernst Effects in High Curie-Temperature Iron-Based Heusler Compounds},
  author={Mende, Felix and Noky, Jonathan and Guin, Satya N and Fecher, Gerhard H and Manna, Kaustuv and Adler, Peter and Schnelle, Walter and Sun, Yan and Fu, Chenguang and Felser, Claudia},
  journal={Advanced Science},
  volume={8},
  number={17},
  pages={2100782},
  year={2021},
  publisher={Wiley Online Library}
}

@article{nayak2016large,
  title={Large anomalous Hall effect driven by a nonvanishing Berry curvature in the noncolinear antiferromagnet Mn3Ge},
  author={Nayak, Ajaya K and Fischer, Julia Erika and Sun, Yan and Yan, Binghai and Karel, Julie and Komarek, Alexander C and Shekhar, Chandra and Kumar, Nitesh and Schnelle, Walter and K{\"u}bler, J{\"u}rgen and others},
  journal={Science advances},
  volume={2},
  number={4},
  pages={e1501870},
  year={2016},
  publisher={American Association for the Advancement of Science}
}

@article{ghimire2020competing,
  title={Competing magnetic phases and fluctuation-driven scalar spin chirality in the kagome metal YMn6Sn6},
  author={Ghimire, Nirmal J and Dally, Rebecca L and Poudel, L and Jones, DC and Michel, D and Magar, N Thapa and Bleuel, M and McGuire, Michael A and Jiang, JS and Mitchell, JF and others},
  journal={Science Advances},
  volume={6},
  number={51},
  pages={eabe2680},
  year={2020},
  publisher={American Association for the Advancement of Science}
}

@article{feng2022giant,
  title={Giant and robust anomalous Nernst effect in a polycrystalline topological ferromagnet at room temperature},
  author={Feng, Zili and Minami, Susumu and Akamatsu, Shuhei and Sakai, Akito and Chen, Taishi and Nishio-Hamane, Daisuke and Nakatsuji, Satoru},
  journal={Advanced Functional Materials},
  volume={32},
  number={49},
  pages={2206519},
  year={2022},
  publisher={Wiley Online Library}
}

@article{negi2025breakdown,
  title={Breakdown of Wiedemann--Franz Law and Large Transverse Thermoelectric Conductivity in the Topological Semimetal, TaSb2},
  author={Negi, Pranav and Wu, Shuai and Panigrahi, Tanmay Kumar and Samanta, Kartik and Pancholi, Abhinav and Pathak, Aloke Kumar and Das, Subarna and Ghosh, Tanmoy and Pan, Yu and Roychowdhury, Subhajit},
  journal={Chemistry of Materials},
  year={2025},
  volume={37},
  number={20},
  pages={8383--8391},
  publisher={ACS Publications}
}

@article{morali2019fermi,
  title={Fermi-arc diversity on surface terminations of the magnetic Weyl semimetal Co3Sn2S2},
  author={Morali, Noam and Batabyal, Rajib and Nag, Pranab Kumar and Liu, Enke and Xu, Qiunan and Sun, Yan and Yan, Binghai and Felser, Claudia and Avraham, Nurit and Beidenkopf, Haim},
  journal={Science},
  volume={365},
  number={6459},
  pages={1286--1291},
  year={2019},
  publisher={American Association for the Advancement of Science}
}

@article{xia2019leveraging,
  title={Leveraging electron-phonon interaction to enhance the thermoelectric power factor in graphene-like semimetals},
  author={Xia, Yi and Park, Junsoo and Ozoli{\c{n}}{\v{s}}, Vidvuds and Wolverton, Chris},
  journal={Physical Review B},
  volume={100},
  number={20},
  pages={201401},
  year={2019},
  publisher={APS}
}

@article{li2024electron,
  title={Electron--Phonon Interaction Mediated Gigantic Enhancement of Thermoelectric Power Factor Induced by Topological Phase Transition},
  author={Li, Zhi and Pal, Koushik and Lee, Huiju and Wolverton, Chris and Xia, Yi},
  journal={Nano letters},
  volume={24},
  number={19},
  pages={5816--5823},
  year={2024},
  publisher={ACS Publications}
}

@article{xia2019high,
  title={High Thermoelectric Power Factor in Intermetallic Co Si Arising from Energy Filtering of Electrons by Phonon Scattering},
  author={Xia, Yi and Park, Junsoo and Zhou, Fei and Ozoli{\c{n}}{\v{s}}, Vidvuds},
  journal={Physical Review Applied},
  volume={11},
  number={2},
  pages={024017},
  year={2019},
  publisher={APS}
}

@article{guo2017large,
  title={Large anomalous Nernst and spin Nernst effects in the noncollinear antiferromagnets Mn 3 X (X= Sn, Ge, Ga)},
  author={Guo, Guang-Yu and Wang, Tzu-Cheng},
  journal={Physical Review B},
  volume={96},
  number={22},
  pages={224415},
  year={2017},
  publisher={APS}
}

@article{yang2017topological,
  title={Topological Weyl semimetals in the chiral antiferromagnetic materials Mn3Ge and Mn3Sn},
  author={Yang, Hao and Sun, Yan and Zhang, Yang and Shi, Wu-Jun and Parkin, Stuart SP and Yan, Binghai},
  journal={New Journal of Physics},
  volume={19},
  number={1},
  pages={015008},
  year={2017},
  publisher={IOP Publishing}
}

@article{yadav2023topological,
  title={Topological transverse thermoelectrics: A new way forward?},
  author={Yadav, Shubham and Kumar, BM Anil and Guin, Satya N},
  journal={The Innovation},
  pages={100530},
  volume={4},
  number={6},
  year={2023},
  publisher={Elsevier}
}

@article{chang20183d,
  title={3D charge and 2D phonon transports leading to high out-of-plane ZT in n-type SnSe crystals},
  author={Chang, Cheng and Wu, Minghui and He, Dongsheng and Pei, Yanling and Wu, Chao-Feng and Wu, Xuefeng and Yu, Hulei and Zhu, Fangyuan and Wang, Kedong and Chen, Yue and others},
  journal={Science},
  volume={360},
  number={6390},
  pages={778--783},
  year={2018},
  publisher={American Association for the Advancement of Science}
}

@article{kim2012effect,
  title={Effect of thermal disorder on high figure of merit in PbTe},
  author={Kim, Hyoungchul and Kaviany, Massoud},
  journal={Physical Review B—Condensed Matter and Materials Physics},
  volume={86},
  number={4},
  pages={045213},
  year={2012},
  publisher={APS}
}

@article{duhan2025strong,
  title={Strong acoustic phonon suppression leads to ultralow thermal conductivity and enhanced thermoelectric performance in Ba Cu Gd Te 3},
  author={Duhan, Jyoti and Wolverton, Chris and Pal, Koushik},
  journal={Physical Review Applied},
  volume={24},
  number={4},
  pages={L041003},
  year={2025},
  publisher={APS}
}

@article{roychowdhury2021enhanced,
  title={Enhanced atomic ordering leads to high thermoelectric performance in AgSbTe2},
  author={Roychowdhury, Subhajit and Ghosh, Tanmoy and Arora, Raagya and Samanta, Manisha and Xie, Lin and Singh, Niraj Kumar and Soni, Ajay and He, Jiaqing and Waghmare, Umesh V and Biswas, Kanishka},
  journal={Science},
  volume={371},
  number={6530},
  pages={722--727},
  year={2021},
  publisher={American Association for the Advancement of Science}
}

@article{xia2020particlelike,
  title={Particlelike phonon propagation dominates ultralow lattice thermal conductivity in crystalline Tl 3 VSe 4},
  author={Xia, Yi and Pal, Koushik and He, Jiangang and Ozoli{\c{n}}{\v{s}}, Vidvuds and Wolverton, Chris},
  journal={Physical Review Letters},
  volume={124},
  number={6},
  pages={065901},
  year={2020},
  publisher={APS}
}

@article{pathak2024deciphering,
  title={Deciphering Pauling's Third Rule: Uncovering Strong Anharmonicity and Exceptionally Low Thermal Conductivity in TlAgSe for Thermoelectrics},
  author={Pathak, Riddhimoy and Paul, Sayan and Das, Subarna and Das, Anustoop and Pati, Swapan K and Biswas, Kanishka},
  journal={Angewandte Chemie},
  volume={136},
  number={43},
  pages={e202408908},
  year={2024},
  publisher={Wiley Online Library}
}

@article{ochs2021computationally,
  title={Computationally guided discovery of axis-dependent conduction polarity in NaSnAs crystals},
  author={Ochs, Andrew M and Gorai, Prashun and Wang, Yaxian and Scudder, Michael R and Koster, Karl and Moore, Curtis E and Stevanovic, Vladan and Heremans, Joseph P and Windl, Wolfgang and Toberer, Eric S and others},
  journal={Chemistry of Materials},
  volume={33},
  number={3},
  pages={946--951},
  year={2021},
  publisher={ACS Publications}
}

@article{ochs2024synergizing,
  title={Synergizing a Large Ordinary Nernst Effect and Axis-Dependent Conduction Polarity in Flat Band KMgBi Crystals},
  author={Ochs, Andrew M and Fecher, Gerhard H and He, Bin and Schnelle, Walter and Felser, Claudia and Heremans, Joseph P and Goldberger, Joshua E},
  journal={Advanced Materials},
  volume={36},
  number={2},
  pages={2308151},
  year={2024},
  publisher={Wiley Online Library}
}

@article{nelson2023axis,
  title={Axis dependent conduction polarity in the air-stable semiconductor, PdSe 2},
  author={Nelson, Ryan A and Deng, Ziling and Ochs, Andrew M and Koster, Karl G and Irvine, Cullen T and Heremans, Joseph P and Windl, Wolfgang and Goldberger, Joshua E},
  journal={Materials Horizons},
  volume={10},
  number={9},
  pages={3740--3748},
  year={2023},
  publisher={Royal Society of Chemistry}
}

@article{shin2016surface,
  title={The surface-to-volume ratio: A key parameter in the thermoelectric transport of topological insulator Bi 2 Se 3 nanowires},
  author={Shin, Ho Sun and Hamdou, Bacel and Reith, Heiko and Osterhage, Hermann and Gooth, Johannes and Damm, Christine and Rellinghaus, Bernd and Pippel, Eckhard and Nielsch, Kornelius},
  journal={Nanoscale},
  volume={8},
  number={28},
  pages={13552--13557},
  year={2016},
  publisher={Royal Society of Chemistry}
}

@article{fu2007topological,
  title={Topological insulators in three dimensions},
  author={Fu, Liang and Kane, Charles L and Mele, Eugene J},
  journal={Physical review letters},
  volume={98},
  number={10},
  pages={106803},
  year={2007},
  publisher={APS}
}

@article{kane2005quantum,
  title={Quantum spin Hall effect in graphene},
  author={Kane, Charles L and Mele, Eugene J},
  journal={Physical review letters},
  volume={95},
  number={22},
  pages={226801},
  year={2005},
  publisher={APS}
}

@article{chen2009experimental,
  title={Experimental realization of a three-dimensional topological insulator, Bi2Te3},
  author={Chen, YL and Analytis, James G and Chu, J-H and Liu, ZK and Mo, S-K and Qi, Xiao-Liang and Zhang, HJ and Lu, DH and Dai, Xi and Fang, Zhong and others},
  journal={science},
  volume={325},
  number={5937},
  pages={178--181},
  year={2009},
  publisher={American Association for the Advancement of Science}
}

@article{szczech2011enhancement,
  title={Enhancement of the thermoelectric properties in nanoscale and nanostructured materials},
  author={Szczech, Jeannine R and Higgins, Jeremy M and Jin, Song},
  journal={Journal of Materials Chemistry},
  volume={21},
  number={12},
  pages={4037--4055},
  year={2011},
  publisher={Royal Society of Chemistry}
}

@article{poudel2008high,
  title={High-thermoelectric performance of nanostructured bismuth antimony telluride bulk alloys},
  author={Poudel, Bed and Hao, Qing and Ma, Yi and Lan, Yucheng and Minnich, Austin and Yu, Bo and Yan, Xiao and Wang, Dezhi and Muto, Andrew and Vashaee, Daryoosh and others},
  journal={science},
  volume={320},
  number={5876},
  pages={634--638},
  year={2008},
  publisher={American Association for the Advancement of Science}
}

@article{burkov2016topological,
  title={Topological semimetals},
  author={Burkov, AA},
  journal={Nature materials},
  volume={15},
  number={11},
  pages={1145--1148},
  year={2016},
  publisher={Nature Publishing Group UK London}
}

@article{he2019fermi,
  title={The Fermi surface geometrical origin of axis-dependent conduction polarity in layered materials},
  author={He, Bin and Wang, Yaxian and Arguilla, Maxx Q and Cultrara, Nicholas D and Scudder, Michael R and Goldberger, Joshua E and Windl, Wolfgang and Heremans, Joseph P},
  journal={Nature materials},
  volume={18},
  number={6},
  pages={568--572},
  year={2019},
  publisher={Nature Publishing Group UK London}
}

@article{goto2025goniopolar,
  title={Goniopolar materials for transverse thermoelectric device},
  author={Goto, Yosuke and Usui, Hidetomo and Murata, Masayuki and Lee, Chul-Ho},
  journal={JSAP Review},
  volume={94},
  pages={250409},
  year={2025},
  publisher={The Japan Society of Applied Physics}
}

@article{weng2015quantum,
  title={Quantum anomalous Hall effect and related topological electronic states},
  author={Weng, Hongming and Yu, Rui and Hu, Xiao and Dai, Xi and Fang, Zhong},
  journal={Advances in Physics},
  volume={64},
  number={3},
  pages={227--282},
  year={2015},
  publisher={Taylor \& Francis}
}

@article{semwal2025concerted,
  title={Concerted rattling-induced strong anharmonicity and phonon coherence lead to ultralow glasslike thermal conductivity in TlAgTe},
  author={Semwal, Shantanu and Xia, Yi and Wolverton, Chris and Pal, Koushik},
  journal={Physical Review B},
  volume={112},
  number={21},
  pages={214109},
  year={2025},
  publisher={APS}
}

@article{haldar2025conduction,
  title={Conduction Band Convergence and Modular Nanostructures: Driving High Thermoelectric Performance in n-Type PbSe},
  author={Haldar, Indrajit and Taneja, Vaishali and Goyal, Naveen and Ubaid, Mohammad and Sarkar, Debattam and Kedia, Dinesh Kumar and Saurabh, Kumar and Singh, Surjeet and Pal, Koushik and Ravishankar, N and others},
  journal={Angewandte Chemie International Edition},
  volume={64},
  number={36},
  pages={e202510305},
  year={2025},
  publisher={Wiley Online Library}
}

@article{biswas2011strained,
  title={Strained endotaxial nanostructures with high thermoelectric figure of merit},
  author={Biswas, Kanishka and He, Jiaqing and Zhang, Qichun and Wang, Guoyu and Uher, Ctirad and Dravid, Vinayak P and Kanatzidis, Mercouri G},
  journal={Nature chemistry},
  volume={3},
  number={2},
  pages={160--166},
  year={2011},
  publisher={Nature Publishing Group UK London}
}

@article{pal2018bonding,
  title={Bonding hierarchy gives rise to high thermoelectric performance in layered zintl compound baau2p4},
  author={Pal, Koushik and He, Jiangang and Wolverton, C},
  journal={Chemistry of Materials},
  volume={30},
  number={21},
  pages={7760--7768},
  year={2018},
  publisher={ACS Publications}
}

@article{pal2021microscopic,
  title={Microscopic mechanism of unusual lattice thermal transport in TlInTe2},
  author={Pal, Koushik and Xia, Yi and Wolverton, Chris},
  journal={npj Computational Materials},
  volume={7},
  number={1},
  pages={5},
  year={2021},
  publisher={Nature Publishing Group UK London}
}

@article{jana2017intrinsic,
  title={Intrinsic rattler-induced low thermal conductivity in Zintl type TlInTe2},
  author={Jana, Manoj K and Pal, Koushik and Warankar, Avinash and Mandal, Pankaj and Waghmare, Umesh V and Biswas, Kanishka},
  journal={Journal of the American Chemical Society},
  volume={139},
  number={12},
  pages={4350--4353},
  year={2017},
  publisher={ACS Publications}
}

@article{pei2012band,
  title={Band engineering of thermoelectric materials},
  author={Pei, Yanzhong and Wang, Heng and Snyder, G Jeffrey},
  journal={Advanced materials},
  volume={24},
  number={46},
  pages={6125--6135},
  year={2012},
  publisher={Wiley Online Library}
}

@article{slade2020contrasting,
  title={Contrasting SnTe--NaSbTe2 and SnTe--NaBiTe2 thermoelectric alloys: high performance facilitated by increased cation vacancies and lattice softening},
  author={Slade, Tyler J and Pal, Koushik and Grovogui, Jann A and Bailey, Trevor P and Male, James and Khoury, Jason F and Zhou, Xiuquan and Chung, Duck Young and Snyder, G Jeffrey and Uher, Ctirad and others},
  journal={Journal of the American Chemical Society},
  volume={142},
  number={28},
  pages={12524--12535},
  year={2020},
  publisher={ACS Publications}
}

@article{osterhage2014thermoelectric,
  title={Thermoelectric properties of topological insulator Bi2Te3, Sb2Te3, and Bi2Se3 thin film quantum wells},
  author={Osterhage, Hermann and Gooth, Johannes and Hamdou, Bacel and Gwozdz, Paul and Zierold, Robert and Nielsch, Kornelius},
  journal={Applied Physics Letters},
  volume={105},
  pages={123117},
  number={12},
  year={2014},
  publisher={AIP Publishing}
}

@article{majhi2017emergence,
  title={Emergence of a weak topological insulator from the BixSey family},
  author={Majhi, Kunjalata and Pal, Koushik and Lohani, Himanshu and Banerjee, Abhishek and Mishra, Pramita and Yadav, Anil K and Ganesan, R and Sekhar, BR and Waghmare, Umesh V and Anil Kumar, PS},
  journal={Applied Physics Letters},
  volume={110},
  number={16},
  year={2017},
  pages={162102},
  publisher={AIP Publishing}
}

@article{fu2020topological,
  title={Topological thermoelectrics},
  author={Fu, Chenguang and Sun, Yan and Felser, Claudia},
  journal={Apl Materials},
  volume={8},
  pages={040913},
  number={4},
  year={2020},
  publisher={AIP Publishing}
}

@article{eschbach2017bi1te1,
  title={Bi1Te1 is a dual topological insulator},
  author={Eschbach, Markus and Lanius, Martin and Niu, Chengwang and M{\l}y{\'n}czak, Ewa and Gospodari{\v{c}}, Pika and Kellner, Jens and Sch{\"u}ffelgen, Peter and Gehlmann, Mathias and D{\"o}ring, Sven and Neumann, Elmar and others},
  journal={Nature communications},
  volume={8},
  number={1},
  pages={14976},
  year={2017},
  publisher={Nature Publishing Group UK London}
}

@article{zheng2016transport,
  title={Transport evidence for the three-dimensional Dirac semimetal phase in ZrT e 5},
  author={Zheng, Guolin and Lu, Jianwei and Zhu, Xiangde and Ning, Wei and Han, Yuyan and Zhang, Hongwei and Zhang, Jinglei and Xi, Chuanying and Yang, Jiyong and Du, Haifeng and others},
  journal={Physical Review B},
  volume={93},
  number={11},
  pages={115414},
  year={2016},
  publisher={APS}
}

@article{neupane2014observation,
  title={Observation of a three-dimensional topological Dirac semimetal phase in high-mobility Cd3As2},
  author={Neupane, Madhab and Xu, Su-Yang and Sankar, Raman and Alidoust, Nasser and Bian, Guang and Liu, Chang and Belopolski, Ilya and Chang, Tay-Rong and Jeng, Horng-Tay and Lin, Hsin and others},
  journal={Nature communications},
  volume={5},
  number={1},
  pages={3786},
  year={2014},
  publisher={Nature Publishing Group UK London}
}

@book{mott2012electronic,
  title={Electronic processes in non-crystalline materials},
  author={Mott, Nevill Francis and Davis, Edward A},
  year={2012},
  publisher={OUP Oxford}
}

@article{sun2013effects,
  title={Effects of Ni, Pd, and Pt substitutions on thermoelectric properties of CoSi alloys},
  author={Sun, Hui and Lu, Xu and Morelli, Donald T},
  journal={Journal of electronic materials},
  volume={42},
  number={7},
  pages={1352--1357},
  year={2013},
  publisher={Springer}
}

@article{ren2005effects,
  title={Effects of Ge and B substitution on thermoelectric properties of CoSi},
  author={Ren, WL and Li, CC and Zhang, LT and Ito, K and Wu, JS},
  journal={Journal of alloys and compounds},
  volume={392},
  number={1-2},
  pages={50--54},
  year={2005},
  publisher={Elsevier}
}

@article{skoug2009high,
  title={High thermoelectric power factor in alloys based on CoSi},
  author={Skoug, E and Zhou, C and Pei, Y and Morelli, DT},
  journal={Applied Physics Letters},
  volume={94},
  number={2},
  pages={022115},
  year={2009},
  publisher={AIP Publishing}
}

@inproceedings{sakai2007filling,
  title={Filling dependence of thermoelectric power in transition-metal monosilicides},
  author={Sakai, A and Yotsuhashi, S and Adachi, H and Ishii, F and Onose, Y and Tomioka, Y and Nagaosa, N and Tokura, Y},
  booktitle={2007 26th International Conference on Thermoelectrics},
  pages={256--259},
  year={2007},
  organization={IEEE}
}

@article{vaqueiro2009powder,
  title={A powder neutron diffraction study of the metallic ferromagnet Co3Sn2S2},
  author={Vaqueiro, Paz and Sobany, Gerard G},
  journal={Solid state sciences},
  volume={11},
  number={2},
  pages={513--518},
  year={2009},
  publisher={Elsevier}
}

@article{xiang2020large,
  title={Large transverse thermoelectric figure of merit in a topological Dirac semimetal},
  author={Xiang, JunSen and Hu, SiLe and Lyu, Meng and Zhu, WenLiang and Ma, ChaoYang and Chen, ZiYu and Steglich, Frank and Chen, GenFu and Sun, PeiJie},
  journal={Science China Physics, Mechanics \& Astronomy},
  volume={63},
  number={3},
  pages={237011},
  year={2020},
  publisher={Springer}
}

@article{xie2024topological,
  title={Topological Electronic Transition Contributing to Improved Thermoelectric Performance in p-Type Mg3Sb2- xBix Solid Solutions},
  author={Xie, Sen and Wan, Xiaolin and Wu, Yasong and Li, Chunxia and Yan, Fan and Ouyang, Yujie and Ge, Haoran and Li, Xianda and Liu, Yong and Wang, Rui and others},
  journal={Advanced Materials},
  volume={36},
  number={26},
  pages={2400845},
  year={2024},
  publisher={Wiley Online Library}
}

@article{xie2020band,
  title={Band inversion induced multiple electronic valleys for high thermoelectric performance of SnTe with strong lattice softening},
  author={Xie, Guizhen and Li, Zhi and Luo, Tingting and Bai, Hui and Sun, Jinchang and Xiao, Yu and Zhao, Li-Dong and Wu, Jinsong and Tan, Gangjian and Tang, Xinfeng},
  journal={Nano Energy},
  volume={69},
  pages={104395},
  year={2020},
  publisher={Elsevier}
}

@article{wang2018nontrivial,
  title={Nontrivial thermoelectric behavior in cubic SnSe driven by spin-orbit coupling},
  author={Wang, Hong-Xiang and Mao, Li-Sha and Tan, Xiaojian and Liu, Guo-Qiang and Xu, Jingtao and Shao, Hezhu and Hu, Haoyang and Jiang, Jun},
  journal={Nano Energy},
  volume={51},
  pages={649--655},
  year={2018},
  publisher={Elsevier}
}

@article{xiao2018approaching,
  title={Approaching topological insulating states leads to high thermoelectric performance in n-type PbTe},
  author={Xiao, Yu and Wang, Dongyang and Qin, Bingchao and Wang, Jinfeng and Wang, Guangtao and Zhao, Li-Dong},
  journal={Journal of the American Chemical Society},
  volume={140},
  number={40},
  pages={13097--13102},
  year={2018},
  publisher={ACS Publications}
}

@article{uchida2022thermoelectrics,
  title={Thermoelectrics: from longitudinal to transverse},
  author={Uchida, Ken-ichi and Heremans, Joseph P},
  journal={Joule},
  volume={6},
  number={10},
  pages={2240--2245},
  year={2022},
  publisher={Elsevier}
}

@article{perumal2016high,
  title={High performance thermoelectric materials and devices based on GeTe},
  author={Perumal, Suresh and Roychowdhury, Subhajit and Biswas, Kanishka},
  journal={Journal of Materials Chemistry C},
  volume={4},
  number={32},
  pages={7520--7536},
  year={2016},
  publisher={Royal Society of Chemistry}
}

@article{feng2024transverse,
  title={Transverse thermoelectric materials: Recent advances and challenges},
  author={Feng, Tao and Zhou, Zihan and Wang, Panshuo and Liao, Zilong and Wang, Yupeng and Zhao, Haoran and Zhang, Wenqing and Liu, Weishu},
  journal={Next Energy},
  volume={3},
  pages={100105},
  year={2024},
  publisher={Elsevier}
}

@article{xu2017topological,
  title={Topological insulators for thermoelectrics},
  author={Xu, Ning and Xu, Yong and Zhu, Jia},
  journal={npj Quantum Materials},
  volume={2},
  number={1},
  pages={51},
  year={2017},
  publisher={Nature Publishing Group UK London}
}

@article{kim2017bridge,
  title={The bridge between the materials and devices of thermoelectric power generators},
  author={Kim, Hee Seok and Liu, Weishu and Ren, Zhifeng},
  journal={Energy \& Environmental Science},
  volume={10},
  number={1},
  pages={69--85},
  year={2017},
  publisher={Royal Society of Chemistry}
}

@article{bell2008cooling,
  title={Cooling, heating, generating power, and recovering waste heat with thermoelectric systems},
  author={Bell, Lon E},
  journal={science},
  volume={321},
  number={5895},
  pages={1457--1461},
  year={2008},
  publisher={American Association for the Advancement of Science}
}

@article{snyder2008complex,
  title={Complex thermoelectric materials},
  author={Snyder, G Jeffrey and Toberer, Eric S},
  journal={Nature materials},
  volume={7},
  number={2},
  pages={105--114},
  year={2008},
  publisher={Nature Publishing Group UK London}
}

@article{pan2024topological,
  title={Topological materials for high performance transverse thermoelectrics},
  author={Pan, Yu and He, Bin and Wang, Honghui and Felser, Claudia},
  journal={Next Energy},
  volume={2},
  pages={100103},
  year={2024},
  publisher={Elsevier}
}

@article{adachi2025fundamentals,
  title={Fundamentals and advances in transverse thermoelectrics},
  author={Adachi, Hiroto and Ando, Fuyuki and Hirai, Takamasa and Modak, Rajkumar and Grayson, Matthew A and Uchida, Ken-ichi},
  journal={Applied Physics Express},
  volume={18},
  number={9},
  pages={090101},
  year={2025},
  publisher={IOP Publishing}
}

@article{toriyama2024topological,
  title={Are topological insulators promising thermoelectrics?},
  author={Toriyama, Michael Y and Snyder, G Jeffrey},
  journal={Materials Horizons},
  volume={11},
  number={5},
  pages={1188--1198},
  year={2024},
  publisher={Royal Society of Chemistry}
}

@article{antonius2016temperature,
  title={Temperature-induced topological phase transitions: Promoted versus suppressed nontrivial topology},
  author={Antonius, Gabriel and Louie, Steven G},
  journal={Physical review letters},
  volume={117},
  number={24},
  pages={246401},
  year={2016},
  publisher={APS}
}

@article{brousseau2020temperature,
  title={Temperature dependence of the topological phase transition of BiTeI from first principles},
  author={Brousseau-Couture, V{\'e}ronique and Antonius, Gabriel and C{\^o}t{\'e}, Michel},
  journal={Physical Review Research},
  volume={2},
  number={2},
  pages={023185},
  year={2020},
  publisher={APS}
}

@article{querales2020towards,
  title={Towards temperature-induced topological phase transition in SnTe: A first-principles study},
  author={Querales-Flores, Jos{\'e} D and Aguado-Puente, Pablo and Dangi{\'c}, {\DJ}or{\dj}e and Cao, Jiang and Chudzinski, Piotr and Todorov, Tchavdar N and Gr{\"u}ning, Myrta and Fahy, Stephen and Savi{\'c}, Ivana},
  journal={Physical Review B},
  volume={101},
  number={23},
  pages={235206},
  year={2020},
  publisher={APS}
}

@article{song2025field,
  title={A Field Guide to Materials with Axis-Dependent Conduction Polarity},
  author={Song, Tianze and Goldberger, Joshua E},
  journal={Chemistry of Materials},
  volume={37},
  number={19},
  pages={7518--7533},
  year={2025},
  publisher={ACS Publications}
}

@article{wang2020anisotropic,
  title={Anisotropic scattering in the goniopolar metal Na Sn 2 As 2},
  author={Wang, Yaxian and Narang, Prineha},
  journal={Physical Review B},
  volume={102},
  number={12},
  pages={125122},
  year={2020},
  publisher={APS}
}

@article{wang2020chemical,
  title={The chemical design principles for axis-dependent conduction polarity},
  author={Wang, Yaxian and Koster, Karl G and Ochs, Andrew M and Scudder, Michael R and Heremans, Joseph P and Windl, Wolfgang and Goldberger, Joshua E},
  journal={Journal of the American Chemical Society},
  volume={142},
  number={6},
  pages={2812--2822},
  year={2020},
  publisher={ACS Publications}
}

@article{zhao2014ultralow,
  title={Ultralow thermal conductivity and high thermoelectric figure of merit in SnSe crystals},
  author={Zhao, Li-Dong and Lo, Shih-Han and Zhang, Yongsheng and Sun, Hui and Tan, Gangjian and Uher, Ctirad and Wolverton, Christopher and Dravid, Vinayak P and Kanatzidis, Mercouri G},
  journal={nature},
  volume={508},
  number={7496},
  pages={373--377},
  year={2014},
  publisher={Nature Publishing Group UK London}
}

@article{sun2012atomically,
  title={Atomically thick bismuth selenide freestanding single layers achieving enhanced thermoelectric energy harvesting},
  author={Sun, Yongfu and Cheng, Hao and Gao, Shan and Liu, Qinghua and Sun, Zhihu and Xiao, Chong and Wu, Changzheng and Wei, Shiqiang and Xie, Yi},
  journal={Journal of the American Chemical Society},
  volume={134},
  number={50},
  pages={20294--20297},
  year={2012},
  publisher={ACS Publications}
}

@article{FIESCHI1954259,
title = {Thermodynamical theory of galvanomagnetic and thermomagnetic phenomena III: Explicit expressions for the measurable effects in isotropic metals},
journal = {Physica},
volume = {20},
number = {1},
pages = {259-273},
year = {1954},
issn = {0031-8914},
doi = {https://doi.org/10.1016/S0031-8914(54)80040-1},
url = {https://www.sciencedirect.com/science/article/pii/S0031891454800401},
author = {R. Fieschi and S.R. {De Groot} and P. Mazur}
}

@article{price1955cxxxv,
  title={CXXXV. Ambipolar thermodiffusion of electrons and holes in semiconductors},
  author={Price, PJ},
  journal={The London, Edinburgh, and Dublin Philosophical Magazine and Journal of Science},
  volume={46},
  number={382},
  pages={1252--1260},
  year={1955},
  publisher={Taylor \& Francis}
}

@article{goldsmid1956thermal,
  title={The thermal conductivity of bismuth telluride},
  author={Goldsmid, HJ},
  journal={Proceedings of the Physical Society. Section B},
  volume={69},
  number={2},
  pages={203--209},
  year={1956}
}

@article{paranjape1960theory,
  title={Theory of the Ettingshausen effect in semiconductors},
  author={Paranjape, BV and Levinger, JS},
  journal={Physical Review},
  volume={120},
  number={2},
  pages={437},
  year={1960},
  publisher={APS}
}

@article{hall1879new,
  title={On a new action of the magnet on electric currents},
  author={Hall, Edwin H and others},
  journal={American Journal of Mathematics},
  volume={2},
  number={3},
  pages={287--292},
  year={1879}
}

@book{peltier1834nouvelles,
  title={Nouvelles exp{\'e}riences sur la caloricit{\'e} des courans {\'e}lectriques},
  author={Peltier, Jean Charles Athanase},
  year={1834},
  publisher = {Crochard}
}

@article{wang2020bi8se9,
  title={Bi8se9: Effective reduction of bipolar diffusion via increasing band gap},
  author={Wang, Rui and Jia, Fei and Li, Xuan and Chen, Ling and Wu, Li-Ming},
  journal={Crystal Growth \& Design},
  volume={20},
  number={5},
  pages={3555--3560},
  year={2020},
  publisher={ACS Publications}
}

@article{goupil2016thermodynamics,
  title={Thermodynamics and thermoelectricity},
  author={Goupil, Christophe and Ouerdane, Henni and Zabrocki, Knud and Seifert, Wolfgang and Hinsche, Nicki F and M{\"u}ller, Eckhard},
  journal={Continuum Theory and Modeling of Thermoelectric Elements},
  pages={1--74},
  year={2016},
  publisher={Wiley Online Library}
}

@article{laughlin1981quantized,
  title={Quantized Hall conductivity in two dimensions},
  author={Laughlin, Robert B},
  journal={Physical Review B},
  volume={23},
  number={10},
  pages={5632},
  year={1981},
  publisher={APS}
}

@article{dyakonov1971current,
  title={Current-induced spin orientation of electrons in semiconductors},
  author={Dyakonov, Mikhail I and Perel, VI},
  journal={Physics Letters A},
  volume={35},
  number={6},
  pages={459--460},
  year={1971},
  publisher={Elsevier}
}

@article{karplus1954hall,
  title={Hall effect in ferromagnetics},
  author={Karplus, Robert and Luttinger, JM},
  journal={Physical Review},
  volume={95},
  number={5},
  pages={1154},
  year={1954},
  publisher={APS}
}

@article{shilpa2023systematic,
  title={A systematic review of thermoelectric peltier devices: Applications and limitations},
  author={Shilpa, M and Raheman, Md and Aabid, Abdul and Baig, Muneer and Veeresha, R and Kudva, Nagesh},
  journal={Fluid Dynamics \& Materials Processing},
  volume={19},
  number={1},
  pages={187},
  year={2023},
  publisher={Tech Science Press}
}

@article{razeghi2025giant,
  title={Giant Anomalous Ettingshausen Effect and Hybrid Longitudinal-Transverse Thermoelectric Cooling in a Nanoscale Magnetic Weyl Semimetal},
  author={Razeghi, Mohammadali and Spiece, Jean and Fonck, Valentin and Zhang, Yao and Rohde, Michael and Joris, Rikkie and Dobson, Phillip S and Weaver, Jonathan MR and Pereira, Lino MC and Granville, Simon and others},
  journal={ACS nano},
  volume={19},
  number={46},
  pages={39725--39734},
  year={2025},
  publisher={ACS Publications}
}

@article{harman1964oriented,
  title={Oriented single-crystal Bismuth Nernst-Ettingshausen refrigerators},
  author={Harman, TC and Honig, JM and Fischler, S and Paladino, AE and Button, M Jane},
  journal={Applied Physics Letters},
  volume={4},
  number={4},
  pages={77--79},
  year={1964},
  publisher={American Institute of Physics}
}

@article{leduc1888modifications,
  title={Modifications de la conductibilit{\'e} calorifique du bismuth dans un champ magn{\'e}tique},
  author={Leduc, A},
  journal={J. Phys. Theor. Appl.},
  volume={7},
  number={1},
  pages={519--525},
  year={1888}
}

@article{stephan1972righi,
  title={Righi-Leduc Effect in Indium},
  author={Stephan, CH and Maxfield, BW},
  journal={Physical Review B},
  volume={6},
  number={8},
  pages={2893},
  year={1972},
  publisher={APS}
}

@article{whitsett1961righi,
  title={Righi-Leduc Effect in Mercuric Selenide},
  author={Whitsett, Charles R},
  journal={Journal of Applied Physics},
  volume={32},
  number={10},
  pages={2257--2260},
  year={1961},
  publisher={American Institute of Physics}
}

@article{sichel1971righi,
  title={The Righi-Leduc effect in Pb-In and In-Pb alloys},
  author={Sichel, EK and Serin, B},
  journal={Physics Letters A},
  volume={37},
  number={2},
  pages={123--124},
  year={1971},
  publisher={Elsevier}
}

@article{madon2016anomalous,
  title={Anomalous and planar Righi-Leduc effects in Ni 80 Fe 20 ferromagnets},
  author={Madon, B and Pham, Do Ch and Wegrowe, J-E and Lacour, Daniel and Hehn, M and Polewczyk, V and Anane, A and Cros, V},
  journal={Physical Review B},
  volume={94},
  number={14},
  pages={144423},
  year={2016},
  publisher={APS}
}

@article{zhou2020identification,
  title={Identification of spin effects in the anomalous righi--leduc effect in ferromagnetic metals},
  author={Zhou, Da-Kun and Xu, Qing-Lian and Yu, Xiao-Qin and Zhu, Zhen-Gang and Su, Gang},
  journal={Scientific Reports},
  volume={10},
  number={1},
  pages={11732},
  year={2020},
  publisher={Nature Publishing Group UK London}
}

@article{zhou2022anomalous,
  title={Anomalous thermal Hall effect and anomalous Nernst effect of CsV 3 Sb 5},
  author={Zhou, Xuebo and Liu, Hongxiong and Wu, Wei and Jiang, Kun and Shi, Youguo and Li, Zheng and Sui, Yu and Hu, Jiangping and Luo, Jianlin},
  journal={Physical Review B},
  volume={105},
  number={20},
  pages={205104},
  year={2022},
  publisher={APS}
}

@article{shi2025toward,
  title={Toward efficient thermoelectric materials and devices: advances, challenges, and opportunities},
  author={Shi, Xiao-Lei and Li, Nan-Hai and Li, Meng and Chen, Zhi-Gang},
  journal={Chemical Reviews},
  volume={125},
  number={16},
  pages={7525--7724},
  year={2025},
  publisher={ACS Publications}
}

@article{ahmad2024largely,
  title={Largely enhanced thermoelectric performance in p-type Bi 2 Te 3-based materials through entropy engineering},
  author={Ahmad, Abid and Zhu, Bin and Wang, Zhongbin and Gui, Zhigang and Wang, Wu and Wang, Tiancheng and Yu, Yong and Huang, Li and He, Jiaqing},
  journal={Energy \& Environmental Science},
  volume={17},
  number={2},
  pages={695--703},
  year={2024},
  publisher={Royal Society of Chemistry}
}

@article{mao2021thermoelectric,
  title={Thermoelectric cooling materials},
  author={Mao, Jun and Chen, Gang and Ren, Zhifeng},
  journal={Nature Materials},
  volume={20},
  number={4},
  pages={454--461},
  year={2021},
  publisher={Nature Publishing Group UK London}
}

@article{qian2025thermoelectric,
  title={Thermoelectric cooling goes cooler},
  author={Qian, Sheng and Fu, Chenguang and Zhu, Tiejun},
  journal={National Science Open},
  volume={4},
  number={3},
  pages={20250009},
  year={2025},
  publisher={China Science Publishing \& Media Ltd. and EDP Sciences}
}

@article{uchida2024hybrid,
  title={Hybrid Transverse Magneto-Thermoelectric Cooling in Artificially Tilted Multilayers},
  author={Uchida, Ken-ichi and Hirai, Takamasa and Ando, Fuyuki and Sepehri-Amin, Hossein},
  journal={Advanced Energy Materials},
  volume={14},
  number={3},
  pages={2302375},
  year={2024},
  publisher={Wiley Online Library}
}

@article{hirai2024hybridizing,
  title={Hybridizing anomalous Nernst effect in artificially tilted multilayer based on magnetic topological material},
  author={Hirai, Takamasa and Ando, Fuyuki and Sepehri-Amin, Hossein and Uchida, Ken-ichi},
  journal={Nature Communications},
  volume={15},
  number={1},
  pages={9643},
  year={2024},
  publisher={Nature Publishing Group UK London}
}

@article{ando2025multifunctional,
  title={Multifunctional composite magnet realizing record-high transverse thermoelectric generation},
  author={Ando, Fuyuki and Hirai, Takamasa and Alasli, Abdulkareem and Sepehri-Amin, Hossein and Iwasaki, Yutaka and Nagano, Hosei and Uchida, Ken-ichi},
  journal={Energy \& Environmental Science},
  volume={18},
  number={9},
  pages={4068--4079},
  year={2025},
  publisher={Royal Society of Chemistry}
}

@article{murata2024zero,
  title={Zero-magnetic-field operation of ordinary-Nernst-effect-based transverse thermoelectric module using embedded permanent magnets},
  author={Murata, Masayuki and Hirai, Takamasa and Seki, Takeshi and Uchida, Ken-ichi},
  journal={Applied Physics Letters},
  volume={124},
  number={19},
  pages={193901},
  year={2024},
  publisher={AIP Publishing}
}

@article{chen2025design,
  title={Design of anomalous Nernst thermoelectric generators for giant power output},
  author={Chen, Mengzhao and Qian, Sheng and Gao, Ziheng and Liu, Shuo and Han, Shen and Zhu, Tiejun and Sun, Yan and Fu, Chenguang},
  journal={The Innovation},
  volume={6},
  pages={100995},
  number={9},
  year={2025},
  publisher={Elsevier}
}

@article{li2021monomaterial,
  title={A monomaterial Nernst thermopile with hermaphroditic legs},
  author={Li, Xiaokang and Zhu, Zengwei and Behnia, Kamran},
  journal={Advanced Materials},
  volume={33},
  number={20},
  pages={2100751},
  year={2021},
  publisher={Wiley Online Library}
}

@article{de2025miniaturized,
  title={A Miniaturized Transverse Thermoelectric Device for Efficient Energy Harvesting and its Capacitive Storage},
  author={De Sousa, J Alejandro and S{\'a}nchez-Villegas, Emma and Avci, Can Onur},
  journal={Advanced Materials Technologies},
  volume={10},
  number={22},
  pages={e00770},
  year={2025},
  publisher={Wiley Online Library}
}

@article{odagiri2024bipolar,
  title={Bipolar transverse thermopower and low thermal conductivity for an anomalous Nernst-type heat flux sensor in GdCo alloys},
  author={Odagiri, Miho and Imaeda, Hiroto and Yagmur, Ahmet and Kurokawa, Yuichiro and Sumi, Satoshi and Awano, Hiroyuki and Tanabe, Kenji},
  journal={Scientific reports},
  volume={14},
  number={1},
  pages={17205},
  year={2024},
  publisher={Nature Publishing Group UK London}
}

@article{sakuraba2016potential,
  title={Potential of thermoelectric power generation using anomalous Nernst effect in magnetic materials},
  author={Sakuraba, Yuya},
  journal={Scripta Materialia},
  volume={111},
  pages={29--32},
  year={2016},
  publisher={Elsevier}
}

@article{yang2017scalable,
  title={Scalable Nernst thermoelectric power using a coiled galfenol wire},
  author={Yang, Zihao and Codecido, Emilio A and Marquez, Jason and Zheng, Yuanhua and Heremans, Joseph P and Myers, Roberto C},
  journal={AIP Advances},
  volume={7},
  number={9},
  pages={095017},
  year={2017},
  publisher={AIP Publishing}
}

@article{harman1963theory,
  title={Theory of the infinite stage Nernst-Ettingshausen refrigerator},
  author={Harman, TC},
  journal={Advanced energy conversion},
  volume={3},
  number={4},
  pages={667--676},
  year={1963},
  publisher={Elsevier}
}

@article{polash2021infinite,
  title={Infinite-stage Nernst-Ettingshausen cryocooler for practical applications},
  author={Polash, M Mobarak Hossain and Vashaee, Daryoosh},
  journal={Physical Review Applied},
  volume={15},
  number={1},
  pages={014011},
  year={2021},
  publisher={APS}
}

@article{chiba2022single,
  title={Single-material-based anomalous Nernst thermopile driven by solar heating and radiative cooling},
  author={Chiba, Naoki and Ishii, Satoshi and Modak, Rajkumar and Uchida, Ken-ichi},
  journal={Applied Physics Express},
  volume={15},
  number={9},
  pages={093001},
  year={2022},
  publisher={IOP Publishing}
}

@article{tanaka2023roll,
  title={Roll-to-Roll Printing of Anomalous Nernst Thermopile for Direct Sensing of Perpendicular Heat Flux},
  author={Tanaka, Hirokazu and Higo, Tomoya and Uesugi, Ryota and Yamagata, Kazuto and Nakanishi, Yosuke and Machinaga, Hironobu and Nakatsuji, Satoru},
  journal={Advanced Materials},
  volume={35},
  number={38},
  pages={2303416},
  year={2023},
  publisher={Wiley Online Library}
}

@article{lou2026recent,
  title={Recent advances and challenges in thermoelectrics toward near-room-temperature and high-temperature applications},
  author={Lou, Qianhui and Gong, Jiayan and Zang, Zizheng and Qian, Sheng and Liu, Yu and Zhang, Yu and Wang, Yuechu and Han, Shen and Fu, Chenguang and Zhu, Tiejun},
  journal={Chemical Society Reviews},
  year={2026},
  volume={55},
  pages={ 2265--2324 },
  publisher={Royal Society of Chemistry}
}

@article{qu2024synergistic,
  title={Synergistic suppression of bipolar effect and lattice thermal conductivity leading to high average figure of merit in Bi0. 4Sb1. 6Te3 materials through alloying with AgSbTe2},
  author={Qu, Xiang and Chen, Xiangbin and Yu, Tian and Qi, Ning and Chen, Zhiquan},
  journal={ACS Applied Materials \& Interfaces},
  volume={16},
  number={45},
  pages={62347--62357},
  year={2024},
  publisher={ACS Publications}
}

@article{pan2021thermoelectric,
  title={Thermoelectric properties of novel semimetals: a case study of YbMnSb2},
  author={Pan, Yu and Fan, Feng-Ren and Hong, Xiaochen and He, Bin and Le, Congcong and Schnelle, Walter and He, Yangkun and Imasato, Kazuki and Borrmann, Horst and Hess, Christian and others},
  journal={Advanced Materials},
  volume={33},
  number={7},
  pages={2003168},
  year={2021},
  publisher={Wiley Online Library}
}

@article{chen2021leveraging,
  title={Leveraging bipolar effect to enhance transverse thermoelectricity in semimetal Mg2Pb for cryogenic heat pumping},
  author={Chen, Zhiwei and Zhang, Xinyue and Ren, Jie and Zeng, Zezhu and Chen, Yue and He, Jian and Chen, Lidong and Pei, Yanzhong},
  journal={Nature Communications},
  volume={12},
  number={1},
  pages={3837},
  year={2021},
  publisher={Nature Publishing Group UK London}
}

@article{hu2026fermiology,
  title={Fermiology-driven goniopolar transverse thermoelectricity in kagome metals},
  author={Hu, Haihua and Ju, Yiwei and Cheng, Erjian and Feng, Xiaolong and Sun, Fei and Lou, Rui and Schnelle, Walter and Koban, Ralf and Wang, Honghui and Fedorov, Alexander and others},
  journal={Nature Materials},
  pages={1--8},
  year={2026},
  publisher={Nature Publishing Group UK London}
}

@article{qian2025anomalous,
  title={Anomalous Nernst Effect and Transverse Thermoelectric Conversion in Co2MnAl-Based Heusler Magnets},
  author={Qian, Sheng and Chen, Mengzhao and Zhu, Tiejun and Fu, Chenguang},
  journal={Annalen der Physik},
  volume={537},
  number={11},
  pages={e00336},
  year={2025},
  publisher={Wiley Online Library}
}

@article{kubler2016weyl,
  title={Weyl points in the ferromagnetic Heusler compound Co2MnAl},
  author={K{\"u}bler, J and Felser, C},
  journal={Europhysics Letters},
  volume={114},
  number={4},
  pages={47005},
  year={2016},
  publisher={EDP Sciences, IOP Publishing and Societ{\`a} Italiana di Fisica}
}

@article{hsiung2014surface,
  title={Surface-dominated transport and enhanced thermoelectric figure of merit in topological insulator Bi1. 5Sb0. 5Te1. 7Se1. 3},
  author={Hsiung, Te-Chih and Mou, Chung-Yu and Lee, Ting-Kuo and Chen, Yang-Yuan},
  journal={Nanoscale},
  volume={7},
  number={2},
  pages={518--523},
  year={2014},
  publisher={The Royal Society of Chemistry}
}

@article{mishra2024majorana,
  title={Majorana thermoelectrics and refrigeration},
  author={Mishra, Sachiraj and Das, Ritesh and Benjamin, Colin},
  journal={Journal of Applied Physics},
  volume={136},
  number={23},
  pages={234401},
  year={2024},
  publisher={AIP Publishing}
}

@article{uusnakki2026initial,
  title={Initial demonstration of a quantum heat engine based on dissipation-engineered superconducting circuits},
  author={Uusn{\"a}kki, Tuomas and M{\"o}rstedt, Timm and Teixeira, Wallace and Rasola, Miika and M{\"o}tt{\"o}nen, Mikko},
  pages={6054},
  journal={Nature Communications},
  year={2026},
  publisher={Nature Publishing Group UK London}
}

@article{yang2026axis,
  title={Axis-Dependent Conduction Polarity: Design Principles and High-Throughput Discovery of Transverse Thermoelectrics},
  author={Yang, Zan and He, Xinyi and Usui, Hidetomo and Kamiya, Toshio and Katase, Takayoshi},
  journal={Journal of the American Chemical Society},
  volume={148},
  number={6},
  pages={6704--6715},
  year={2026},
  publisher={ACS Publications}
}

@article{manna2018colossal,
  title={From colossal to zero: controlling the anomalous Hall effect in magnetic Heusler compounds via Berry curvature design},
  author={Manna, Kaustuv and Muechler, Lukas and Kao, Ting-Hui and Stinshoff, Rolf and Zhang, Yang and Gooth, Johannes and Kumar, Nitesh and Kreiner, Guido and Koepernik, Klaus and Car, Roberto and others},
  journal={Physical Review X},
  volume={8},
  number={4},
  pages={041045},
  year={2018},
  publisher={APS}
}

@article{wang2012enhanced,
  title={Enhanced thermoelectric properties of Bi2 (Te1- x Se x) 3-based compounds as n-type legs for low-temperature power generation},
  author={Wang, Shanyu and Tan, Gangjian and Xie, Wenjie and Zheng, Gang and Li, Han and Yang, Jihui and Tang, Xinfeng},
  journal={Journal of Materials Chemistry},
  volume={22},
  number={39},
  pages={20943--20951},
  year={2012},
  publisher={The Royal Society of Chemistry}
}

@article{zheng2014high,
  title={High-temperature mechanical and thermoelectric properties of p-type Bi0. 5Sb1. 5Te3 commercial zone melting ingots},
  author={Zheng, Yun and Xie, Hongyao and Shu, Shengcheng and Yan, Yonggao and Li, Han and Tang, Xinfeng},
  journal={Journal of electronic materials},
  volume={43},
  number={6},
  pages={2017--2022},
  year={2014},
  publisher={Springer}
}

@article{roychowdhury2015tailoring,
  title={Tailoring of electronic structure and thermoelectric properties of a topological crystalline insulator by chemical doping},
  author={Roychowdhury, Subhajit and Shenoy, U Sandhya and Waghmare, Umesh V and Biswas, Kanishka},
  journal={Angewandte Chemie International Edition},
  volume={54},
  number={50},
  pages={15241--15245},
  year={2015},
  publisher={Wiley Online Library}
}

@article{roychowdhury2016effect,
  title={Effect of potassium doping on electronic structure and thermoelectric properties of topological crystalline insulator},
  author={Roychowdhury, Subhajit and Sandhya Shenoy, U and Waghmare, Umesh V and Biswas, Kanishka},
  journal={Applied Physics Letters},
  volume={108},
  pages={193901},
  number={19},
  year={2016},
  publisher={AIP Publishing}
}

@article{muzaffar2020antisite,
  title={Antisite Defect-Enhanced Thermoelectric Performance of Topological Crystalline Insulators},
  author={Muzaffar, Muhammad Usman and Zhang, Shunhong and Cui, Ping and He, Jiaqing and Zhang, Zhenyu},
  journal={Advanced Functional Materials},
  volume={30},
  number={35},
  pages={2003162},
  year={2020},
  publisher={Wiley Online Library}
}

@article{deng2024ordered,
  title={Ordered grain boundary reconstruction induces high-efficiency thermoelectric power generation in SnTe},
  author={Deng, Qian and Zhang, Fujie and Yang, Xiaoyu and Li, Ruiheng and Xia, Chengliang and Nan, Pengfei and Chen, Yue and Ge, Binghui and Ang, Ran and He, Jiaqing},
  journal={Energy \& Environmental Science},
  volume={17},
  number={24},
  pages={9467--9478},
  year={2024},
  publisher={The Royal Society of Chemistry}
}

@article{meng2017grain,
  title={Grain boundary engineering for achieving high thermoelectric performance in n-type skutterudites},
  author={Meng, Xianfu and Liu, Zihang and Cui, Bo and Qin, Dandan and Geng, Huiyuan and Cai, Wei and Fu, Liangwei and He, Jiaqing and Ren, Zhifeng and Sui, Jiehe},
  journal={Advanced Energy Materials},
  volume={7},
  number={13},
  pages={1602582},
  year={2017},
  publisher={Wiley Online Library}
}

@article{lin2020expression,
  title={Expression of interfacial Seebeck coefficient through grain boundary engineering with multi-layer graphene nanoplatelets},
  author={Lin, Yue and Wood, Maxwell and Imasato, Kazuki and Kuo, Jimmy Jiahong and Lam, David and Mortazavi, Anna N and Slade, Tyler J and Hodge, Stephen A and Xi, Kai and Kanatzidis, Mercouri G and others},
  journal={Energy \& Environmental Science},
  volume={13},
  number={11},
  pages={4114--4121},
  year={2020},
  publisher={The Royal Society of Chemistry}
}

@article{kim2024selective,
  title={Selective charge carrier transport and bipolar conduction in an inorganic/organic bulk-phase composite: Optimization for low-temperature thermoelectric performance},
  author={Kim, Cham and Kim, Taewook and Cho, Jaehun},
  journal={ACS applied materials \& interfaces},
  volume={16},
  number={4},
  pages={5036--5049},
  year={2024},
  publisher={ACS Publications}
}

@article{matsushita2017thermoelectric,
  title={Thermoelectric properties of 3D topological insulator: Direct observation of topological surface and its gap opened states},
  author={Matsushita, Stephane Yu and Huynh, Khuong Kim and Yoshino, Harukazu and Tu, Ngoc Han and Tanabe, Yoichi and Tanigaki, Katsumi},
  journal={Physical Review Materials},
  volume={1},
  number={5},
  pages={054202},
  year={2017},
  publisher={APS}
}

@article{zhai2018tunable,
  title={Tunable optimum temperature range of high-performance zone melted bismuth-telluride-based solid solutions},
  author={Zhai, Renshuang and Wu, Yehao and Zhu, Tie-Jun and Zhao, Xin-Bing},
  journal={Crystal Growth \& Design},
  volume={18},
  number={8},
  pages={4646--4652},
  year={2018},
  publisher={ACS Publications}
}

@article{hu2015tuning,
  title={Tuning multiscale microstructures to enhance thermoelectric performance of n-type Bismuth-Telluride-based solid solutions},
  author={Hu, Lipeng and Wu, Haijun and Zhu, Tiejun and Fu, Chenguang and He, Jiaqing and Ying, Pingjun and Zhao, Xinbing},
  journal={Advanced Energy Materials},
  volume={5},
  number={17},
  pages={1500411},
  year={2015},
  publisher={Wiley Online Library}
}

@article{hu2018synergistic,
  title={Synergistic compositional--mechanical--thermal effects leading to a record high zT in n-type V2VI3 alloys through progressive hot deformation},
  author={Hu, Lipeng and Zhang, Yang and Wu, Haijun and Liu, Yamei and Li, Junqin and He, Jian and Ao, Weiqin and Liu, Fusheng and Pennycook, Stephen John and Zeng, Xierong},
  journal={Advanced Functional Materials},
  volume={28},
  number={35},
  pages={1803617},
  year={2018},
  publisher={Wiley Online Library}
}

@article{liu2025lattice,
  title={Lattice plainification and band engineering lead to high thermoelectric cooling and power generation in n-type Bi2Te3 with mass production},
  author={Liu, Dongrui and Bai, Shulin and Wen, Yi and Peng, Jiayi and Liu, Shibo and Shi, Haonan and Li, Yichen and Hong, Tao and Liang, Huiqiang and Qin, Yongxin and others},
  journal={National Science Review},
  volume={12},
  number={2},
  pages={nwae448},
  year={2025},
  publisher={Oxford University Press}
}

@article{peng2025lattice,
  title={Lattice Plainification and Intercalation Advances Power Generation and Thermoelectric Cooling in n-type Bi2 (Te, Se) 3},
  author={Peng, Jiayi and Liu, Dongrui and Bai, Shulin and Wen, Yi and Liang, Huiqiang and Su, Lizhong and Qian, Xin and Wang, Dongyang and Gao, Xiang and Ding, Zhihai and others},
  journal={Advanced Energy Materials},
  volume={15},
  number={18},
  pages={2404653},
  year={2025},
  publisher={Wiley Online Library}
}

@article{guo2026sharpened,
  title={Sharpened Dirac Cone and Elevated Carrier Mobility in Bi4Te3 Engineered by a High [Bi2] Bilayer Content for Thermoelectrics},
  author={Guo, Chao and Jia, Fei and Wu, Yu-Qian and Liu, Xin-Yi and Liu, Yi-Meng and Chen, Ling and Wu, Li-Ming},
  journal={Angewandte Chemie International Edition},
  volume={65},
  number={11},
  pages={e24705},
  year={2026},
  publisher={Wiley Online Library}
}

@article{sakai2025berry,
  title={Berry curvature driven transverse thermoelectric generation in topological magnets},
  author={Sakai, Akito and Nakatsuji, Satoru},
  journal={Science and Technology of Advanced Materials},
  volume={26},
  number={1},
  pages={2554047},
  year={2025},
  publisher={Taylor \& Francis}
}

@article{wang2023large,
  title={Large magneto-transverse and longitudinal thermoelectric effects in the magnetic Weyl semimetal TbPtBi},
  author={Wang, Honghui and Zhou, Zizhen and Ying, Jianjun and Xiang, Ziji and Wang, Rui and Wang, Aifeng and Chai, Yisheng and He, Mingquan and Lu, Xu and Han, Guang and others},
  journal={Advanced Materials},
  volume={35},
  number={2},
  pages={2206941},
  year={2023},
  publisher={Wiley Online Library}
}

@article{li2025thermomagnetic,
  title={Thermomagnetic properties in topological materials},
  author={Li, Peng and Jiang, Binbin and Qiu, Pengfei and Shi, Xun},
  journal={Advanced Functional Materials},
  volume={35},
  number={42},
  pages={2506631},
  year={2025},
  publisher={Wiley Online Library}
}

@article{wang2026observation,
  title={Observation of the amplified transverse thermoelectric signals by reduced-dimensionality transport anisotropy},
  author={Wang, Honghui and He, Bin and Feng, Xiaolong and Hu, Haihua and Koban, Ralf and Schnelle, Walter and Cheng, Erjian and Pan, Yu and Zhou, Xiaoyuan and Felser, Claudia},
  journal={Nature Communications},
  volume={17},
  number={1},
  pages={9607},
  year={2026},
  publisher={Nature Publishing Group UK London}
}

@article{greenaway1965band,
  title={Band structure of bismuth telluride, bismuth selenide and their respective alloys},
  author={Greenaway, David Leslie and Harbeke, G{\"u}nther},
  journal={Journal of Physics and Chemistry of Solids},
  volume={26},
  number={10},
  pages={1585--1604},
  year={1965},
  publisher={Elsevier}
}

@article{jandl1994thermogalvanomagnetic,
  title={Thermogalvanomagnetic properties of Sn-doped Bi95Sb5 and its application for solid state cooling},
  author={Jandl, P and Birkholz, U},
  journal={Journal of applied physics},
  volume={76},
  number={11},
  pages={7351--7366},
  year={1994},
  publisher={American Institute of Physics}
}

@inproceedings{nakamura1997transport,
  title={Transport coefficients of InSb in a strong magnetic field},
  author={Nakamura, Hiroaki and Ikeda, Kazuaki and Yamaguchi, Satarou},
  booktitle={XVI ICT'97. Proceedings ICT'97. 16th International Conference on Thermoelectrics (Cat. No. 97TH8291)},
  pages={142--146},
  year={1997},
  organization={IEEE}
}

@article{jovovic2008low,
  title={Low temperature thermal, thermoelectric, and thermomagnetic transport in indium rich Pb1- xSnxTe alloys},
  author={Jovovic, Vladimir and Thiagarajan, Suraj Joottu and Heremans, Joseph P and Komissarova, Tatiana and Khokhlov, Dmitri and Nicorici, A},
  journal={Journal of Applied Physics},
  volume={103},
  number={5},
  year={2008},
  pages={053710},
  publisher={AIP Publishing}
}

@article{pathak2025impact,
  title={Impact of pressure on metavalent bonding in BiTe influencing electronic topological transitions},
  author={Pathak, Riddhimoy and Joseph, Anjana and Dutta, Prabir and Joseph, Boby and Duhan, Jyoti and Chandra, Sushmita and Narayana, Chandrabhas and Pal, Koushik and Biswas, Kanishka},
  journal={Angewandte Chemie},
  volume={137},
  number={14},
  pages={e202422652},
  year={2025},
  publisher={Wiley Online Library}
}
\bibliographystyle{rsc} 

\end{document}